\documentclass[12pt]{iopart}

\usepackage{iopams}
\usepackage[dvips]{graphicx}
\usepackage{mathrsfs}
\usepackage{cite}

\def\openone{\leavevmode\hbox{\small1\kern-4.2pt\normalsize1}}

\newcommand{\bea}{\begin{eqnarray}}
\newcommand{\ea}{\end{eqnarray}}

\begin{document}

\title[]{Flux-charge duality and topological quantum phase fluctuations in quasi-one-dimensional superconductors}

\author{Andrew J. Kerman}

\address{MIT Lincoln Laboratory, Lexington, MA, USA}

\ead{ajkerman@ll.mit.edu}

\begin{abstract}
It has long been thought that macroscopic phase coherence breaks down in effectively lower-dimensional superconducting systems even at zero temperature due to enhanced topological quantum phase fluctuations. In quasi-1D wires, these fluctuations are described in terms of ``quantum phase-slip" (QPS): tunneling of the superconducting order parameter for the wire between states differing by $\pm$2$\pi$ in their relative phase between the wire's ends. Over the last several decades, many deviations from conventional bulk superconducting behavior have been observed in ultra-narrow superconducting nanowires, some of which have been identified with QPS. While at least some of the observations are consistent with existing theories for QPS, other observations in many cases point to contradictory conclusions or cannot be explained by these theories. Hence, a unified understanding of the nature of QPS, and its relationship to the various observations has yet to be achieved. In this paper we present a new model for QPS which takes as its starting point an idea originally postulated by Mooij and Nazarov [Nature Physics {\bf 2}, 169 (2006)]: that \textit{flux-charge duality}, a classical symmetry of Maxwell's equations, can be used to relate QPS to the well-known Josephson tunneling of Cooper pairs. Our model provides an alternative, and qualitatively different, conceptual basis for QPS and the phenomena which arise from it in experiments, and it appears to permit for the first time a unified understanding of observations across several different types of experiments and materials systems.
\end{abstract}

\vspace{2pc}

\maketitle

%
\section{Introduction}

Topologically-charged fluctuations in field theories appear in many areas of physics, such as structure formation in the early universe \cite{williams99,antunes}, magnetic ordering in Ising \cite{chui} and Heisenberg \cite{Hmag} systems, liquid crystals \cite{Lcrystals}, superfluid Helium \cite{berezinskii,KT,kosterlitz,minnhagen,williams87,shenoy,packard,packard2}, dilute atomic Bose-Einstein condensates \cite{atom1D,atom2Ddalibard,atom2Dcornell}, and superconductors \cite{williams99,VAPsc,bezryRev,kivelson,HTS,fisher}. In systems described by classical fields, thermal fluctuations of this type are often used to describe a corresponding thermodynamic phase transition where the field becomes ordered (or disordered), such as the Berezinskii-Kosterlitz-Thouless (BKT) vortex unbinding transition \cite{berezinskii,KT,kosterlitz,minnhagen}, the Lambda transition in liquid $^4$He \cite{williams87,shenoy}, and the interfacial roughening transition \cite{chui}.

The importance of topologically-charged fluctuations is dramatically increased in systems which are effectively lower-dimensional, often realized experimentally using superfluids or superconductors, where the phase of their macroscopic order parameter functions as the field in which topological defects are embedded. Examples include superconducting thin films \cite{VAPsc,fisher,samban,baturinaSIT} and narrow wires \cite{bezryRev}, lattice planes in high-T$_{\rm C}$ superconductors \cite{kivelson,HTS}, and superfluid Helium or dilute atomic Bose-Einstein condensates in confining potentials with quasi-2D \cite{berezinskii,KT,kosterlitz,atom2Ddalibard,atom2Dcornell} or quasi-1D \cite{packard,packard2,atom1D} character. In quasi-1D systems, whose transverse dimension is $\lesssim\xi$, the relevant coherence length, topological fluctuations are known as ``phase slips", and can be viewed conceptually as the passage of a quantized vortex line \textit{through} the 1D system. They were first discussed by Anderson in the context of neutral superfluid Helium flow through narrow channels \cite{anderson}, and by Little for persistent charged supercurrents in closed superconducting loops \cite{little}. During the course of such an excitation, the amplitude of the order parameter fluctuates to zero in a short segment of the channel of length $\sim\xi$, allowing the phase difference between the wire's ends $\Delta\phi$ to change by $\pm 2\pi$, in some cases accompanied by a quantized change in the supercurrent flow. In the presence of an external force $F$, this process (averaged over many phase-slip events) results in Ohm's-law behavior with a particle current proportional to $F$, rather than the ballistic acceleration expected for the superfluid state. For a charged superfluid this corresponds to finite electrical resistance, as was discussed in detail by Langer, Ambegaokar, McCumber, and Halperin (LAMH) \cite{LA,MH} and others \cite{kleinert}, for quasi-1D superconductors near their critical temperature $T_{\rm C}$ where the order parameter is close to zero. In subsequent experiments \cite{webb,newbower} on $\sim$0.2-0.5 $\mu$m-diameter crystalline Sn ``whiskers" which validated these ideas, finite resistances were observed to persist over a measurable temperature interval \textit{below} the mean-field $T_{\rm C}$.

These early works on quasi-1D systems considered only \textit{classical} processes, in which thermal fluctuations provide the free energy required to suppress the order parameter locally. However, in 1986 Mooij and co-workers suggested that an analogous \textit{quantum} process might exist, similar to macroscopic quantum tunneling (MQT) in Josephson junctions (JJs) \cite{MQTJJ,pop2010,pop2012,manucharyan2012}, by which the macroscopic system tunnels coherently between states whose $\Delta\phi$ differ by $\pm2\pi$ \cite{mooijQPS}. Just like the thermal phase slips discussed by Little \cite{little} and LAMH \cite{LA,MH}, such a process would depend exponentially on the wire's cross-sectional area, via the free energy required to suppress the order parameter over a length $\xi$. However, it would rely not on thermal energy but rather on some as yet unpecified (and presumably weak) source of quantum phase fluctuations, and thus it was presumed that extremely narrow wires would be required to observe it. Shortly thereafter, using lithographically defined, $\sim$~50 nm-wide superconducting Indium wires, Giordano measured finite resistance that persisted much farther below $T_{\rm C}$ than for wider wires \cite{giordano}, in the form of a crossover from the temperature scaling predicted by LAMH near $T_{\rm C}$ to a much slower temperature dependence farther from it. Using a heuristic argument in which the thermal energy scale in LAMH theory was replaced with a hypothesized \textit{quantum} energy scale, Giordano interpreted this observation as a crossover from thermal to quantum phase fluctuations, and was able to obtain a reasonable fit to his data. Many other experiments have since been carried out using different materials systems, which also exhibited some form of anomalous non-LAMH resistance below $T_{\rm C}$ \cite{gioPbIn,Pbwires,lau,altomare,zgirski,bezryRev,Tiwires} (though rarely in the form of a clearly evident crossover), and many authors have used Giordano's basic intuition as the basis for interpreting $R$ vs. $T$ data \cite{lau,altomare,zgirski,Tiwires,bezryRev,bezryNP}. In addition, a pioneering microscopic theory for QPS was later developed by Golubev, Zaikin, and co-workers (GZ) \cite{GZPRL,GZPRB} which appeared to validate Giordano's general idea, identifying his hypothesized quantum energy scale for QPS as the superconducting gap $\Delta$.

However, in other recent experiments using extremely narrow Pb \cite{granular,otherSIT}, Nb \cite{Nbwires}, and MoGe \cite{bezrySIT,bezryNP,bezryRev} nanowires $\lesssim$ 10 nm wide, the anomalous low-$T$ resistance previously identified directly with QPS was often completely \textit{absent}. This is difficult to explain within Giordano's hypothesis, given that the strength of QPS should increase exponentially as the wire cross-section is decreased. In response to these remarkable observations, it was then suggested that the observed deviations from LAMH temperature scaling may be explained purely in terms of a combination of LAMH phase slips and granularity \cite{granular,duan} and/or inhomogeneity \cite{inhomog} of the wires, rather than by QPS. On the other hand, the same MoGe nanowires which showed no evidence for QPS in $R$ vs. $T$ measurements \textit{did} exhibit low-$T$ anomalous resistance near their apparent critical current. These observations were made with techniques identical to those used to identify QPS phenomena in Josephson junctions \cite{MQTJJ}, and were consistent with a quantum energy scale for the phase fluctuations \cite{bezryNP,aref} just as Giordano had suggested, even though no evidence for this was seen in the $R$ vs. $T$ data for the same wires. Also striking was an apparent \textit{complete destruction} of superconductivity as $T\rightarrow0$ in other nanowires having a normal-state resistance $R_{\rm n}\gtrsim R_{\rm Q}$, where $R_{\rm Q}\equiv h/4e^2$ is known as the superconducting quantum of resistance \cite{tinkSIT,bezrySIT,bezryRev,otherSIT}. Although theories exist which predict insulating \cite{buchler,khlebSIT,meidan} or metallic \cite{GZPRL,GZPRB,demler,MIT} states in 1D as $T\rightarrow 0$, it is unclear whether any can explain a $T=0$ critical point at $R_{\rm n}\sim R_{\rm Q}$. Overall then, although some promising agreement between experimental and theoretical results has been obtained, there is still no consensus on how to self-consistently explain all of the observations, or on the precise role and nature of QPS in the phenomena observed.

In 2006, Mooij and Nazarov (MN) \cite{mooijfluxchg} made what may turn out to be a conceptual leap forward: they postulated that a classical symmetry known as \textit{flux-charge duality} \cite{beasleydual,likharevdual,kadindual,secondary,panyukov,bloch,pistolesi,hekking,ABAC} can be used to connect QPS with Josephson tunneling (JT), the well-known process in which Cooper pairs penetrate through a thin insulating barrier separating two superconducting electrodes, and establish macroscopic phase coherence between them. Based on this idea, MN posited the existence of a quantum phase slip potential energy $U_{\rm ps}(q)=-E_{\rm S}\cos{q}$, dual to the Josephson potential $U_{\rm J}(\phi)=-E_{\rm J}\cos\phi$. Here, $\phi$ and $q$ are known in the JJ literature as the phase and quasicharge, $E_{\rm J}$ is the well-known Josephson energy, and $E_{\rm S}$ is a new energy scale for QPS, which MN left as an input parameter. This mirrors the duality between the characteristic inductive energy of a wire $E_{\rm L}\equiv\Phi_0^2/2L_{\rm w}$ (where $L_{\rm w}$ is the wire's inductance) and the charging energy of a JJ given by: $e^2/2C_{\rm J}$ (where $C_{\rm J}$ is the junction capacitance). From their elegant hypothesis, MN generated a phenomenology of QPS dual to that of JJs, including a dual set of classical nonlinear equations for $q$, and a dual class of circuits involving 1D superconducting nanowires, what they called ``phase-slip junctions" (PSJs) \cite{mooijfluxchg,QPSosc,QPSdev}. Based on these ideas, several groups have recently performed new types of experiments \cite{astafiev,astafievNbN,zorin,arutyunovSR,arutyunovTiQPS}, in some cases directly realizing these dual circuits \cite{astafiev,astafievNbN,zorin,arutyunovTiQPS}, and providing the most direct evidence yet seen for QPS in continuous wires \footnote[1]{Note that granular wires, which consist of superconducting islands separated by insulating barriers, are effectively one-dimensional JJ arrays, whose phase-slip processes are well-understood \cite{doniach,glazman,pop2010,pop2012,manucharyan2012}.}.

In this work, we describe a new and alternative theory for QPS which takes MN's intuition as a starting point, and which may be able to shed light on a number of the outstanding questions related to QPS. We begin in section~\ref{s:QPSintro} with an introduction to the original intuition of Mooij and co-workers \cite{mooijQPS} for QPS, and its relation to equivalent phenomena in JJs. Section~\ref{s:duality} describes flux-charge duality, in preparation for section~\ref{s:QPS} where we build on this to construct a model for the origin of the basic QPS phenomenon, and use it to calculate the phase-slip energy $E_{\rm S}$. Our result for this quantity is qualitatively different from previous theories, in that it centrally involves the dielectric permittivity due to bound, polarizable charges in and around the superconductor, a quantity which does not appear in this way in previous theories for QPS. In our model this permittivity plays the role of an effective mass for ``fluxons", fictitious dynamical particles dual to Cooper pairs whose motion ``through" a 1D wire corresponds to a quantum phase slip event, just as Cooper pair motion through an insulating barrier corresponds to a JT event.

In section \ref{s:tline}, we build on these results to construct a distributed, nonlinear transmission line model of a quasi-1D superconducting nanowire. We show that in the presence of QPS, its dynamical equations for quasi-classical phase evolution in one spatial and one time dimension (1+1D) can be cast into a form identical to the static Maxwell-London equations in two spatial dimensions (2D), and from this we establish a direct analogy between the dynamics of electric flux penetration into a superconductor in 1+1D and the classical statistical mechanics governing magnetic flux penetration in 2D. We then use this analogy to predict macroscopic topological phase excitations in 1+1D we call \textit{type II phase slips}, which are the \textit{electric} analog of magnetic vortices in a type II superconductor, and which have a characteristic length scale $\lambda_{\rm E}$ we call the \textit{electric penetration depth}. These II phase slips are ``secondary" macroscopic quantum processes \cite{secondary}, in the sense that they arise as a collective effect out of the ``primary", microscopic QPS process, just as Bloch oscillations arise as a collective effect out of JT in lumped JJs \cite{bloch,schon,pistolesi,panyukov,secondary,hekking}.

In section \ref{s:expt}, we introduce a simple model for the interaction of these type II phase slips with the nanowire's electromagnetic environment, as well as a lumped circuit model for that environment similar to that used previously for JJs \cite{kautz}. We use this in conjunction with our transmission line model to calculate $R$ vs. $T$ for four experimental cases from different research groups, using different superconducting materials, chosen in particular because they cannot simultaneously be described by models that attribute anomalous resistance above that predicted by LAMH directly to a QPS ``rate" at finite temperature \cite{giordano,lau,bezryRev,bezryNP,Tiwires,zgirski}. By contrast, our model can approximately reproduce all four experimental curves, with input parameters either fixed at accepted or measured values, or (for parameters that are not known) chosen with eminently reasonable values. The key additional ingredient in our model which allows it to explain a wider range of phenomena in $R$ vs. $T$ curves is the additional length scale $\lambda_{\rm E}$, which itself has a temperature dependence. Next, we show how our model provides also a new interpretation of the quantum temperatures observed in MoGe nanowires by Bezryadin \cite{bezryNP,aref}, giving for the first time (to our knowledge) a quantitative potential explanation of the measured values. An important element of our explanation is the effect of a low environmental impedance at high frequencies, which provides damping for quantum phase fluctuations, and makes a description in terms of a quasi-classical phase appropriate. Related ideas were discussed previously by MN \cite{mooijfluxchg}, and also in the context of JJs \cite{bloch,schon,secondary,panyukov,pistolesi,hekking}. Lastly, in this section we show that our model is consistent with all four of the very recent, direct measurements of QPS, made by several different groups and using different materials \cite{astafiev,astafievNbN,zorinPRB,arutyunovTiQPS}. The electric penetration length $\lambda_{\rm E}$ also plays a crucial role in this agreement, since for two of these cases \cite{zorinPRB,arutyunovTiQPS} we find that $\lambda_{\rm E}$ is much shorter than the wire length. In this regime, the resulting behavior is not that of a lumped element, and our theory predicts that the Coulomb-blockade voltage $V_{\rm C}$ (the quantity observed in these two experiments) is independent of the wire length, in constrast to $E_{\rm S}$ which is by definition proportional to it.

Finally, in section \ref{s:SIT}, we suggest an alternative explanation for the observed destruction of superconductivity when $R_{\rm n}\gtrsim R_{\rm Q}$ \cite{bezrySIT}. Whereas most previous attempts to understand this apparent insulating behavior as $T\rightarrow0$ have been built on the idea of a \textit{dissipative} phase transition \cite{GZPRL,GZPRB,buchler,khlebSIT}, we hypothesize instead a \textit{disorder}-driven transition, with virtual type II phase slip-anti phase slip pairs as the fundamental quantum excitation. This picture is analogous to the so-called ``dirty Boson" model for quantum vortex-antivortex pair unbinding in quasi-2D superconductors \cite{fisher}, which has been used to explain an apparent superconductor-to-insulator transition (SIT) in highly-disordered thin films \cite{baturinaSIT,samban,gapins}. In this context, we discuss the interesting case of a SIT observed in microstructured 2D superconductors which essentially consist of a network of quasi-1D nanowires, and describe how this may be an intermediate case between the observed transitions in uniform 2D films and 1D wires. In section~\ref{s:conclusion} we summarize, and make some concluding remarks on the implications of our model for applications of QPS to future devices. \ref{a:vars} contains tables of selected variables and abbreviations used in the paper. \ref{a:RvsT}, \ref{a:moge}, and \ref{a:ES} provide details on the microscopic parameter values used to obtain the results in figs.~\ref{fig:RvsT} and~\ref{fig:bezryresults}, and table~\ref{tab:ES}. Lastly, \ref{a:circuits} provides some details on PSJ circuits which are dual to well-known JJ-based superconducting devices.

\section{The nature of QPS}\label{s:QPSintro}

\begin{figure}[t]
\begin{center}
\resizebox{0.8\linewidth}{!}{\includegraphics{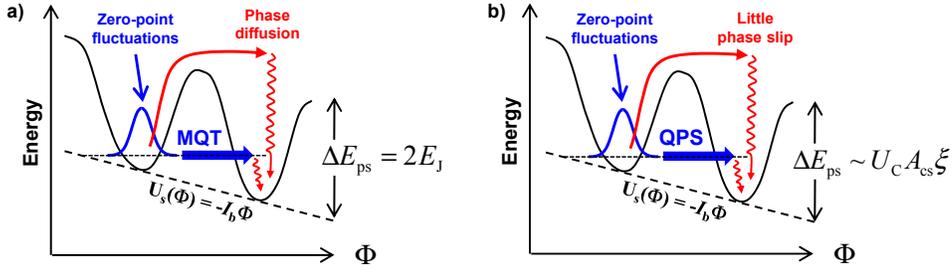}}
\caption{Early intuition for quantum phase-slip based on macroscopic quantum tunneling in Josephson junctions. (a) Schematic of the effective potential for the quasiflux $\Phi$ (equivalently, the gauge-invariant phase difference across the junction) of a JJ with an applied bias current $I_{\rm b}$. The barrier is due to the Josephson potential energy, and the ``tilt" comes from the free energy contribution $U_{\rm S}(\Phi)=-I_{\rm b}\Phi$ associated with a current source. In the superconducting state, the so-called ``phase particle", with ``position" $\Phi$, is localized in a given potential well. Thermal activation of the phase particle over the barrier (solid red arrow) followed by retrapping in the adjacent (or a nearby) potential well due to electrical damping (red wavy arrow) is known as phase diffusion \cite{kautz}, and produces a finite voltage and corresponding effective resistance even in the superconducting state. In the presence of zero-point fluctuations of the JJ's plasma oscillation (associated with its Josephson inductance and the junction's capacitance), the system can also tunnel through the potential barrier into the adjacent well, a phenomenon known as macroscopic quantum tunneling \cite{MQTJJ}. Although this is in principle a coherent, reversible process, in conjunction with nonzero damping (short, wavy red arrow) it can also result in an average escape rate for the phase particle and a corresponding voltage. (b) abstract potential envisioned for a quasi-1D superconducting wire as a function of its quasiflux $\Phi$ (gauge-invariant phase difference between the wire's ends), where the potential barrier is taken to be the condensation energy of a length $\xi$ of the wire, the minimum energy required to establish a localized null in the order parameter. Little or LAMH phase slips correspond to the the system surmounting this barrier due to a thermal fluctuation and then being retrapped (presumably also by damping). The original intuition of Mooij and co-workers \cite{mooijQPS} was that a phenomenon equivalent to MQT could also occur in a continuous wire, in the presence of a source of quantum phase fluctuations.}
\label{fig:MQTQPS}
\end{center}
\end{figure}

The qualitative picture of QPS originally put forth by Mooij and co-workers \cite{mooijQPS} is illustrated in fig.~\ref{fig:MQTQPS}, built on an analogy to macroscopic quantum tunneling (MQT) in JJs. For the JJ case, the quantum Hamiltonian is \cite{schon,bloch}:

\begin{equation}
\hat{H}_{\rm JJ}=\frac{\hat{Q}^2}{2C_{\rm J}}+E_{\rm J}\left[1-\cos\left(2\pi\frac{\hat\Phi}{\Phi_0}\right)\right]-I_{\rm b}\hat\Phi\label{eq:JJham}
\end{equation}

\noindent where $I_{\rm b}$ is an external bias current, and $[\hat\Phi,\hat{Q}]=\rmi\hbar$. The quantities $\hat Q$ and $\hat\Phi$ have units of charge and flux, and will be defined precisely below. We will refer to them as the quasicharge and quasiflux, respectively, and they are generalizations of the charge that has passed through the junction barrier and the gauge-invariant phase difference across the barrier. The quasiflux $\hat\Phi$ can be viewed as the coordinate of a fictitious particle whose ``mass" is $C_{\rm J}$, and which moves in a so-called ``tilted washboard" potential given by the last two terms in eq.~\ref{eq:JJham}, and illustrated in fig.~\ref{fig:MQTQPS}(a). The corresponding Heisenberg equations of motion for $\hat\Phi$ give the well-known classical, nonlinear behavior of the JJ in the limit where quantum fluctuations of $\hat\Phi$ about its expectation value can be neglected ($E_{\rm J}\gg e^2/2C_{\rm J}$, or equivalently $Z_{\rm J}\ll R_{\rm Q}$ where $Z_{\rm J}\equiv\sqrt{L_{\rm J}/C_{\rm J}}$ is the junction impedance). In this classical limit, the dominant way for the JJ to exhibit a phase-slip (i.e. for the particle to move from one well to the next) is for a thermal or other classical fluctuation to drive the system to an energy above the top of the Josephson barrier, as shown in fig.~\ref{fig:MQTQPS}; in the presence of damping (typically due to a shunt resistor), the particle is then ``re-trapped" in the adjacent (or other nearby) potential well, and this process then repeats stochastically, resulting in a phenomenon known as phase diffusion \cite{kautz}. A similar qualitative picture can be used to understand thermal LAMH phase slips in a quasi-1D superconductor\footnote[1]{Note that in the superconducting case, the condition for quasi-1D refers only to the macroscopic order parameter, and not to the bare energy levels of the conduction electrons, whose density of states is still fully 3D in the regime of interest here (equivalently, the Fermi wavelength $2\pi/k_{\rm F}$ is much smaller than the wire's transverse dimensions, so that there are many single-electron conduction channels near the Fermi energy in the metal).}, shown in fig.~\ref{fig:MQTQPS}(b). In this case, however, the classical potential energy as a function of $\Phi$ contains within it the physics originally described by Little \cite{little} and LAMH \cite{LA,MH}, such that each point on the horizontal axis represents a quasistationary solution of the Ginsburg-Landau (GL) equations for a wire with fixed $\Phi$ across it, and the point of maximum energy where $\Phi\approx\Phi_0/2$ is the so-called saddle-point solution also discussed in the context of superconducting weak links \cite{likharev}.

In both the JJ and quasi-1D wires, for purely classical fluctuations, the phase-slip rate can be written \cite{kramers,AHattempt,grabert}:

\begin{equation}
\Gamma_{\rm ps}=\Omega_{\rm ps}\exp\left[-\frac{\delta E_{\rm ps}}{k_{\rm B}T}\right]\label{eq:PSRate}
\end{equation}

\noindent where $\delta E_{\rm ps}$ is a classical energy barrier, which for JJs is simply $2E_{\rm J}$. For LAMH phase slips, the energy barrier is given by the total condensation energy of a length $\xi$ of the wire with cross-sectional area $A_{\rm cs}$ \cite{LA,MH,kleinert,giordano,lau,bezryRev,bezryNP}, up to a numerical factor:

\begin{eqnarray}
\delta E_{\rm LAMH}&\sim & U_{\rm C}A_{\rm cs}\xi\nonumber\\
&\sim&\frac{1}{L_\xi}\left(\frac{\Phi_0}{2\pi}\right)^2\label{eq:LAMH}
\end{eqnarray}

\noindent where $U_{\rm C}$ is the superconductor's condensation energy density, which goes to zero as $T\rightarrow T_{\rm C}$. In the second line $L_\xi$ is the kinetic inductance of a length $\xi$ of wire, such that the barrier can also be viewed as the energy cost to put $\Phi_0/2\pi$ across that length. The quantity $\Omega_{\rm ps}$ in eq.~\ref{eq:PSRate} is known as the attempt frequency \cite{kramers,AHattempt,grabert}, a term derived from the idea of an effective classical particle making multiple ``attempts" to surmount the energy barrier, originally used in treatments of Brownian motion and chemical reactions \cite{kramers}. In the JJ case, the attempt frequency is derived from the Josephson inductance and the effective capacitance and resistance shunting the junction; for example, for an undamped junction it is simply the oscillation frequency derived from its Josephson inductance and shunt capacitance (known as the junction plasma frequency). In LAMH's treatment of quasi-1D wires, the attempt frequency is derived from time-dependent GL theory \cite{LA,MH}; however, the exponential dependence of the phase-slip rate on the energy barrier and $T_{\rm C}$ makes it difficult to quantitatively compare this theory with experiment.

Just as with an actual massive particle in a confining potential like that shown in fig.~\ref{fig:MQTQPS}, at low enough temperature zero-point fluctuations become important; for the JJ this appears in the form of macroscopic quantum tunneling (MQT), in which these quantum fluctuations allow the system to tunnel \textit{through} the barrier \cite{MQTJJ}. In the absence of damping and in the limit of low bias current, this tunneling is completely coherent and reversible, and can be described purely in terms of superpositions of the stationary energy eigenstates of the system (known as the Wannier-Stark ladder \cite{WSladder}); if the current is turned on suddenly, the resulting coherent dynamics are known as Bloch oscillations \cite{bloch}. If the system is damped, on the other hand, it can relax irreversibly to the ground state of the adjacent well after tunneling (indicated by the short, wavy red line in fig.~\ref{fig:MQTQPS}), giving up its energy to the reservoir associated with the damping, and the process can then be repeated. Since in these dynamics $C_{\rm J}$ plays the role of a mass, $\hat{Q}$ a momentum, and $\hat{Q}^2/2C_{\rm J}$ the resulting kinetic energy, one can easily identify the source of quantum phase fluctuations in the JJ system: the finite junction capacitance $C_{\rm J}$ results in an energy cost to localize the position $\hat\Phi$, due to the corresponding fluctuations in its conjugate momentum $\hat{Q}$.

Figure~\ref{fig:MQTQPS}(b) shows the analogous picture suggested by Mooij and co-workers \cite{mooijQPS} to motivate QPS: in the presence of quantum zero-point phase fluctuations, even a continuous superconducting wire (if it is narrow enough, so that the energy barrier is low enough) can undergo a form of MQT. The question is, what is the source of these quantum phase fluctuations in a continuous superconducting wire? Giordano's identification of a crossover in $R$ vs. $T$ curves for very thin wires prompted him to suggest a quantum phase slip ``rate" analogous to the thermal phase slip rate that produces LAMH-type resistance, but with the thermal energy $k_{\rm B}T$ replaced by this other, manifestly quantum energy scale for zero-point phase fluctuations (or ``quantum temperature" $T_{\rm Q}$ as it would be described in the language of JJs \cite{grabert,MQTJJ,bezryNP,aref})\footnote[1]{The idea of a ``rate" implies irreversibility and therefore a continuum of states that functions as a dissipative reservoir. In a JJ, this dissipation comes from the shunt resistance. However, in cases where an equivalent QPS ``rate" is used to explain a linear resistance of continuous wires in the $I_{\rm b}\rightarrow0$ limit \cite{giordano,lau,bezryNP,altomare,Tiwires,Tiwires,zgirski}, no source of dissipation is explicitly mentioned, which in our view is problematic. In the absence of dissipation as $I_{\rm b}\rightarrow0$, the tilted washboard potential would exhibit no quantum phase slip ``rate" or measurable resistance, but simply the set of stationary energy eigenstates known as the Wannier-Stark ladder \cite{WSladder}. Subsequent theories have predicted \textit{nonlinear} resistances due to QPS even at $T=0$ \cite{GZPRL,buchler}, but these necessarily go to zero as $I_{\rm b}\rightarrow0$, in contrast to the linear resistances observed in experiments. In our model, as we will see in section~\ref{s:expt}, linear, phase-slip-induced resistances arise only due to \textit{thermal} processes in the presence of an explicitly dissipative electromagnetic environment.}. In his original work \cite{giordano}, and subsequent treatments based on it \cite{lau,saito,chang,duan,bezryRev}, this quantum phase fluctuation energy scale was taken to be $\sim\hbar/\tau_{\rm GL}$, where $\tau_{\rm GL}\equiv\pi\hbar/[8k_{\rm B}(T_{\rm C}-T)]$ is the GL relaxation time. The microscopic theory of GZ \cite{GZPRL,GZPRB}, although it did not posit the existence of a linear phase-slip resistance at $T=0$, did in fact give an energy scale $\sim\Delta\sim\hbar/\tau_{GL}$ for the quantum phase fluctuations, in qualitative agreement with Giordano's original intuition.

In this paper, using MN's hypothesis of flux-charge duality between quantum phase slip and Josephson tuneling as a starting point, we construct an alternative model for QPS in which the energy scale for quantum phase fluctuations is capacitive in nature, just like the charging energy for JJs, but with the capacitance here arising from the polarizable, bound electrons both inside and near the wire; the effective permittivity of this polarizable environment is then the background upon which the fluctuating electric fields associated with QPS occur. In preparation for describing this model, we first give some background on flux-charge duality, the principle on which it is based.

\section{Flux-charge duality}\label{s:duality}

Flux-charge duality is a classical symmetry of Maxwell's equations\footnote[7]{See, for example, ref.~\cite{nori_dual}.} which is best known in the context of planar lumped-element circuits \cite{beasleydual,likharevdual,kadindual,secondary,panyukov,bloch,pistolesi,hekking,ABAC}, where it manifests itself in the invariance of the equations of motion under the transformation shown in  fig.~\ref{fig:duality}(a), and is also connected to the relationship between right-handed and left-handed metamaterials made from lumped circuit elements \cite{metam}. In the more general continuous case, it can be made apparent by defining the quantities:

\begin{eqnarray}
Q(\Sigma)&\equiv&\oint_\sigma \rmd t({\bi H}\cdot{\bi d\bsigma})=\int_\Sigma \rmd t({\bi J}_{\rm Q}\cdot{\bi d\bSigma}),\;\;\;\;\;\;{\bi J}_{\rm Q}={\bi J}+\frac{\rmd{\bi D}}{\rmd t}\label{eq:quasicharge}\\
\Phi(\Gamma)
&\equiv&\oint_\Gamma \rmd t({\bi E}\cdot{\bi d\bGamma}),\;\;\;\;\;\;{\bi E}=-\nabla V-\frac{\rmd{\bi A}}{\rmd t}\label{eq:quasiflux}
\end{eqnarray}

\noindent where $Q(\Sigma)$ is associated with a surface $\Sigma$ (bounded by a closed curve $\sigma$) and $\Phi(\Gamma)$ with a curve $\Gamma$, as illustrated in fig.~\ref{fig:duality}(b). These quantities reduce to the so-called ``branch variables" in the Lagrangian description of electric circuits described in refs. \cite{yurke,devoret} if $\Gamma$ in fig.~\ref{fig:duality}(b) connects the two ends of the branch. Figures ~\ref{fig:duality}(c) and (d) illustrate the duality between these quantities, such that equations \ref{eq:quasicharge} and \ref{eq:quasiflux} can both be interpreted as arising from a sum of ``free" and ``bound" current densities:

\begin{figure}[t]
\begin{center}
\resizebox{0.7\linewidth}{!}{\includegraphics{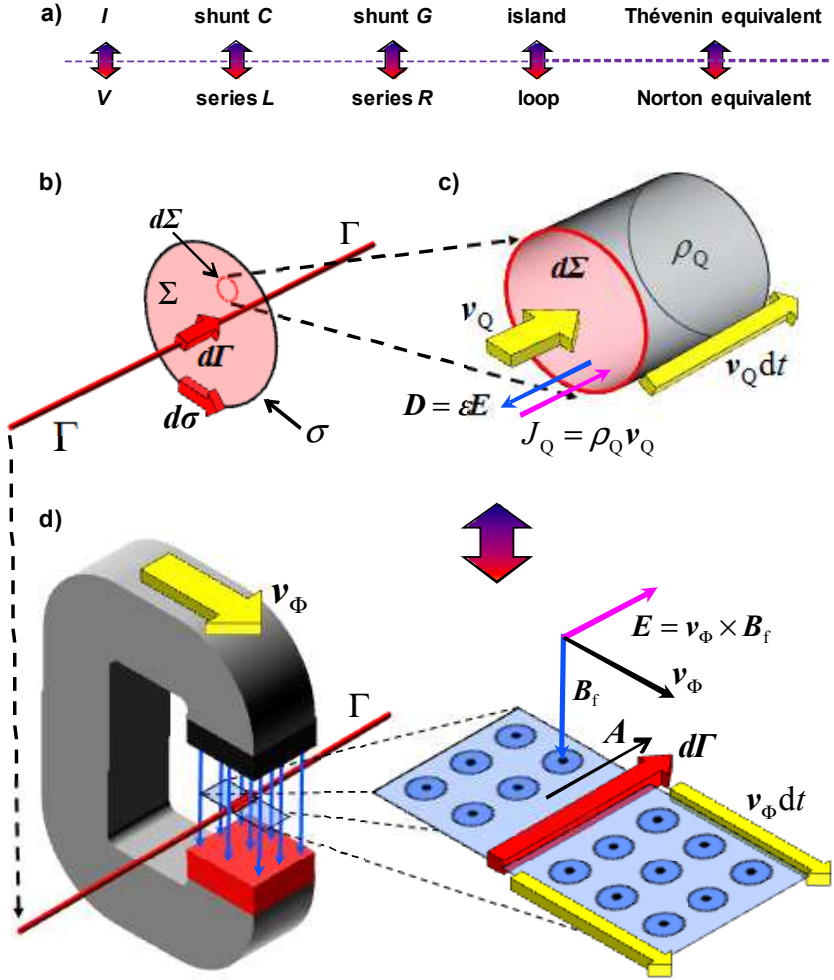}}
\caption{Flux-charge duality. (a) tabulates the duality transformation for planar, lumped-element circuits, while (b)-(d) illustrate the continuous case. (b) Illustration of the vector quantities used in the definitions of $Q(\Sigma)$ and $\Phi(\Gamma)$ [c.f., eqs.~\ref{eq:quasicharge}-\ref{eq:quasiflux}]. (c) The free current density $\rho_{\rm Q}{\bi v}_{\rm Q}$ is the motion of free charge density $\rho_{\rm Q}$ at a velocity ${\bf v}_{\rm Q}$, through a surface area element ${\bi d}\bSigma$. The bound current density $\rmd{\bi D}/\rmd t$ is the displacement current density on ${\bi \bSigma}$. (d) An example of ``free" flux density, using a permanent magnet moving at velocity ${\bi v}_\Phi$ relative to the stationary curve $\Gamma$, such that the associated free flux ``current" is: ${\bi E}={\bi v}_\Phi\times{\bi B}_{\rm f}$. In this construction, ${\bi E}\cdot{\bi d\bGamma}$ is precisely the flux per unit time passing through a segment ${\bi d\bGamma}$. The bound flux ``current" density $-\rmd{\bi A}/\rmd t$ is associated with time-varying currents flowing along $\Gamma$, and the associated induced emfs from Faraday's law. Although the case of a moving magnet is somewhat artificial, any electric field in a medium can be broken into these two components: one associated with bound charges, and the other with induced emfs from time varying currents (free charges).}
\label{fig:duality}
\end{center}
\end{figure}

\begin{eqnarray}
\phantom{{\bi J}_\Phi\equiv}{\bi J}_{\rm Q}&=&\underbrace{\phantom{\frac{|}{|}}\rho_{\rm Q}{\bi v}_{\rm Q}\phantom{\frac{|}{|}}}_{\footnotesize\textrm{free charge}}+\underbrace{\phantom{\frac{|}{|}}\frac{\rmd{\bi D}}{\rmd t}\phantom{\frac{|}{|}}}_{\footnotesize\textrm{bound charge}}\label{eq:JQ}\\
{\bi J}_\Phi\equiv{\bi E}&=&\underbrace{\phantom{\frac{|}{|}}{\bi v}_\Phi\times{\bi B}_{\rm f}\phantom{\frac{|}{|}}}_{\footnotesize\textrm{``free" flux}}-\underbrace{\phantom{\frac{|}{|}}\frac{\rmd{\bi A}}{\rmd t}\phantom{\frac{|}{|}}}_{\footnotesize\textrm{``bound" flux}}\label{eq:Jphi}
\end{eqnarray}

\noindent Here, $\rho_{\rm Q}$ is an ordinary density of free charge moving at velocity $v_{\rm Q}$, and ${\bi B}_{\rm f}$ is a magnetic flux density moving at velocity ${\bi v}_\Phi$. Using the London gauge ${\bi A}=-\Lambda\rho_{\rm Q}{\bi v}_{\rm Q}$ for a superconductor (where the London coefficient is $\Lambda=\mu_0\lambda^2$ with $\lambda$ the magnetic penetration depth) and ${\bi D}=\epsilon{\bi E}$ for an insulator, yields:

\begin{eqnarray}
\textrm{superconductor:}\;\;\;\Lambda\frac{\rmd{\bi J}}{\rmd t}={\bi E}\;\rightarrow\;\;L_{\rm k}\frac{\rmd^2Q(\Sigma)}{\rmd t^2}&=&V_\Gamma\label{eq:london}\\
\phantom{h}\phantom{h}& \phantom{\updownarrow} \nonumber\\
\;\;\;\;\;\;\;\;\;\;\textrm{insulator:}\;\;\;\;\;\epsilon\frac{\rmd{\bi E}}{\rmd t}={\bi J}\;\rightarrow\;\;\;C\frac{\rmd^2\Phi(\Gamma)}{\rmd t^2}&=&I_\Sigma\label{eq:displ}
\end{eqnarray}

\noindent where on the right side $V_\Gamma$ is the voltage difference between the two ends of $\Gamma$ and $I_\Sigma$ is the current flowing through $\Sigma$. Equation~\ref{eq:london} for the superconductor is none other than London's first equation, according to which $Q$ moves ballistically under the action of a force $V$, and with an effective mass given by the kinetic inductance $L_{\rm k}$; correspondingly, eq.~\ref{eq:displ} is Maxwell's equation for the displacement current in an insulator, which can be viewed as ballistic acceleration of $\Phi$ under the action of a ``force" $I$, with an effective mass given by the capacitance $C$. Therefore, at the classical level of the Maxwell-London equations, superconductors and insulators are dual to each other.

We now arrive at the proposed duality between a JJ and a PSJ, first suggested by MN (though here we have arrived at it in a different way). We start by considering only the lumped-element case, as was done by MN. This will be generalized to the fully distributed case starting with section~\ref{s:tline} below. As shown in fig.~\ref{fig:JJPSJ}, a JJ consists of two superconducting islands of Cooper pairs separated by an insulating potential barrier, while a PSJ can be viewed as two \textit{insulating} ``islands" of flux quanta (henceforth referred to as ``fluxons") separated by a \textit{superconducting} potential barrier. If we place the surface $\Sigma$ inside the insulating barrier of a JJ [Fig.~\ref{fig:JJPSJ}(a)] with junction capacitance $C_{\rm J}$, and the curve $\Gamma$ inside a superconducting nanowire [Fig.~\ref{fig:JJPSJ}(b)] of kinetic inductance $L_{\rm k}$ (neglecting its geometric inductance), we have:

\begin{figure}[t]
\begin{center}
\resizebox{0.7\linewidth}{!}{\includegraphics{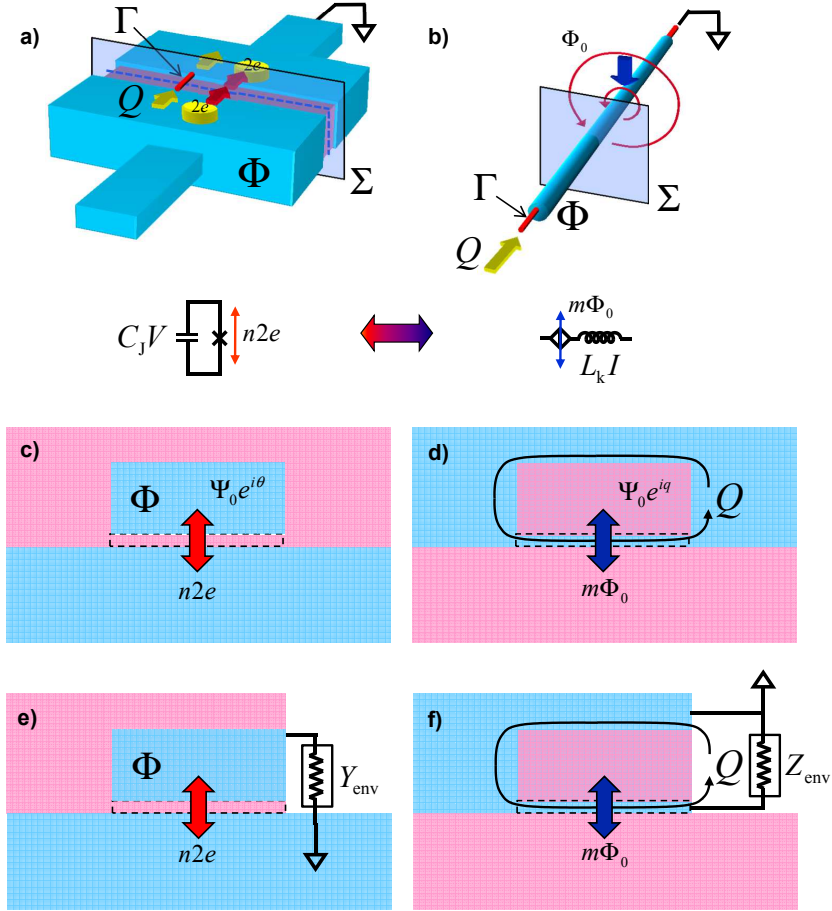}}
\caption{Flux-charge duality, Josephson tunneling, and quantum phase slip. Superconductor is shown in blue, and insulator in red. (a) and (b) illustrate the geometry of the surface $\Sigma$ and curve $\Gamma$ which are used to define the quasicharge $Q$ and quasiflux $\Phi$ in the text. (c) schematic of a JJ, consisting of an insulating tunnel barrier between a superconducting island and ``ground" (this is also known as a charge qubit). (d) schematic of a PSJ, consisting of a superconducting nanowire tunnel barrier between an insulating island and ``ground" (which for fluxons is an insulator). Note the closed superconducting loop around the insulating island in this case, which is known as a phase-slip qubit \cite{PSqubit}. In (e) and (f) we add an electromagnetic environment, in terms of an admittance $Y_{\rm env}$ for the JJ or an impedance $Z_{\rm env}$ for the PSJ, such that the tunnel barrier between the island and ground in each case is shunted by a dissipative element.}
\label{fig:JJPSJ}
\end{center}
\end{figure}

\begin{eqnarray}
\textrm{  JJ}:\;\;Q&=&\underbrace{\phantom{|}n2e\phantom{|}}_{\footnotesize\textrm{free}}+\underbrace{\phantom{|}C_{\rm J}V\phantom{|}}_{\footnotesize\textrm{bound}}\;\;\;\;\Phi=\frac{\Phi_0}{2\pi}\theta+\oint_\Gamma{\bi A}\cdot{\bi d\bGamma}=m\Phi_0+L_{\rm J}I\label{eq:QJJ}\\
\textrm{PSJ}:\;\;\Phi&=&\underbrace{\phantom{|}m\Phi_0\phantom{|}}_{\footnotesize\textrm{free}}+\underbrace{\phantom{|}L_{\rm k}I\phantom{|}}_{\footnotesize\textrm{bound}}\;\;\;\;Q=Q_{\rm f}+\int_\Sigma{\bi D}\cdot{\bi d\bSigma}=n2e+C_{\rm k}V\label{eq:PhiPSJ}
\end{eqnarray}

\noindent For the JJ, $C_{\rm J}V$ is the charge on the capacitance $C_{\rm J}$ of the junction barrier induced by a voltage difference $V$ across it, and $n$ is the number of Cooper pairs that have passed through it. The quantity $Q$ appearing in eqs.~\ref{eq:JJham} and~\ref{eq:QJJ} is then a dimensional version of the so-called junction quasicharge \cite{schon,bloch,pistolesi,panyukov,hekking}. The quantity $\Phi$ appearing in eqs.~\ref{eq:JJham} and~\ref{eq:QJJ} for the JJ also consists of two terms, the first of which is due to the phase difference $\theta$ between the order parameters of the two superconducting electrodes, plus a second term due to magnetic fields inside the junction. As shown on the far right of eq.~\ref{eq:QJJ}, it can also be written as the sum of the contributions from the kinetic flux induced by a current $I$ flowing through the Josephson inductance $L_{\rm J}$, and the passage of $m$ (discrete) fluxons through the junction. This quantity is then a dimensional version of the gauge-invariant phase difference across the junction \cite{orlando} (also referred to as the ``quasiphase" in ref.~\cite{QPSdev}). Henceforth, we will refer to $\Phi$ as the ``quasiflux". For the PSJ in eq.~\ref{eq:PhiPSJ}, dual statements to those for the JJ apply: the quantity $L_{\rm k}I$ is the total ``bound" flux of a nanowire having kinetic inductance $L_{\rm k}$ associated with a current $I$, and $m$ is the discrete number of fluxons that have passed through the wire. The wire's quasicharge $Q$ is a sum of the total free charge $Q_{\rm f}$ that has passed through the wire, plus a term associated with electric fields on the wire's so-called ``kinetic capacitance" $C_{\rm k}$ (the dual of Josephson inductance) \cite{mooijfluxchg}. Kinetic capacitance was suggested by MN as a formal consequence of the assumed flux-charge duality between the JJ and PSJ, and we discuss in section~\ref{s:QPS} below how our model for QPS gives an intuitive interpretation of its origin.

For thick enough superconducting wires, the only way for $m$ to be nonzero is if some part of the wire was in the normal state at some time, as occurs in an LAMH phase slip over a length of wire $\sim\xi$, the GL coherence length. These events are dissipative, produce a measurable voltage pulse, and can be associated with passage of a fluxon through the null in the superconducting order parameter at a localized, measurable position and time. By contrast, the dual to JT, which we want to identify with QPS, would necessarily be coherent, delocalized fluxon \textit{tunneling} through the entire length of wire, such that no information about where the phase-slip occurred exists. Just as in a JJ, where localizing a Cooper pair tunneling event would cost electrostatic energy, localizing a fluxon tunneling event in a PSJ would cost kinetic-inductive energy.

\section{Quantum phase slip}\label{s:QPS}

We now describe our model for QPS, whose basic intuition is contained in fig.~\ref{fig:duality}(d): Fluctuations of the phase difference between the ends of a wire correspond to fluxon ``currents" passing ``through" the wire, which are none other than electric fields along it. The effective mass associated with this fluxon motion is then an electric permittivity, which determines a ``kinetic" (electrodynamic) energy cost for phase fluctuations. This is the crucial new energy scale which allows us to define QPS in our model, in conjunction with the appropriate ``confining" potential energy $U(\Phi)$ for $\Phi$ (the ``phase particle") whose classical minima define the mean-field superconducting state [c.f., fig.~\ref{fig:MQTQPS}(b)]. If the zero-point quantum fluctuations about this state are sufficiently strong, they can produce (macroscopic) quantum tunneling between adjacent minima of the potential, which in the absence of damping gives exactly the behavior postulated by MN \cite{mooijfluxchg}.

Before exploring the implications of this idea, however, we must first define more precisely what we mean by the electric permittivity inside the wire relevant for quantum phase fluctuations along it. We do this in the context of the simplest (Drude) model of a metal, consisting of a gas of nearly free conduction electrons of mass $m_{\rm e}$ and density $n_{\rm e}$, superimposed on a background of fixed ions of density $n_i$; the permittivity inside the metal at frequency $\omega$ in this model is:

\begin{equation}
\epsilon(\omega)=\epsilon_{\rm b}(\omega)+\frac{\rmi\sigma(\omega)}{\omega}\label{eq:epsilon}
\end{equation}

\noindent where the complex conductivity $\sigma(\omega)$ and background permittivity $\epsilon_{\rm b}(\omega)$ are:

\begin{eqnarray}
\sigma(\omega)&\equiv&\frac{\sigma_0}{1-\rmi\omega\tau_{\rm s}}\label{eq:sigma}\\
\epsilon_{\rm b}(\omega)&\equiv&\epsilon_0+n_{\rmi}\alpha(\omega)\label{eq:epsilonB}
\end{eqnarray}

\noindent Here, $\sigma_0\equiv n_ee^2\tau_{\rm s}/m_{\rm e}$ is the DC conductivity for a scattering time $\tau_{\rm s}$ of conduction electrons, and $\alpha(\omega)$ is the polarizability of each ion. The contribution of this ionic background to the permittivity, sometimes known as ``core polarization" \cite{kukkonen,ashcroft}, can be viewed as arising from interband transitions, and can be as large as $\sim10\epsilon_0$ in simple noble metals \cite{ehrenreich}, and even much higher in materials with polarizable, low-lying electronic excited states \cite{choiCore} like the highly-disordered materials typically used for QPS studies\footnote[1]{This may seem reminiscent of ref.~\cite{astafiev}, in which the proximity of the host material to a metal-insulator transition (presumably accompanied by a large polarizability) was emphasized as important for achieving strong QPS. An interesting consequence of our model, by contrast, will turn out to be that a large permittivity \textit{suppresses} QPS.}. It can be difficult to measure at high frequencies ($\omega\tau_{\rm s}\gg1$), however, since it is superposed with the large, negative contribution from the metal's inductive (free carrier) response in this regime [c.f., eq.~\ref{eq:sigma}].

Taking this limit $\omega\tau_{\rm s}\gg1$, and making the replacements $m_{\rm e}\rightarrow2m_{\rm e}, e\rightarrow2e,n_{\rm e}\rightarrow n_{\rm s}$ we arrive at the simplest possible model for a superconductor, in which Cooper pairs of mass $2m_{\rm e}$, charge $2e$, and density $n_{\rm s}$ move without resistance; the permittivity is then:

\begin{equation}
\epsilon(\omega)\approx\epsilon_{\rm b}\left[1-\frac{\Omega_{\rm p}^2}{\omega^2}\right]\label{eq:SCeps}
\end{equation}

\noindent where we have defined the quantity:

\begin{equation}
\Omega_{\rm p}\equiv\sqrt{\frac{1}{\Lambda\epsilon_{\rm in}}}\label{eq:CPplasma}
\end{equation}

\noindent known as the Cooper pair plasma frequency \cite{mooijschon,orlando}, with $\Lambda\equiv m_{\rm e}/(2n_{\rm s}e^2)$ the London coefficient \cite{orlando}. Formally, this is the oscillation frequency of the Cooper-paired electrons relative to the ion cores, with an effective (kinetic) inductance due to their mass, and an effective capacitance due to $\epsilon_{\rm b}$. Now, in real superconductors this frequency is essentially always larger than the superconducting gap, such that real excitation of this mode would break Cooper pairs and thus be strongly damped; however, in our model it is rather the zero-point fluctuations of this plasma oscillation with which we are concerned, and which will result in QPS.

\begin{figure}[t]
\begin{center}
\resizebox{1.0\linewidth}{!}{\includegraphics{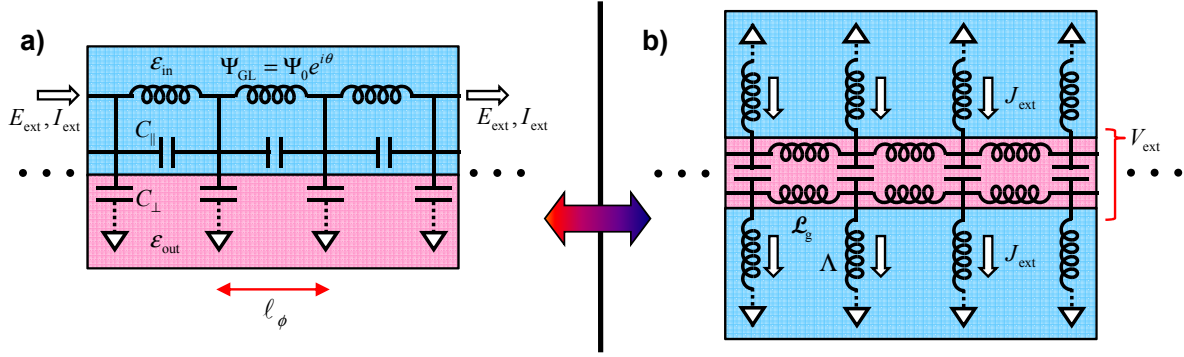}}
\caption{Dual models of PSJs and JJs I: schematic. (a) quasi-1D Ginsburg-Landau superconductor with order parameter $\Psi_{\rm GL}\equiv\Psi_0e^{\rmi\theta(x)}$ and corresponding nonlinear series inductance, electric permittivity due to bound charges $\epsilon_{\rm in}$, and distributed shunt capacitance of the surrounding dielectric $\epsilon_{\rm out}$. Dotted lines to ground indicate the fact that while at low frequencies the electric field lines of propagating modes along the wire (Mooij-Sch\"{o}n modes \cite{mooijschon}) would typically terminate at a distant, physical ground plane, at high frequencies the fields are confined closer to the wire [c.f., eq.~\ref{eq:MScap}]. (b) dual model for a JJ, where the insulating barrier has both a shunt capacitance and series geometric inductance (associated with magnetic fields inside the barrier). The shunt inductors indicate the kinetic inductivity of the superconducting electrodes, and the dotted lines indicate a frequency dependence of the field penetration into the electrodes for propagating modes along the junction (Fiske modes \cite{fiske}). Throughout this work, to facilitate comparison between these two cases, we take one dimension of the junction barrier as fixed, and consider only changes in the length of the junction in the other dimension.}
\label{fig:tline1}
\end{center}
\end{figure}

Our model for a quasi-1D superconducting wire is shown schematically in fig.~\ref{fig:tline1}(a), and for comparison the dual model for a JJ is shown in fig.~\ref{fig:tline1}(b). We discretize the system along one dimension, at a length scale $l_\phi$ to be discussed below. The shaded blue kinetic inductors indicate the usual mean-field GL theory\footnote[1]{Although GL theory is in general valid only very close to $T_{\rm C}$, the materials currently used for QPS experiments are all in the dirty, local, type-II limit where it is a good approximation all the way to $T=0$ (see, for example, ref.~\cite{tinkham}).} with order parameter $\Psi_{\rm GL}=\Psi_0e^{\rmi\theta}$. The capacitors $C_{||}$ and $C_\perp$ indicate schematically the distributed permittivities $\epsilon_{\rm in}$ and $\epsilon_{\rm out}$ for electric fields inside and outside the superconductor, respectively. Note that here $\epsilon_{\rm in}$ describes \textit{only the bound-electron response}, corresponding to the first term in eq.~\ref{eq:SCeps}, which then appears in parallel with the free (superconducting) component with kinetic inductivity $\Lambda=\mu_0\lambda^2$, corresponding to the second term in eq.~\ref{eq:SCeps}. The semiclassical plasma modes of such a quasi-1D system were discussed in the seminal work of Mooij and Sch\"{o}n (MS) \cite{mooijschon} for a wire of circular cross-section embedded in an insulating medium of permittivity $\epsilon_{\rm out}$. The dispersion relation for these modes can be written in the form:

\begin{equation}
\omega(k)=\Omega_{\rm p}\sqrt{1+\left(\frac{1}{k\Lambda_{\rm 1D}}\right)^2}\label{eq:omegak}
\end{equation}

\noindent where $k$ is the wavenumber and $\Lambda_{\rm 1D}$ is a quasi-1D Coulomb screening length which can be expressed in our discretized model in terms of the discrete capacitors shown in fig.~\ref{fig:tline1}(a) thus:

\begin{eqnarray}
\frac{\Lambda_{\rm 1D}}{l_\phi}&=&\sqrt\frac{C_{||}}{C_\perp}\label{eq:lambda1D}\\
C_{||}&=&\frac{\epsilon_{\rm in}A_{\rm cs}}{l_\phi}(kl_\phi)^2\\
C_\perp&=&2\pi r_0\epsilon_{\rm out}\frac{K_1(kr_0)}{K_0(kr_0)}(kl_\phi)\label{eq:MScap}
\end{eqnarray}

\noindent where $K_{\rm n}(y)$ are the modified Bessel functions of order $n$ and argument $y$, and in the continuum limit $(kl_\phi\ll1)$ these results in conjunction with fig.~\ref{fig:tline1}(a) agree with ref.~\cite{mooijschon}\footnote[7]{With the exception that MS took $\epsilon_{\rm in}=\epsilon_0$ in ref.~\cite{mooijschon}.}. Equation~\ref{eq:lambda1D} is familiar from the physics of 1D JJ arrays, defining the length scale over which the Coulomb interaction between charges is screened out by the distributed shunt capacitances $C_\perp$. On short length scales where $k\Lambda_{\rm 1D}\gg1$ this shunt capacitance has a negligible effect, and eq.~\ref{eq:omegak} reduces to the bulk plasma frequency $\Omega_{\rm p}$ [c.f., eq.~\ref{eq:CPplasma}]. In the opposite limit where $k\Lambda_{\rm 1D},kr_0\ll1$, $C_\perp$ dominates and eq.~\ref{eq:MScap} reduces to an approximately wavelength-independent capacitance per length: $\mathcal{C}_\perp=C_\perp/l_\phi\approx2\pi\epsilon_{\rm out}/\ln[1/kr_0]$. Correspondingly, eq.~\ref{eq:omegak} reduces to an approximately linear dispersion relation with a fixed wave propagation velocity known as the Mooij-Sch\"{o}n velocity $v_{s}=1/\sqrt{\mathcal{L}_{\rm k}\mathcal{C}_\perp}$ and a linear impedance $Z_{\rm L}=\sqrt{\mathcal{L}_{\rm k}/\mathcal{C}_\perp}$, where $\mathcal{L}_{\rm k}=\Lambda/A_{\rm cs}$ is the kinetic inductance per length.

\begin{figure}[t]
\begin{center}
\resizebox{0.92\linewidth}{!}{\includegraphics{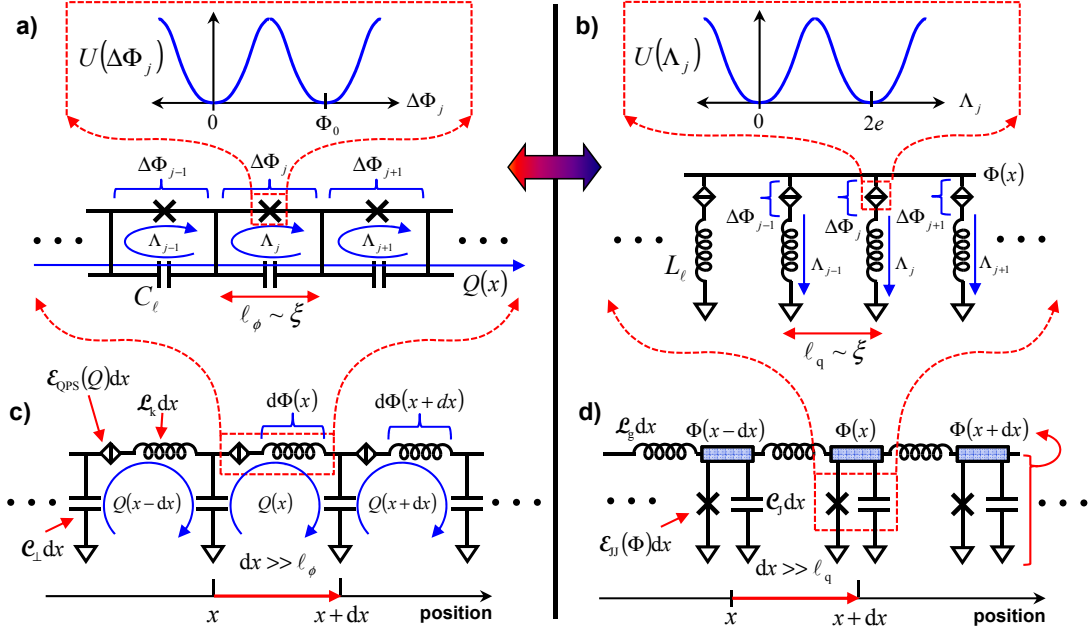}}
\caption{Dual models of PSJs and JJs II: nonlinear transmission lines. (a) discrete model of weak QPS on short length scales, where each ``link" of characteristic length $l_\phi\sim\xi$ is treated as a parallel plasma oscillator composed of a nonlinear inductor with a single-valued, $\Phi_0$-periodic potential $U(\Delta\Phi_{\rm J})$ (the ordinary GL superconductor), and the capacitance $C_l$ [eq.~\ref{eq:QPScap}] associated with potential differences along the wire. Zero-point fluctuations of this oscillator (occurring independently for each length $l_\phi$) generate QPS via tunneling between wells of the periodic effective potential $U(\Delta\Phi_{\rm J})$. The quantum variables associated with QPS in the $j^{\rm th}$ link are its loop charge $\Lambda_{\rm J}$ and quasiflux $\Delta\Phi_{\rm J}$, with $[\Delta\Phi_{\rm J},\Lambda_{\rm k}]=\rmi\hbar\delta_{jk}$. At these short length scales, the quasicharge $Q(x)$ is assumed to be uniform along $x$. (b) The dual short-length-scale model of a JJ, in which each length $l_{\rm q}\sim\xi$ of the barrier becomes an independent \textit{series} plasma oscillator (note that we consider the junction to be short in one of its two areal dimensions, so that it can be viewed as a 1D system). This oscillator is composed of a nonlinear capacitance (the barrier capacitance, modified by Cooper pair tunneling, to produce a $2e$-periodic effective potential energy $U(\Lambda_{\rm J})$ for the loop charges), and an effective kinetic inductance $L_l$ of the nearby region inside the electrodes. Josephson tunneling can then be viewed as arising from zero-point fluctuations (occurring independently for each length $l_{\rm q}\sim\xi$) of these oscillators. At short length scales $\Phi(x)$ is assumed to be $x$-independent (magnetic fields in the $\mathcal{L}_{\rm g}$ are neglected). In (c), the distributed shunt capacitance $\mathcal{C}_\perp$ now allows $Q$ to be a function of position along the wire, and in (d) the distributed series inductance $\mathcal{L}_{\rm g}$ similarly allows $\Phi$ to vary spatially. To describe the physics at longer length scales (and lower energy scales) the ground state energy densities $\mathcal{E}_{\rm QPS}(Q)$ and $\mathcal{E}_{\rm JJ}(\Phi)$ of the discrete models (a) and (b) are incorporated into the nonlinear transmission lines shown in (c) and (d), respectively, as classical potential energies for the long-wavelength dynamics of $Q(x,t)$ and $\Phi(x,t)$. Both of these models are described by the sine-Gordon equation in an appropriate semi-classical limit, which for the PSJ is when $Z_{\rm L}=\sqrt{\mathcal{L}_{\rm k}/\mathcal{C}_\perp}\gg R_{\rm Q}$, and for the JJ when $Z_{\rm J}=\sqrt{\mathcal{L}_{\rm g}/\mathcal{C}_{\rm J}}\ll R_{\rm Q}$. }
\label{fig:tline2}
\end{center}
\end{figure}

We assume that for an individual QPS event occurring far from the ends of the wire, all of its dynamics are contained within a length $l_\phi$. We further assume that QPS is sufficiently ``weak" (in a manner to be defined more precisely below) that we can neglect the interactions between multiple QPS events which would otherwise result from the shunt capacitances $C_\perp$. Note that in making this assumption we are only neglecting the possibility that two QPS events occur within $\Lambda_{\rm 1D}$ of each other, since at distances beyond this their Coulomb interaction will already be screened out. This assumption about the short-length-scale physics of QPS allows us to associate with each segment a single effective parallel capacitor $C_l$, as shown in fig.~\ref{fig:tline2}(a), which contains contributions from electric fields both inside and outside the wire:

\begin{equation}
C_l\equiv \left[C_{||}+\frac{C_\perp}{2}\right]_{kl_\phi\rightarrow1}\label{eq:QPScap}
\end{equation}

\noindent This definition is based on the requirement that in the $l_\phi\ll r_0$ limit we should require that: $C_l\rightarrow\epsilon_{\rm in}A_{\rm cs}/l_\phi$, the simple parallel-plate capacitance for a length $l_\phi$. In this limit, the electric field is almost completely confined within the wire, whereas in the opposite limit $l_\phi\gg r_0$ most of the field is outside the wire. Note that the relative participation of these two regions is also affected by the relative size of $\epsilon_{\rm in}$ and $\epsilon_{\rm out}$, since the higher permittivity material will tend to ``attract" the electric flux associated with QPS. In neglecting the shunt capacitance to the environment on short length scales $\sim l_\phi$, we are also by construction neglecting the spatial variation of the wire's quasicharge $Q(x)$ on these length scales, since $-\partial_xQ\equiv\rho_\perp$, the polarization charge per length stored on $\mathcal{C}_\perp$. This is dual to the usual lumped-element treatments of JT \cite{JJmicro,orlando}, where in calculating the microscopic Josephson coupling the gauge-invariant phase difference across the junction is assumed not to vary spatially across the junction area. This corresponds to neglecting the geometrical inductance inside the Josephson barrier and therefore the magnetic fields generated in it by currents, which is valid for JJs much smaller than the Josephson penetration depth $\lambda_{\rm J}$ \cite{orlando}.

As indicated in fig.~\ref{fig:tline2}(a), we also associate with each segment of the wire a nonlinear kinetic inductor (indicated by a JJ symbol). For the $j^{\rm th}$ segment this inductor has a quasiflux variable $\Delta\Phi_j$ defined by: $\Delta\Phi_j=\oint_{(j-1)l_\phi}^{jl_\phi}\nabla\Phi(x)\rmd x$, such that the quasiflux at the end of the $j^{\rm th}$ segment defined relative to the end of the wire is: $\Phi_j\equiv\sum_{k=1}^j\Delta\Phi_k$. We take the boundary conditions for a single, isolated QPS event in the $j^{\rm th}$ segment to be: $\Delta\Phi_k=0,\;\forall k\neq j$, such that $\Phi(x)$ during the event is fixed everywhere along the wire but inside that segment\footnote[1]{Note that this is a different boundary condition than used for the calculation of the thermal phase-slip energy barrier by LAMH \cite{LA,MH}, where a fixed phase difference across the wire was assumed (more precisely, a fixed $V=0$). Here, we allow the phase across a segment in which an isolated QPS event occurs (and therefore across the wire's ends) to vary \textit{freely}, which essentially corresponds to the absence of any phase damping (the effects of damping due to the electromagnetic environment will be considered in sections~\ref{s:tline} and~\ref{s:expt} below). This is dual to the implicit assumption used in the calculation of the Josephson coupling for a JJ that there is no charge damping.}. We can then treat the kinetic inductor of each segment in terms of a local potential energy $U(\Delta\Phi_j)$ (i.e. the kinetic-inductive energy evaluated as a function of fixed $\Delta\Phi_j$). This function is $\Phi_0$-periodic, with a minimum whenever $\Delta\Phi_j$ is an integer multiple of $\Phi_0$, very similar to a JJ [c.f., eq.~\ref{eq:JJham}] (although $U(\Delta\Phi_j)$ becomes less and less like a simple cosine as $l_\phi$ increases beyond $\xi$ \cite{likharev}).

The model of fig.~\ref{fig:tline2}(a) is similar to a 1D JJ array, in the so-called ``nearest-neighbor" limit \cite{doniach,choi} which applies on length scales much longer than the Coulomb screening length [c.f., eq.~\ref{eq:lambda1D}]. In this case it is advantageous to use a loop variable representation, rather than a node variable representation \cite{yurke,devoret}, since in the latter case the interactions between node charges are highly nonlocal. We define the loop charges $\hat\Lambda_j$ as shown in the figure, which are the canonical momenta for the position variables $\Delta\hat\Phi_j$ such that $[\Delta\hat\Phi_j,\hat\Lambda_k]=\rmi\hbar\delta_{j,k}$. In this representation, the classical Euclidean action of the system is:

\begin{equation}
\mathcal{S}=\sum_j\int_0^{\hbar\beta}\rmd\tau\left[\frac{[\Lambda_j-Q(x)]^2}{2C_l}+U(\Delta\Phi_j)\right]\label{eq:S0eqn}
\end{equation}

\noindent where $\tau\equiv \rmi t$, $\beta\equiv1/k_{\rm B}T$, and we are primarily interested in the $\beta\rightarrow\infty$ limit. Equation~\ref{eq:S0eqn} describes the motion of independent fictitious particles with positions $\Delta\Phi_j$ and mass $C_l$, under the influence of the periodic kinetic-inductive potential $U(\Delta\Phi_j)$:

\begin{eqnarray}
U(\Delta\Phi_j)&\equiv&\int_0^{\Delta\Phi_j}I(\Delta\Phi^\prime)\rmd(\Delta\Phi^\prime)\nonumber\\*
&\approx&V_{\rm 1D}\left[1-\cos\phi_j+\frac{l_\phi^2}{15\xi^2}\left(\frac{3}{4}-\cos\phi_j+\frac{\cos2\phi_j}{4}\right)\right]\label{eq:Uphi}
\end{eqnarray}

\noindent where $I(\Delta\Phi)$ is the current-phase relation for each segment, which we take from the theory of Aslamazov and Larkin \cite{AL} to yield the result on the second line, in which the quantity $V_{\rm 1D}\equiv A_{\rm cs}\Phi_0^2/2\pi\Lambda l_\phi$ can be viewed as a 1D superfluid stiffness \cite{kivelson}, and $\phi_j\equiv2\pi\Delta\Phi_j/\Phi_0$. Equation~\ref{eq:Uphi} holds approximately for short lengths up to $l_\phi\sim\xi$. For longer lengths, $U(\Delta\Phi_j)$ can be evaluated numerically using the results of ref.~\cite{likharev}. The QPS contribution to the ground state can be evaluated in this simplified model by seeking stationary, topologically nontrivial paths connecting the endpoints: \{$\Delta\Phi_j(\tau),\tau$\}$=$\{$m\Phi_0,0$\} and \{$(m\pm1)\Phi_0,\hbar\beta$\}, where $m$ is an integer. In the $\beta\rightarrow\infty$ limit, these are known as vacuum instantons \cite{JQL}, and the corresponding solution is well known in the semiclassical approximation (where $S_0\gg1$) in the case of a simple cosine potential\footnote[7]{We have numerically evaluated the correction to this (and subsequent results) due to a nonsinusoidal $I(\Delta\Phi)$ for segment lengths up to $l_\phi\approx3.48\xi$, where the current-phase relation becomes multivalued and there is no longer a classical Euclidean path connecting the relevant endpoints \cite{likharev}; we find only corrections at the $\sim$10\% level, irrelevant at the crude level of approximation being used here.}, having total action:

\begin{figure}[t]
\begin{center}
\resizebox{0.9\linewidth}{!}{\includegraphics{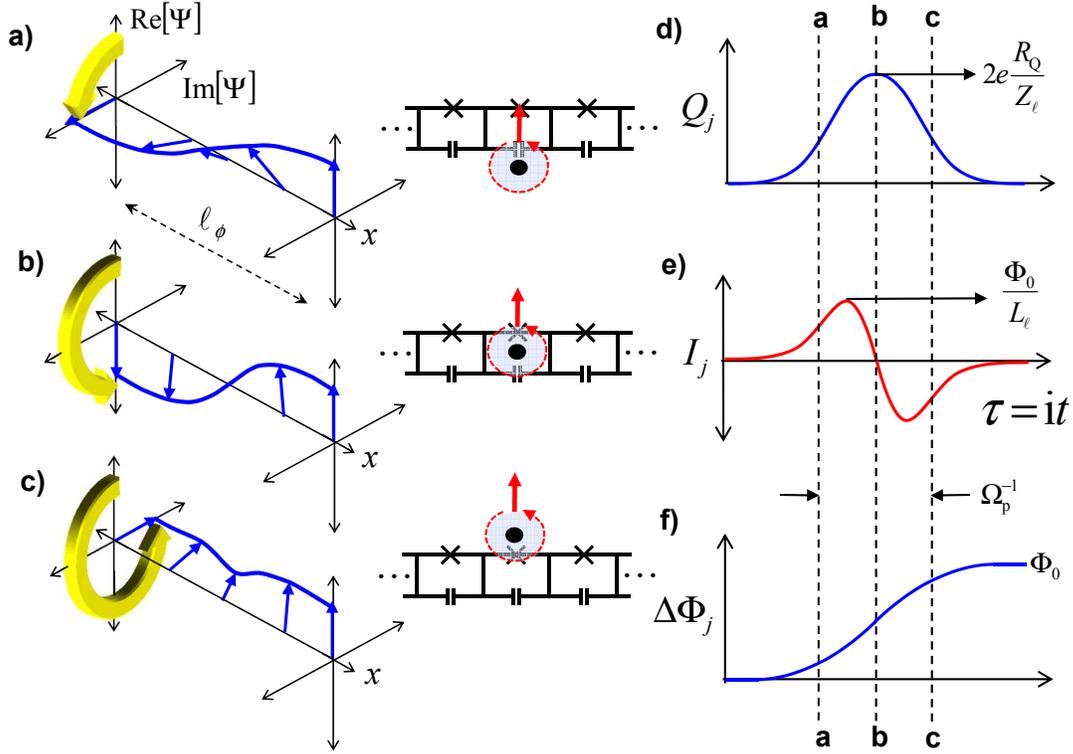}}
\caption{Schematic picture of quantum phase slip in our model. Panels (a)-(c) show the wire's order parameter along the $j^{\rm th}$ link of length $l_\phi$ at three different times. Panels (d)-(f) plot the (lumped) link quantities as a function of time, with the times corresponding to (a), (b), and (c) marked by the vertical dashed lines. (a) Over a length $l_\phi$, a transient current flows, charging up $C_l$ (the corresponding displacement current makes the total current zero, and no net quasicharge moves along the wire), such that $\Delta\Phi_j$ winds up; This can be viewed as a fluxon beginning to pass through the wire; (b) At the ``core" of the QPS, the current is zero, the charge on $C_l$ has reached a maximum, and a gauge-invariant phase difference of $\pi$ appears between the wire's ends; this can be viewed as a fluxon (virtually) inside the wire; (c) The current reverses, discharging $C_l$. The wire returns to its initial state, with a net quasiflux evolution between the wire's ends of $\Phi_0$, corresponding to passage (tunneling) of a fluxon through the wire. }
\label{fig:QPS}
\end{center}
\end{figure}

\begin{equation}
S_0\approx8\frac{V_{\rm 1D}}{\hbar\tilde\Omega_{\rm p}}\;\;\;\;\;\tilde\Omega_{\rm p}\equiv\Omega_{\rm p}\sqrt{\frac{C_{||}}{C_l}}\label{eq:SQPS}
\end{equation}

\noindent where $\Omega_{\rm p}$ is the bulk Cooper pair plasma frequency \cite{mooijschon,orlando} defined above [c.f., eq.~\ref{eq:CPplasma}] and $\tilde\Omega_{\rm p}$ is the corresponding plasma frequency for the length scale $l_\phi$, including the effect of fields outside of the wire. The Euclidean time dynamics of the order parameter corresponding to this solution are illustrated in fig.~\ref{fig:QPS}.

The frequency $\tilde\Omega_{\rm p}$ is in general greater than the gap frequency, so that any \textit{classical} oscillations at $\tilde\Omega_{\rm p}$ would be essentially those of a normal metal; however, such classical dynamics would occur only at very high energy. Here, we are concerned instead with zero-temperature, quantum fluctuation corrections to the ground state of the superconductor, such that the characteristic time over which the system can virtually occupy energy states near the top of the barrier ($\sim\hbar/V_{\rm 1D}$) is much shorter than the characteristic decay time for the order parameter ($\sim\tau_{\rm GL}$, the GL relaxation time). In this limit, we can neglect the dissipation (corresponding to breaking of Cooper pairs) that would inevitably occur on longer timescales. This situation is analogous, for example, to the perturbative treatment of Josephson tunneling within the BCS theory of superconductivity, which can be understood as arising through virtual excitation of quasiparticles, which are also dissipative degrees of freedom \cite{martinisJJ}. Another example is the case of Raman transitions between discrete ground states in an atomic system via an electronic excited state (or even multiple excited states) with a short lifetime $\Gamma_{\rm e}^{-1}$; the excited state is occupied only virtually for a time: $\Delta_{\rm e}^{-1}\ll\Gamma_{\rm e}^{-1}$ where $\Delta_{\rm e}$ is the detuning of a driving field from resonance with the optical transition between ground and excited states, such that spontaneous scattering into the radiation continuum via the excited state (the equivalent of electrical dissipation in our case) can be neglected. In both examples the decay of excited states can be approximately neglected when compared to the coherent, low-energy process of interest, and the excited state can be ``adiabatically eliminated" \cite{adiabatic} to produce an effective potential energy for the ground state\footnote[1]{An exception to this is when degrees of freedom external to the quantum system of interest have excited states which are populated, and whose stored energy can be exchanged with the system. In the present context of quantum circuits, this corresponds to a resistive electromagnetic environment. For the purposes of QPS in our model, there are three possible sources of such dissipation: (i) the intrinsic resistance of the metal at $\tilde\Omega_{\rm p}$, whose effect we can neglect compared to its inductive response as long as $\tilde\Omega_{\rm p}\tau_{\rm s}\gg 1$ [c.f., eq.~\ref{eq:sigma}]; (ii) the transverse radiation continuum in the medium surrounding the wire with impedance $\lesssim 377\Omega$, which has negligible coupling to QPS since $l_\phi$ is orders of magnitude smaller than the wavelength corresponding to $\tilde\Omega_{\rm p}$ in this medium; and (iii) the propagating plasma oscillation modes on the wire, which are excluded by construction from the model of fig.~\ref{fig:tline2}(a) since the loop charges $\Lambda_i$ do not interact. We will add back in the effect of these modes when we consider distributed systems in section~\ref{s:tline}.}.

The resulting approximate expression (when $S_0\gg1$) for the ground-state energy per unit length\footnote[7]{There will, of course, be higher energy bands in this potential as well, corresponding to excited states of the Cooper pair plasma oscillation; however, these will be extremely short-lived, since at such high energies the Cooper pairs will no longer be bound.} can be written in terms of the action $S_0$ \cite{JQL,instantons,schon}:

\begin{eqnarray}
\mathcal{E}_{\rm QPS}(Q)&\approx&\frac{\hbar\tilde\Omega_{\rm p}}{l_\phi}\left[\frac{1}{2}-\sqrt\frac{2S_0}{\pi}\rme^{-S_0}\cos\left(2\pi\frac{Q}{2e}\right)\right]\label{eq:E0}\nonumber\\
&\equiv&\mathcal{E}_0-\mathcal{E}_{\rm S}\cos{q}
\end{eqnarray}

\noindent where $q\equiv \pi Q/e$ is the dimensionless quasicharge. Using eqs.~\ref{eq:SQPS} and ~\ref{eq:E0}, we can then write the phase-slip energy per unit length as:

\begin{equation}
\mathcal{E}_{\rm S}\equiv\frac{E_{\rm S}}{l}=\frac{2}{l_\phi}\sqrt{\frac{\hbar\tilde\Omega_{\rm p} V_{\rm 1D}}{\pi}}\exp\left[-\frac{8V_{\rm 1D}}{\hbar\tilde\Omega_{\rm p}}\right]\label{eq:ES}
\end{equation}

\noindent This quantity is arguably the central parameter for QPS. It has been identified \cite{PSqubit,mooijfluxchg} with the``rate" of quantum phase slips estimated by Giordano \cite{giordano}, and later calculated by several authors using time-dependent GL theory \cite{saito,chang,duan}, and by GZ using microscopic theory \cite{GZPRL,GZPRB}. In one form or another, it is the essential input parameter to all subsequent theoretical work aimed at deducing the effects of QPS, appearing as the dual of the Josephson energy in lumped-element treatments \cite{tinkSIT,PSqubit,mooijfluxchg,khlebsimple}, and in more recent theories in terms of the so-called ``QPS fugacity" $f\equiv e^{-S_0}$ \cite{khlebSIT,buchler,demler,meidan}. In all of these cases it is either left as an unknown input parameter, or taken from the results of GZ or earlier authors.

Previous results have been based on an action of the form (up to numerical factors): $S_0\sim\delta E_{\rm LAMH}/\Delta$ \cite{giordano,saito,chang,tinkSIT,GZPRL,GZPRB,astafiev} where $\delta E_{\rm LAMH}\sim U_CA_{\rm cs}\xi$ is the free energy barrier originally used by LAMH \cite{LA,MH} for thermal phase slips, and $\Delta$ is the superconducting gap. Since the QPS action $S_0$ can be viewed as the ratio of the potential energy barrier for phase-slips to the energy scale of the quantum phase fluctuations which produce tunneling through that barrier ($S_0\sim$ barrier height $\times$ characteristic quantum fluctuation time), this form is essentially consistent with Giordano's original hypothesis: that the relevant ``kinetic" energy scale for QPS is $\sim\Delta\propto\hbar/\tau_{\rm GL}$. By contrast, in our model the quantum phase fluctuations arise from a qualitatively different source, being associated with a virtual plasma oscillation involving the Cooper pairs and the electric permittivity of the environment in which they are embedded.

This picture of QPS has an appealing symmetry with Josephson tunneling, as illustrated by our model of fig.~\ref{fig:tline2}(c) and the dual model of fig.~\ref{fig:tline2}(d) for JT: in both cases, the source of quantum tunneling can be traced back to the \textit{finite mass of the superconducting electrons}. For the PSJ (JJ), when these electrons are confined inside a sufficiently narrow region around the quasi-1D wire (the slotline formed by the JJ barrier), the corresponding short-wavelength zero-point fluctuations of their plasma modes allow the phase (charge) to undergo tunneling between adjacent potential minima, producing QPS (JT). A crucial point about this confinement for QPS is that the phase-slip energy can become appreciable already at wire diameters still much too large for the zero-point phase fluctuations to have any impact on the Cooper pairing itself, resulting in the coexistence of a pairing (superconducting) energy gap with insulating behavior (i.e., $Q$ is completely localized). This is similar to the case of a Coulomb-blockaded JJ \cite{corlevi,hekking}, and may also be related (albeit more indirectly) to the observation of a local pairing gap in highly-disordered, thin superconducting films on the insulating side of a SIT \cite{gapins}. We discuss the latter point further in section~\ref{s:SIT}.

Our model for lumped-element QPS also provides a natural intuition for the origin of the kinetic capacitance (dual to the Josephson inductance) suggested by MN. Written as a distributed quantity (in units of Farads$\times$length) it is:

\begin{equation}
\mathcal{C}_{\rm k}=\left[\frac{d^2}{dQ^2}\mathcal{E}_{\rm QPS}(Q)\right]^{-1}\equiv\frac{\mathcal{C}_{\rm k0}}{\cos{q}}
\end{equation}

\noindent where $q\equiv\pi Q/e$ and:

\begin{eqnarray}
\mathcal{C}_{\rm k0}&\equiv&\left(\frac{2e}{2\pi}\right)^2\frac{1}{\mathcal{E}_{\rm S}}\\
&\approx&(C_ll_\phi) \times\sqrt{\frac{2}{\pi}}\frac{e^{S_0}}{S_0^{3/2}}\;\;,\hspace{0.5cm}S_0\gg 1\label{eq:Ck}
\end{eqnarray}

\noindent The form of eq.~\ref{eq:Ck} suggests that the kinetic capacitance is simply a remnant of the ``bare", purely geometric series capacitance $C_l$, \textit{renormalized} by QPS. That is, in the limit of very strong QPS ($V_{\rm 1D},S_0\rightarrow0$) the wire acts simply like a dielectric rod whose behavior is governed only by the bound charges associated with the capacitance $C_l$ of each segment; as the superfluid stiffness is increased from zero, the kinetic capacitance increases smoothly from the bare value, eventually increasing exponentially as superconductivity is further strengthened, such that the corresponding QPS energy goes to zero. This is the exact dual of the JT case, where the Josephson inductance of the junction can be viewed as a renormalized ``remnant" of the bare (bulk) kinetic inductivity of the superconducting electrodes.

Another interesting result of the model presented so far is that at a given point in the wire, the QPS amplitude depends not just on the properties of the wire itself, but also on the permittivity of the dielectric medium immediately outside it, according to eq.~\ref{eq:QPScap}. The narrower the wire, and the smaller the ratio $\epsilon_{\rm in}/\epsilon_{\rm out}$, the greater the penetration of QPS electric fields into the region outside the wire\footnote[1]{Of course, this is the case in our model in a sense by construction, since we have fixed the length scale for QPS at $l_\phi$; however, in a truly continuous theory for QPS at short length scales we would not expect this to change qualitatively, since it will never be energetically favorable for QPS to occur with appreciable amplitude over arbitrarily short length scales $\ll\xi$ (equivalently, the potential energy barrier for a fluxon to tunnel through the continuous wire entirely in between two points separated by a distance $\ll\xi$ will be very high).}. This kind of nonlocality is exactly dual to what occurs in a JJ, where the tunneling energy $E_{\rm J}$ depends not just on the properties of the barrier itself, but also on the kinetic inductivity of the ``surrounding" superconductor of the adjacent electrodes. Thus, in the JT (QPS) case, stronger quantum tunneling occurs when the superconducting (insulating) gap of the surrounding medium is large, and the insulating (superconducting) gap of the tunnel barrier is small\footnote[7]{In this description, a large insulating gap of the dielectric surrounding a quasi-1D wire would be associated with a small polarizability and therefore a small $\epsilon_{\rm out}$, just as a large superconducting gap for the electrodes of a JJ is associated with a small kinetic inductivity.}.

Before proceeding to the next section, we discuss briefly the ``weak" QPS assumption which underlies the model of fig.~\ref{fig:tline2}(a). In our derivation of eq.~\ref{eq:ES} above, the assumption that QPS is ``weak" took the form of a semiclassical approximation to the full 1+1D quantum field theory, in which the QPS action $S_0$ was taken to be large. In the usual mapping from 1+1D Euclidean space at $T=0$ to the equivalent 2D classical statistical mechanics problem \cite{susskind,girvinRMP,instantons}, this corresponds to a small fugacity $f=e^{-S_0}$ for the 2D statistical fluctuations corresponding to QPS events in 1+1D. Therefore, these events are rare, their density very low. It is for this reason that the model of fig.~\ref{fig:tline2}(a) is justified, in which simultaneous QPS events in adjacent segments do not interact with each other by construction: such occurrences are ``rare enough" (in Euclidean time) that they contribute negligibly to the partition function. This is a dual statement to the usual perturbative assumption made in the context of JT, which produces the well-known, simple proportionality between the junction's normal state tunneling resistance and its critical current \cite{JJmicro}.

\section{Distributed quantum phase slip junctions}\label{s:tline}

In the previous section, we described our model for QPS on short length scales $l_\phi\sim\xi$, over which electric fields outside of the wire (the wire's shunt capacitance to the environment) were included using a renormalized series capacitance $C_l$ for each discrete segment. We saw that the characteristic (Euclidean) frequency associated with the length scale $l_\phi$ was the renormalized Cooper pair plasma frequency $\tilde\Omega_{\rm p}$. However, we left unspecified the length scale at which lower-energy dynamics would become important, effectively treating the wire as a lumped element. As we will now see, at lower energy scales and longer length scales additional physics will need to be included to treat the fully distributed case.

We make the assumption that a large separation of energy scales exists between that governing QPS at lengths $\sim l_\phi$ and the low-energy dynamics of $Q(x,t)$ we now seek to investigate (we will see below the conditions under which this is justified). Based on this assumption, we treat the phase-slip potential $\mathcal{E}_{\rm QPS}(Q)$ as a purely classical energy which depends only on $Q(t)$ (and not, for example, on $\partial_tQ$). This is analogous to the Born-Oppenheimer approximation often used to treat interatomic interactions, where the microscopic QPS at length scale $\sim\l_\phi$ plays the role analogous to electronic motion, and the slower, lower-energy dynamics of $Q(x,t)$ is analogous to the nuclear motion. It is also the same approximation used in the treatment of classical quasicharge dynamics of lumped Josephson junctions \cite{bloch,panyukov,pistolesi,schon,hekking}. The resulting distributed model for a nanowire is shown in fig.~\ref{fig:tline2}(c), in which $\mathcal{E}_{\rm QPS}(Q)$ is associated with a ``bare" phase slip element in the same way that the Josephson potential $\mathcal{E}_{\rm JJ}(\Phi)$ is associated with a bare Josephson element, as shown in fig.~\ref{fig:tline2}(d). The long-wavelength behavior of the superconducting response is described by the kinetic inductance per length $\mathcal{L}_{\rm k}$, and the distributed shunt capacitance per length $\mathcal{C}_\perp$, where we now assume that the frequencies of interest are low enough that this becomes the wavelength-independent capacitance per length to a nearby ground plane. When QPS is weak ($\mathcal{E}_{\rm QPS}(Q)\rightarrow0$), the wire reduces to a simple, linear transmission line, on which waves propagate at the Mooij-Sch\"{o}n velocity $v_{\rm s}$. In fig.~\ref{fig:tline2}(d) we show the dual to our model, which is simply the nonlinear transmission line (a superconducting slotline) used to describe a long Josephson junction. In the limit of weak Josephson coupling ($\mathcal{E}_{\rm JJ}(\Phi)\rightarrow0$), this becomes a linear transmission line on which waves propagate at the so-called Swihart velocity \cite{swihart} (dual to $v_{\rm s}$).

We now describe the system of fig.~\ref{fig:tline2}(c) in the continuum limit (with the proviso that we only consider length scales $\gg l_\phi$), again using a Euclidean path-integral approach, with partition function \cite{GZPRL,GZPRB,buchler,instantons,girvinRMP}:

\begin{equation}
\mathcal{Z}=\int\mathcal{D}\Psi\exp[-\mathcal{S}(\Psi)]\label{eq:partfunct}
\end{equation}

\noindent where $\mathcal{D}\Psi$ indicates a functional integration over paths in $x,\tau$-space, and the dimensionless Euclidean action is ($\beta\equiv 1/k_{\rm B}T \rightarrow \infty$):

\begin{eqnarray}
\mathcal{S}&=&\frac{1}{\hbar}\int_0^{\hbar\beta} d\tau\int dx
\left \{ \frac{\rho_\perp^2}{2\mathcal{C}_\perp}+\frac{\mathcal{L}_{\rm k}I^2}{2}+\mathcal{E}_{\rm QPS}(Q)\right \}\nonumber\\
&=&\frac{1}{2\pi\mathcal{K}}\int \rmd u\rmd v\left\{\left(\partial_uq\right)^2+\left(\partial_vq\right)^2-\cos{q}\right\}\label{eq:S}
\end{eqnarray}

\noindent In the first line, $I=\partial_tQ$ and $\rho_\perp=-\partial_xQ$ are the current flowing through $\mathcal{L}_{\rm k}$ and linear charge density stored on $\mathcal{C}_\perp$ at the spacetime point $x,\tau$, and for the second line we have defined:

\begin{eqnarray}
\mathcal{K}&\equiv&\frac{R_{\rm Q}}{Z_{\rm L}}\label{eq:K}\\
u&\equiv&\frac{x}{\lambda_{\rm E}}, \;v\equiv\omega_{\rm p}\tau\label{eq:uv}\\
\lambda_{\rm E}^2&\equiv&\frac{\mathcal{C}_{\rm k0}}{\mathcal{C}_\perp}=l_\phi^2\times\left(\frac{\pi }{4\mathcal{K}}\right)^2\sqrt{\frac{2S_0}{\pi}}e^{S_0}\gg l_\phi^2,\Lambda_{\rm 1D}^2\label{eq:lE}\\
\omega_{\rm p}^2&\equiv&\frac{1}{\mathcal{L}_{\rm k}\mathcal{C}_{\rm k0}}=\tilde\Omega_{\rm p}^2\times\sqrt{\frac{\pi}{2}}S_0^{3/2}e^{-S_0}\ll\tilde\Omega^2_{\rm p}\label{eq:wp}
\end{eqnarray}

\noindent The quantities $\lambda_{\rm E}$ and $\omega_{\rm p}$ are dual to the Josephson penetration depth and Josephson plasma frequency in a long JJ, respectively; we hereafter refer to them as the electric penetration depth and phase-slip plasma frequency. Note that $\lambda_{\rm E}$ is defined as a ratio of the effective series kinetic capacitance to the parallel shunt capacitance, and is therefore a kind of Coulomb screening length similar to $\Lambda_{\rm 1D}$ [c.f., eq.~\ref{eq:lambda1D}]; however, as indicated on the right side of the equation, it is exponentially large (for $S_0\gg1$) compared to microscopic quantities. A corresponding relationship exists between the plasma frequencies: $\omega_{\rm p}\ll\tilde\Omega_{\rm p}$. These are precisely the separation of length and energy scales that justify the Born-Oppenheimer approximation underlying the model of fig.~\ref{fig:tline2}(c).


Returning to action of eq.~\ref{eq:S}, the corresponding Euclidean equation of motion is the sine-Gordon equation \cite{instantons}:

\begin{equation}
\nabla_{uv}^2q+\sin{q}=0\label{eq:sineg}
\end{equation}

\noindent where $\nabla_{uv}\equiv\hat{\bi u}\partial_u+\hat{\bi v}\partial_v$ (${\hat{\bi u}}$ and ${\hat{\bi v}}$ are unit vectors) and the dimensionless coordinates $u$ and $v$ were defined in eq.~\ref{eq:uv}. Equation~\ref{eq:sineg} is the exact dual of the usual semiclassical result for a long Josephson junction \cite{orlando} (which is simply eq.~\ref{eq:sineg} with $q$ replaced by $\phi$, the gauge-invariant phase difference across the junction [c.f., fig.~\ref{fig:tline2}(d)]), and is also similar to results for long 1D JJ arrays in the charging limit \cite{havilandCPS,schonCPS,havilandKI,benjacobCPS}. We can therefore infer several things: First, we have the usual propagating modes with dispersion relation: $\omega^2=\omega_{\rm p}^2+(kv_{\rm s})^2$ \cite{orlando}, which are the dual of Fiske modes in long JJs \cite{fiske}, and are also analogous to spin-wave excitations in the corresponding classical 2D XY model \cite{berezinskii,kosterlitz,KT,minnhagen}. We make the usual assumption \cite{buchler} that these Gaussian fluctuations can be factorized out in eq.~\ref{eq:partfunct} such that they simply renormalize the bare parameter values in $\mathcal{S}$, leaving only topologically nontrivial paths to be evaluated. Next, we can infer the existence of a charged soliton \cite{havilandCPS,schonCPS,havilandKI,benjacobCPS}, or so-called ``kink" excitation \cite{instantons} in the field $q(x)$ of size $\sim\lambda_{\rm E}$, with total charge $2e$ (residing on $\mathcal{C}_\perp$), and which can propagate freely without deformation. This is the dual of a Josephson vortex in a long JJ \cite{orlando}, which is a kink in the field $\phi(x)$ of spatial extent $\sim\lambda_{\rm J}$ (the Josephson penetration depth), that carries a total flux $\Phi_0$.

For large enough systems where $\lambda_{\rm E}$ can be used as the ultraviolet cutoff, this 1+1D quantum sine-Gordon model can be mapped to the well-known classical statistical mechanics of 2D magnetic domain interfaces in the 3D Ising model \cite{chui}. Our $q$ maps to the height (in the $z$-direction) of a domain boundary surface between two spin orientations, while the cosine potential ``enforces" the lattice periodicity. The Ising interactions between nearest neighbors in the $x$ and $y$ directions map to the $(\partial_u)^2$ and $(\partial_v)^2$ terms in eq.~\ref{eq:S}. The 3D Ising system undergoes an interfacial roughening transition with increasing temperature $T$ at a critical value $T_{\rm C}\sim J/k_{\rm B}$ (with $J$ the Ising coupling) which has identical universal behavior to the BKT transition in the classical 2D XY model \cite{berezinskii,kosterlitz,KT,minnhagen}. The transition occurs when statistical fluctuations corresponding to localized regions where a step upward or downward occurs in the interface grow to large sizes and proliferate. For our system, this maps to a $T=0$ quantum phase transition at $\mathcal{K}\sim1$ in which virtual soliton-antisoliton pairs unbind, producing charge fluctuations that destroy the insulating state associated with a well-defined $q$ \cite{havilandCPS}.

Our description so far has been well suited to the insulating side of this transition ($\mathcal{K}<1$), where $q$ becomes increasingly well-defined as $\mathcal{K}\rightarrow0$. However, most experiments aiming to observe evidence for QPS have used wires nominally in the superconducting state, about which phase fluctuations can be viewed as a perturbation. Therefore, it makes sense also to examine our system on the superconducting side of the transition ($\mathcal{K}>1$), where $\phi$ becomes increasingly well-defined as $\mathcal{K}\rightarrow\infty$. To do this, it is illustrative to rewrite eq. \ref{eq:sineg} in the following form:

\begin{eqnarray}
\nabla_{uv}\times{\bi q}={\bi j}\label{eq:ampere}\\
\nabla_{uv}\times {\bi j}=-{\bi e}\label{eq:london2}\\
\nabla_{uv}\cdot{\bi j}=0\label{eq:cont}
\end{eqnarray}

\noindent with the definitions:

\begin{eqnarray}
{\bi j}&\equiv&\mathcal{K}\nabla_{uv}\phi\nonumber\\
&=&\frac{\pi}{e}\left[\frac{I}{\omega_{\rm p}}\hat{\bi u}+\rho_\perp\lambda_{\rm E}\hat{\bi v}\right]\label{eq:j}
\end{eqnarray}

\begin{equation}
\left. \begin{array}{cl}
{\bi e}&\equiv(E/E_{\rm C})\;{\hat{\bi z}}\nonumber\label{eq:e}\\
{\bi q}&\equiv q\;\hat{\bi z}\label{eq:d}\nonumber
\end{array} \right\}\;\;{\bi e}=\mathcal{C}_{\rm k0}\int_0^q\frac{\rmd{\bi q}^\prime}{\mathcal{C}_{\rm k}(q^\prime)}
\end{equation}

\noindent where $E$ is the electric field, $E_{\rm C}\equiv e\mathcal{E}_{\rm S}/\pi$ is the critical electric field such that $E/E_{\rm C}=\sin{q}$, and eq.~\ref{eq:cont} follows from continuity. Equations \ref{eq:ampere} and \ref{eq:london2} have an identical form to Amp\`{e}re's law and London's second equation in 2D which govern the equilibrium penetration of a perpendicular magnetic field into a thin, type II superconducting film \cite{orlando}, with the correspondence: ${\bi q}\leftrightarrow{\bi H}$, ${\bi e}\leftrightarrow{\bi B}$, ${\bi j}\leftrightarrow{\bi J}$ and where the right side of eq.~\ref{eq:e} plays the role of the constitutive relation between ${\bi H}$ and ${\bi B}$. These equations, however, describe the dynamical penetration in 1+1D of  \textit{longitudinal electric field} into a superconducting wire\footnote[1]{Note that the $\hat{\bf z}$ direction is purely fictitious here, and defined only to permit the aforementioned analogy. Similarly, the quantity ${\bi j}$ is not to be confused with an actual current density, although it plays the analogous role in eqs. \ref{eq:ampere}-\ref{eq:london2} to the current density in the Maxwell-London equations; its $u$ component is proportional to the total current flowing in the wire at a given spacetime point, and its $v$ component is proportional to the linear charge density $\rho_\perp$ at that point. Formally similar methods for describing electric fields in superconductors in 1+1D were also used in refs.~\cite{ivlev,saito}.}. The analog to the GL $\kappa$ parameter for our 1+1D system is:

\begin{equation}
\kappa_{\rm E}\equiv\frac{\lambda_{\rm E}}{l_\phi}\label{eq:kappa}
\end{equation}

\noindent and the type II limit $\kappa_{\rm E}\gg1$ is automatically satisfied when $S_0\gg 1$ [c.f., eq.~\ref{eq:lE}], a precondition of our analysis.

\begin{figure}[t]
\begin{center}
\resizebox{1.0\linewidth}{!}{\includegraphics{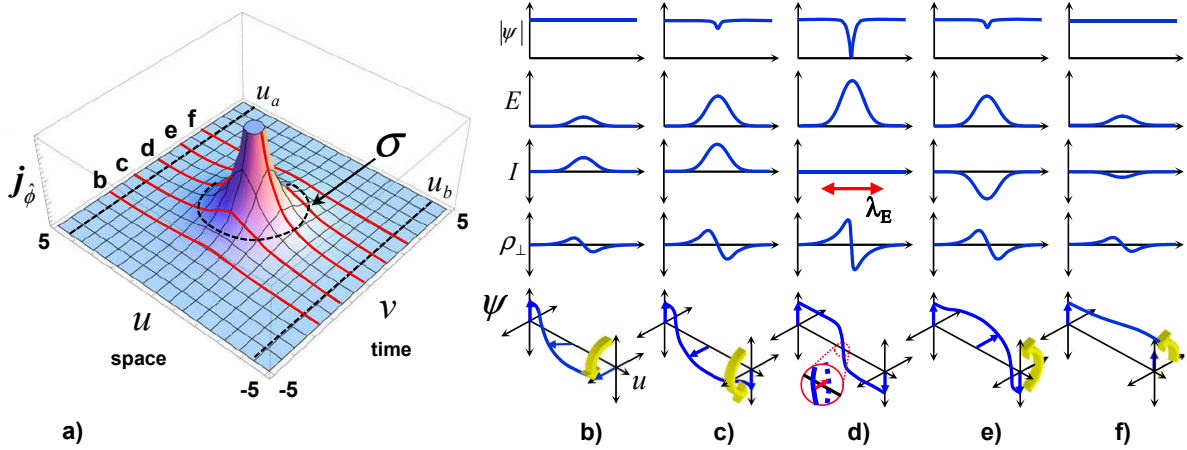}}
\caption{Type II phase slip in a 1D superconductor. In 1+1D, a normal core of size $\sim l_\phi$ is surrounded by circulating ``currents" ${\bi j}$ [eqs.~\ref{eq:j}, \ref{eq:typeIIj}] plotted in (a). Over the course of the event, the quasiflux between the positions $u_a$ and $u_{\rm b}$ evolves by $\pm\Phi_0$. A possible curve $\sigma$ for the line integral of eq.~\ref{eq:phi0} is shown as a solid black line. Panels (b)-(f) show, for the five fixed times ($v$ values) marked by red lines in (a), the corresponding magnitude of the order parameter $|\Psi|$, the local electric field $E$, the local current $I$, the polarization charge per length $\rho_\perp$ on $\mathcal{C}_\perp$, and finally $\Psi$ as a phasor, all as a function of position $u$ along the wire. (b)-(c) On the leading edge of the vortex, current begins to flow over a length $\sim\lambda_{\rm E}$ in the $+\hat{\bi u}$ direction; this begins to charge up $\mathcal{C}_{\rm k}$, producing a gradient in $\rho_\perp$ (the $\hat{\bi v}$ component of ${\bi j}$). (d) Once the total quasiflux across the wire comes close to $\Phi_0/2$, the order parameter can evolve continuously to a state in which the current is zero and there is a null at the center of the vortex of spatial length $\sim l_\phi$ and duration $l_\phi/v_{\rm s}$. At this point the order parameter ``passes through" the u-axis, and the supercurrent reverses. (e)-(f) The current in the $-\hat{\bi u}$ direction then discharges the kinetic capacitance as it ramps down to zero. The null in the order parameter as a function of $u$ at $v=0$ shown in (d), top panel, is effectively a saddle point for the system, closely related to those encountered in long weak links \cite{likharev} and in LAMH phase slips \cite{little,LA,MH}. Our 1+1D solution in $u,v$ for the screening ``currents" ${\bi j}$ surrounding the vortex core corresponds to an instanton \cite{GZPRL,GZPRB,doniach,buchler} in $x,\tau$, and describes the dynamics by which the system tunnels through this energy barrier and passes through the saddle point. This is a macroscopic quantum process that arises out of (microscopic) QPS, whose lumped-element limit is dual to Bloch oscillation in a JJ (which arises in an analogous manner from the microscopic process of JT) \cite{bloch,panyukov,pistolesi,schon,hekking}.}
\label{fig:typeII}
\end{center}
\end{figure}

Interestingly, it turns out that there are 1+1D electric analogs for many well-known features of type II magnetic flux penetration, starting with the magnetic vortex. We call this 1+1D dynamical process, illustrated in fig.~\ref{fig:typeII},  a ``type II phase slip". It is a topologically nontrivial solution to eqs.~\ref{eq:ampere}-\ref{eq:j}, in which a normal core of size $\sim \kappa_{\rm E}^{-1}$ in $u,v$ is surrounded by circulating screening ``currents" ${\bi j}$ [c.f., eq.~\ref{eq:j}] extending out to $\rho\equiv\sqrt{u^2+v^2}\sim1$. In order to include only closed paths in eqs.~\ref{eq:partfunct} and \ref{eq:S}, we must impose the condition (analogous to fluxoid quantization in the 2D magnetic case \cite{orlando}):

\begin{equation}
\oint_\sigma{\bi j}\cdot{\bi d}{\bi s}+\int_\alpha{\bi e}\cdot{\bi d}{\bi a}=\pm2\pi\label{eq:phi0}
\end{equation}

\noindent where $\sigma$ is a closed curve in the $uv$ plane which contains the core and bounds the surface $\alpha$ [fig.~\ref{fig:typeII}(a)]. This condition means that the quasiflux $\Phi_{ab}$ between spatial points $u_a$ and $u_{\rm b}$ on either side of the vortex evolves by $\Phi_0$ (-$\Phi_0$) during the event. Using eqs.~\ref{eq:ampere}-\ref{eq:phi0}, and assuming that far from the core of the phase slip we can write: $\mathcal{C}_{\rm k}(q)\approx\mathcal{C}_{\rm k0}$ and $\mathcal{L}_{\rm k}(I)\approx\mathcal{L}_{\rm k}(0)$ (our 1+1D analog to the usual approximation that far from the core of a magnetic vortex $\Lambda (J)\approx\Lambda(0)$ \cite{orlando}), we obtain [fig.~\ref{fig:typeII}]:

\begin{equation}
\label{eq:typeIIj}
{\bi j}(\rho)=\pm \mathcal{K}K_1\left(\rho\right)\hat{\bi\bphi},\;\;\;\rho\kappa_{\rm E}\gg1
\end{equation}

\noindent where we have also assumed $\hbar\beta\gg\omega_{\rm p}^{-1}$. The resulting Euclidean action for the type II phase slip is then:

\begin{equation}
\mathcal{S}_{\rm II}\approx\frac{\mathcal{K}}{2}K_0\left(\frac{1}{\kappa_{\rm E}}\right)\label{eq:typeIIS}
\end{equation}

\noindent and the action associated with the interaction between type II phase slips separated by $\delta\rho\equiv|\vec\rho_1-\vec\rho_2|$ is:

\begin{eqnarray}
\mathcal{S}_{\rm int}(\delta\rho)&=&\pm\mathcal{K}K_0\left(\delta\rho\right),\;\;\delta\rho\kappa_{\rm E}\gg1\\
&\approx&\mp\mathcal{K}\ln\left(\delta\rho\right), \;\;\delta\rho<1\label{eq:typeIIint}
\end{eqnarray}

\noindent where the sign is negative for a phase slip-anti phase slip pair. The direct analogy between these 1+1D electric results and their 2D magnetic counterparts \cite{orlando} can now be exploited to understand their implications\footnote[1]{This analogy should not be confused with flux-charge duality, in spite of any apparent similarity. In our description, electric fields in 1+1D and magnetic fields in 2D are related by a \textit{Wick rotation} (analytic continuation to imaginary time); a similar relationship exists, for example, between the least-action trajectory of a projectile in 1+1D and the lowest-energy, static solution in 2D for a string suspended at two points. }.

First of all, the quantum mechanics of these vortex objects can be mapped directly to the statistical mechanics of the classical 2D XY model \cite{berezinskii,KT,kosterlitz,minnhagen} (which describes thermodynamic vortex fluctuations in thin superconducting films \cite{VAPsc}, among other things) with effective vortex fugacity: $f=\exp(-S_{\rm II})$ [c.f., eq.~\ref{eq:typeIIS}] and interaction energy: $U_{\rm int}=\hbar\omega_{\rm p}S_{\rm int}(\delta\rho)$ [c.f., eq.~\ref{eq:typeIIint}]. Thus, we expect a BKT vortex-unbinding transition as $\mathcal{K}$ (which corresponds to the temperature of the analogous 2D classical system) is decreased from large values, at $\mathcal{K}\sim1$. The fact that this is the same critical point discussed above in the context of a charged soliton-antisoliton unbinding transition as $\mathcal{K}\sim1$ was approached from below is not an accident; in fact, these are two descriptions of \textit{the same transition}, as discussed in ref.~\cite{chui}. It simply makes more sense to use a vortex representation when $\mathcal{K}>1$ and a charge representation when $\mathcal{K}<1$. The remarkable conceptual similarity between these two representations is an example of Kramers-Wannier duality, originally used in the context of the statistical physics of Ising spin models \cite{KWdual}, and later applied to quantum field theories \cite{savitdual} (a particular example of which is the ``dirty boson" model \cite{fisher} of the 2+1D quantum phase transition in highly disordered superconducting films). In fact, the well-known approximate self-duality for lumped JJs (between the case of high environmental impedance where $q$ is well-defined and low environmental impedance where $\phi$ is well-defined \cite{schmid,schon,grabert}) is a limiting 0+1D example of this same concept.

Before discussing finite wires and comparing our model to experimental observations, we conclude this section with a brief comparison of the established theory of GZ \cite{GZPRL,GZPRB} to what we have presented here so far. The GZ theory is fundamentally a variational calculation, using a microscopic expression for the Euclidean action of the wire (derived from BCS theory). This calculation is also built on a particular ansatz for the form of a QPS event, consisting of two parts: at large distances from the core, the QPS event is simply taken to be the electromagnetic response of the the linear plasma modes of the wire (MS modes) to a topological point defect in 1+1D (i.e., an instanton solution to the linear wave equation for a transmission line, but with an additional delta-like source term in $x$ and $t$); the core is treated separately, and taken to have length and time scales $x_0$ and $\tau_0$ (which are the variational parameters) over which the gap is zero and dissipation is assumed to occur. The result of this calculation, up to numerical factors, is $x_0\sim\xi$ and $\tau_0\sim\hbar/\Delta$, so that:

\begin{equation}
S_{\rm GZ}=A\frac{\delta E_{\rm LAMH}}{\Delta}\propto\frac{R_{\rm Q}}{R_\xi}\label{eq:GZQPS}
\end{equation}

\noindent where $A$ is a material-independent, numerical constant of order unity, and the proportionality on the right side follows from standard BCS relations, with $R_\xi$ the resistance for a length $\xi$ of the wire. Thus, the QPS fluctuation can be interpreted as virtual excitation of the the energy $\delta E_{\rm LAMH}$ for a time $\hbar/\Delta$\footnote[1]{Note that in the GZ theory of ref~\cite{GZPRB}, eq.~\ref{eq:GZQPS} holds when: $l/\xi\ll e^2N_0A_{\rm cs}/\mathcal{C}_\perp$, where $N_0$ is the density of states at the Fermi level. This limit is well-satisfied for all wires in the experiments discussed here.}.

As discussed by GZ and subsequent authors, with a characteristic timescale for QPS of $\tau_0\sim\hbar/\Delta$, the wavelength of MS modes near the corresponding frequency $\tau_0^{-1}$ is much greater than the QPS size, and long enough that these modes are in the region of approximately linear dispersion where there is an approximately wavelength-independent capacitance per unit length $\mathcal{C}_\perp$. Just as is the case with 1D JJ arrays, this shunt capacitance is the source of interactions between QPS events (the currents from two interacting events both charge or discharge the distributed shunt capacitance of the length of wire which separates them). Now, because the distributed shunt capacitance only enters this treatment in the context of the linear MS modes, the long-range QPS interaction is then determined purely by the form of the instanton of the corresponding linear wave equation. This results in a QPS interaction with no natural length scale, falling off purely logarithmically with increasing spacetime separation. This interaction is analogous to that encountered in classical 2D systems of magnetic vortices (in a neutral superfluid) \cite{berezinskii,kosterlitz,KT} or electric charges \cite{minnhagen}, and this brings about an analogy to the BKT transition of the classical 2D XY model\footnote[1]{One important difference is that the QPS fugacity here $y=e^{-S_0}$ is an independent physical parameter from the dimensionless admittance $\mathcal{K}$, whereas in the 2D XY model the two analogous quantities (the vortex fugacity and the temperature) are \textit{not} independent.}~\cite{GZPRL,buchler}. Another consequence of a QPS frequency scale $\tau_0^{-1}\sim\Delta/\hbar$ is the importance of dissipation, and this features prominently in the theory of GZ.

In our model as presented so far, instead of the MS plasma mode dynamics being a linear response to a pointlike defect``source" in 1+1D at the frequency $\tau_0^{-1}$, we describe QPS directly in terms of the zero-point motion of the MS plasma oscillation itself, at a wavelength $l_\phi\sim\xi$ and frequency $\tilde\Omega_{\rm p}$. As described by eqs.~\ref{eq:MScap} and \ref{eq:QPScap}, at these wavelengths charged fluctuations are screened out on the length scale $\Lambda_{\rm 1D}$ (analogous to the well-known Coulomb screening length in 1D JJ arrays \cite{havilandCPS,schonCPS,havilandKI,benjacobCPS}), such that QPS interactions are cut off at distances larger than this. This, in conjunction with the semiclassical approximation $S_0\gg1$, is what allowed us to use the lumped-element model of fig.~\ref{fig:tline2}(c) which neglects interactions between QPS events entirely. These interactions came back in to our problem when we considered the fully distributed case, involving longer length scales $\lambda_{\rm E}\gg\xi\sim l_\phi$ and lower energy scales $\omega_{\rm p}\ll\tilde\Omega_{\rm p}$.

\section{Finite wires and experimental systems}\label{s:expt}

\begin{figure}[t]
\begin{center}
\resizebox{0.85\linewidth}{!}{\includegraphics{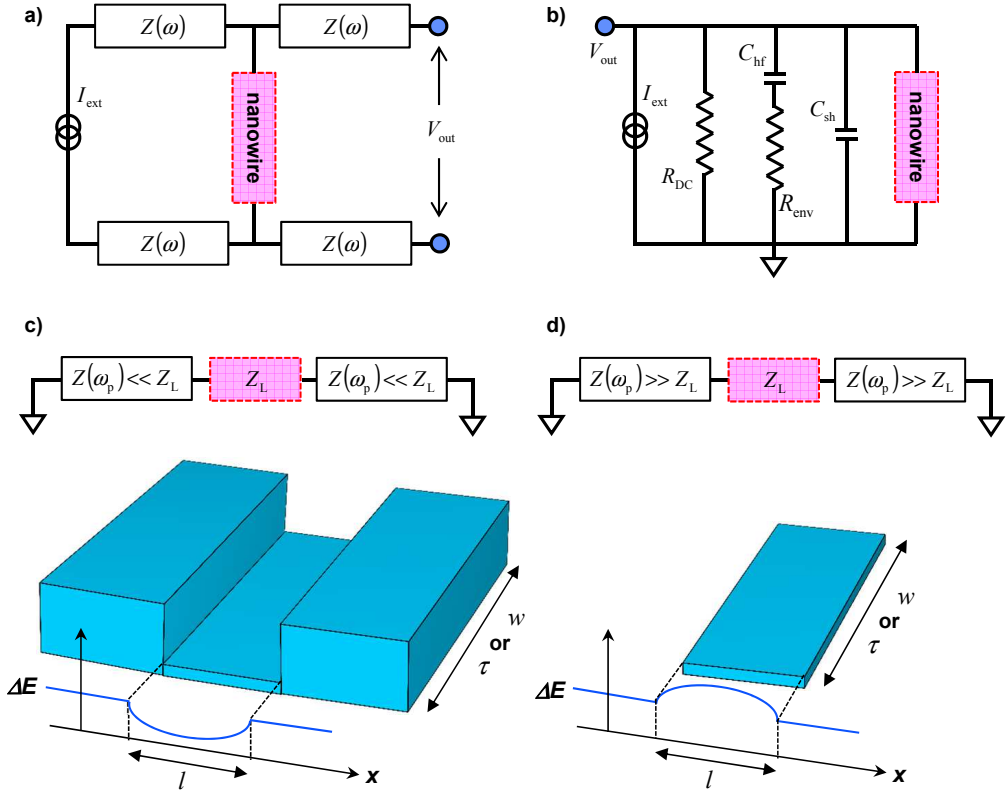}}
\caption{Experimental configuration for typical QPS measurements. (a) four-wire configuration used in typical $R$ vs. $T$ measurements. (b) lumped circuit model of the electromagnetic environment, following ref.~\cite{kautz}. At low frequencies, the wire effectively sees a current source with large DC compliance $R_{\rm DC}$, but at high frequencies lumped parasitics and the characteristic impedance of the measurement connections reduce the effective impedance. This is modeled by a lumped shunt capacitance $C_{\rm sh}$ in parallel with a high-frequency resistance $R_{\rm env}$, which becomes important above the high-pass corner frequency $(R_{\rm env}C_{\rm hf})^{-1}$. (c) in nearly all experiments where specialized techniques are not used to control the high-frequency EM environment, the dominant contribution to this environment is $R_{\rm env}$, which is likely to be $\ll Z_{\rm L}$, the linear impedance of the nanowire. In this limit, the interaction of a type II phase slip with the wire edges can be described in terms of image phase-slips of the same sign, resulting in a repulsion from the wire's ends, and a potential minimum at the center of the wire. The corresponding 2D magnetic case analogous to this is a weak superconducting link between two thick superconducting banks (a Josephson weak-link junction \cite{likharev}) where a magnetic vortex attempting to pass across the junction encounters a potential minimum (a saddle point) at the center of the bridge. (d) If, on the other hand, $R_{\rm env}\gg Z_{\rm L}$, the image phase slips have opposite sign such that the real phase slip is attracted to the wire's ends and a potential maximum occurs in the center of the wire. The analogous 2D magnetic case is that of an isolated superconducting strip \cite{martinisvortex}.}
\label{fig:expt}
\end{center}
\end{figure}

In order to discuss the implications of our work for past and ongoing experiments aimed at observing evidence for QPS, we must first consider boundary conditions appropriate for the electrical connections to nanowires used in actual measurements. We consider the limit where the radiation wavelength corresponding to the characteristic frequency $\omega_{\rm p}$ in the medium surrounding the wire is much larger than the wire length, so that the electromagnetic environment can be treated as a simple, lumped-element boundary condition at the wire's ends. The typical experimental configuration is shown in fig.~\ref{fig:expt}(a): a four-wire resistance measurement, in which the leads are usually designed to have high resistance at the low frequencies associated with quasistatic IV measurements\footnote[7]{Two notable exceptions are the very recent experiments of refs.~\cite{astafiev,astafievNbN,arutyunovSR}, which use qualitatively different measurement techniques.}. Our circuit model for this configuration is similar to that used for JJs \cite{kautz}, and is shown in fig.~\ref{fig:expt}(b). As pointed out in ref.~\cite{kautz}, unless special techniques are used (such as in refs.~\cite{corlevi,zorin,zorinPRB,arutyunovTiQPS}), the lead impedance $Z(\omega)$ is certain to become relatively low ($< Z_0$, the impedance of free space) at high enough frequency, even if $Z(\omega)\gg Z_0$ as $\omega\rightarrow0$. Given that the important frequency for our model is $\omega_{\rm p}$, which will turn out to be relatively high, a crucial feature of the environment model of fig.~\ref{fig:expt}(b) is a low, resistive impedance at high frequency such that: $Z_{\rm env}(\omega_{\rm p})\approx R_{\rm env}\ll Z_{\rm L},R_{\rm Q}$. In this limit, the classical boundary condition at the wire's ends is effectively a short, such that interaction of a type II phase slip with the wire's ends can be described using image phase slips of the same sign \cite{buchler}; this results in a repulsion from the ends and an activation energy barrier for phase slip events $\delta E_{\rm II}(x)$ as a function of the phase slip position $x$ like that shown in fig.~\ref{fig:expt}(c). It is important to note that this is \textit{not} analogous to the 2D magnetic case of an isolated, finite-width superconducting strip as in ref.~\cite{martinisvortex}. Rather, our situation is analogous to a very short superconducting weak link between two large banks, where the link length $l$ is analogous to our wire's length, and the link width $w\gg l$ maps to Euclidean time in 1+1D [fig.~\ref{fig:expt}(c)]. In both of these cases the vortex (type II phase slip) sees a free-energy (Euclidean action) \textit{minimum} at the link (wire) center. In the opposite case where $Z_{\rm env}\gg Z_{\rm L}$, the image vortices have \textit{opposite} sign, such that phase slips are attracted to the edges as shown in fig.~\ref{fig:expt}(d); this is in fact the 1+1D analog to the finite-width superconducting strip of ref.~\cite{martinisvortex}.

For very long wires with $l\gg\lambda_{\rm E}$, the contribution of the environment can naturally be neglected, since even in the high-$Z$ case where the action is lower for phase slips to occur within a distance $\lambda_{\rm E}$ of the two ends [c.f., fig.~\ref{fig:expt}(d)] which then interact predominantly with their images, the statistical weight of such paths in the partition function becomes negligible for long enough wires. However, when $l$ becomes sufficiently smaller than $\lambda_{\rm E}$, the interaction with image phase slips eventually dominates the partition function, such that the environmental impedance alone determines the ground state (as opposed to $Z_{\rm L}$)\footnote[1]{The method of images was also used in ref.~\cite{buchler} to discuss boundary effects; however, in that work it was applied directly to GZ-type \textit{microscopic} quantum phase slip events. By contrast, we have applied this method to our type II phase slips, \textit{macroscopic} quantum processes \cite{secondary} which arise as a consequence of treating microscopic QPS events as dual to Cooper pair tunneling events in lumped JJs. This distinction can be clarified by considering the duals of these two cases: our theory is dual to the usual JJ treatment, where the ``bare" Josephson energy per length is calculated in the lumped limit, neglecting the geometric inductance $\mathcal{L}_{\rm g}$ of the junction. This result is then plugged in to a distributed theory for the ``long" junction, out of which arises the Josephson penetration depth $\lambda_{\rm J}$ \cite{orlando}, to which our $\lambda_{\rm E}$ is dual. The premise of the QPS theory of ref.~\cite{buchler}, on the other hand, is dual to treating a long JJ by directly considering from the beginning the full quantum mechanics of Cooper pair tunneling events in the distributed system [c.f., fig.~\ref{fig:tline2}(d)].}. This is how the crossover occurs in our model to the lumped-element regime (discussed by MN \cite{mooijfluxchg} as the dual of the extensively-studied case of lumped JJs \cite{schmid,schon,panyukov,likharevdual}). By contrast, the length scale which arises in the theories of GZ \cite{GZPRL} and ref.~\cite{buchler} for finite wires is $\hbar v_{\rm s}/k_{\rm B}T$, such that within the approximations used in these works the behavior is always lumped at zero temperature.

These considerations regarding electric field penetration into finite wires in 1+1D have direct analogs in the physics of magnetic vortex penetration in 2D. In fact, as discussed in~\ref{a:environment}, the equilibrium thermodynamics governing type II magnetic flux penetration (in terms of a Gibbs free energy which includes the magnetic work done by or on the field source), has an exact analog in our 1D case (in terms of a Euclidean action which includes the work done by or on the circuit environment). Thus, under appropriate conditions, all of the well-known results concerning type II flux penetration in 2D can be appropriated for our purposes here, in particular the existence of type II phase slip ``lattices" corresponding to spatially and temporally periodic electric field penetration. An example of the current distributions for the two lowest-action type II phase slip lattices, for a wire with $l\ll\lambda_{\rm E}$ in a low-impedance environment ($R_{\rm env}\ll R_{\rm Q},Z_{\rm L}$) corresponding to an effective voltage bias, is shown schematically in fig.~\ref{fig:lowZ}(a). These two lattices can be identified directly with the two lowest energy bands of an approximately lumped phase-slip junction, as shown in fig.~\ref{fig:lowZ}(b), and discussed by MN \cite{mooijfluxchg}. To see this, first consider the total Euclidean action $\mathcal{S}_{\rm II}^{\rm tot}(x)$ of a type II phase slip at position $x$ in the $R_{\rm env}\ll Z_{\rm L},R_{\rm Q}$ limit, and the corresponding classical energy barrier $\delta E_{\rm II} (x)$ ($x=0$ is taken to be the middle of the wire):

\begin{figure}[t]
\begin{center}
\resizebox{0.5\linewidth}{!}{\includegraphics{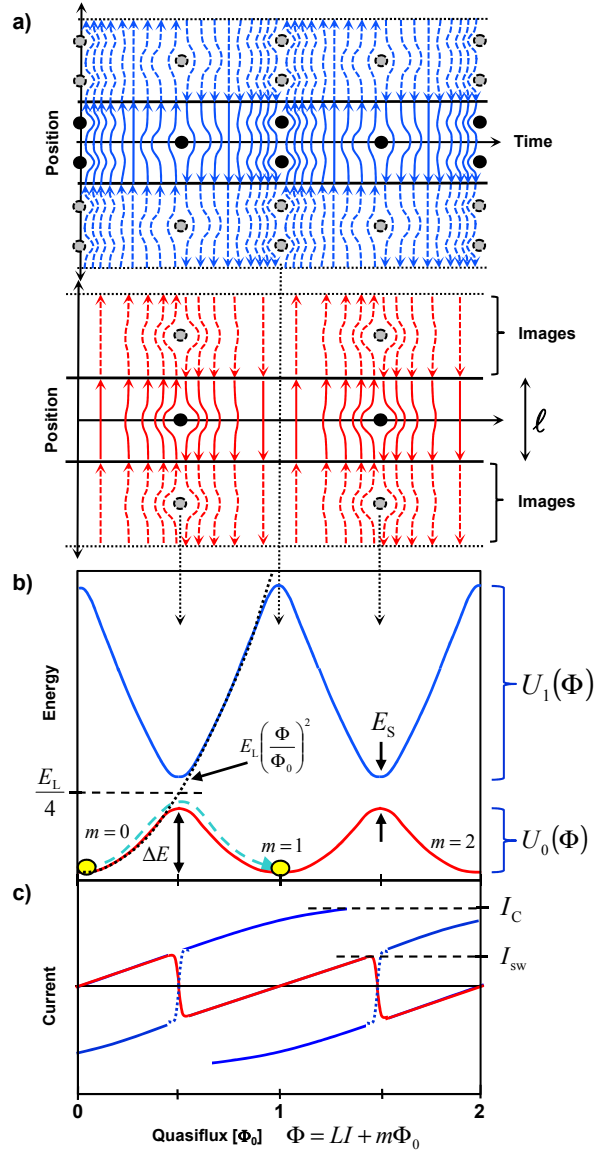}}
\caption{Type II phase-slips when $R_{\rm env}\ll R_{\rm Q},\;Z_{\rm L}$. (a) the two lowest-energy lattices for a constant voltage $V=\dot\Phi$ across a short wire ($l\ll\lambda_{\rm E}$), where the lines/arrows represent ${\bi j}$ [c.f., eq.~\ref{eq:j}], and solid circles the phase-slip cores; dashed lines and shaded circles indicate image phase slips. (b) the two lowest bands $U_0(\Phi)$ and $U_1(\Phi)$, dual to the quasicharge bands of a JJ in a high-$Z$ environment \cite{bloch,panyukov,pistolesi,schon,hekking}. Inductive parabolae with $E=E_{\rm L}(\Phi/\Phi_0-m)^2$ are degenerate at half-integer values of $\Phi/\Phi_0$, where an avoided crossing of width $E_{\rm S}$ (for $E_{\rm S}\ll E_{\rm L}$) occurs between states with $m$ differing by $\pm1$\cite{PSqubit,mooijfluxchg,glazman}. A higher-order interaction also couples states with $m$ differing by $\pm2$ at integer values of $\Phi/\Phi_0$ (the upper level of these crossings is not shown). If $E_{\rm S}\rightarrow 0$, the wire is simply an inductance $L_{\rm k}$ with $E=E_{\rm L}(\Phi/\Phi_0)^2$ (dashed black line). The current distributions shown in (a) correspond to adiabatic evolution along the bands in (b), indicated by the dashed arrow. (c) $I(\Phi)$ for the wire, in which QPS-induced avoided crossings result in a switching current $I_{\rm sw}<I_{\rm C}$ into a voltage state. A constant $V=\dot\Phi$ produces an oscillatory current, as shown by the red line in (c).}
\label{fig:lowZ}
\end{center}
\end{figure}

\begin{eqnarray}
\delta E_{\rm II} (x)&\sim&\hbar\omega_{\rm p}\mathcal{S}_{\rm II}^{\rm tot}(x)\nonumber\\*
&=&\hbar\omega_{\rm p}\Biggl\{\mathcal{S}_{\rm II}+\frac{1}{2}\sum_{k=1}^\infty\Biggl[\mathcal{S}_{\rm int}[2kl]+\frac{1}{2}\mathcal{S}_{\rm int}[(2k-1)l-2x]\nonumber\\*
&\;&\hspace{5cm}+\frac{1}{2}\mathcal{S}_{\rm int}[(2k-1)l+2x]\Biggr]\Biggr\}\label{eq:typeIItot}
\end{eqnarray}

\noindent Here, the first line is valid as long as $\beta^{-1}=k_{\rm B}T\ll\hbar\omega_{\rm p}$, and in the second line the summations are over image phase slips. In the $\lambda_{\rm E}\gg l$ limit we can neglect the $x$-dependence as well as the first (self-energy) term, and replace the sums with an integral, to obtain:

\begin{equation}
\delta E_{\rm II}(\lambda_{\rm E}\gg l)\approx\frac{E_{\rm L}}{4}\left[1+\frac{l}{\lambda_{\rm E}}\frac{2}{\pi}\left(\ln{\frac{l}{\lambda_{\rm E}}}-1\right)\right]\label{eq:deltaEl}
\end{equation}

\noindent where $E_{\rm L}\equiv\Phi_0^2/(2L_{\rm k})$ is the inductive energy of the wire with total kinetic inductance $L_{\rm k}$. Thus, the first term in eq.~\ref{eq:deltaEl} is precisely the kinetic-inductive energy $E_{\rm L}/4$ that would be approximately expected at $\Phi=\Phi_0/2$ from fig.~\ref{fig:lowZ}(b) in the $S_0\gg 1$ limit, as well as from the lumped-element description of MN \cite{mooijfluxchg}, and the second term is the leading-order correction to this result in the small quantity $l/\lambda_{\rm E}$. Since a constant voltage across the wire implies that $\Phi$ evolves at a constant rate, corresponding to motion at constant ``velocity" along the horizontal axis ($\rmd\Phi/\rmd t\equiv V$) of figs.~\ref{fig:lowZ}(b),(c), the type II phase-slip cores can be identified with the avoided crossings that define the energy bands $U_0(\Phi)$ and $U_1(\Phi)$. The crossings shown at half-integer values of $\Phi/\Phi_0$ occur where two states with $m$ differing by 1 are coupled, and correspond to a single phase-slip core in the wire. The crossings at integer values of $\Phi/\Phi_0$ (the upper state of which is $U_2(\Phi)$, not shown in the figure) occur where states with $m$ differing by 2 are coupled, and therefore correspond to the simultaneous presence of two phase slip cores in the wire, as shown in the upper half of (a) at these points. The temporal \textit{current} oscillations [fig.~\ref{fig:lowZ}(d)] that occur in the lowest energy band at fixed \textit{voltage} are the exact dual of Bloch oscillations in a lumped JJ \cite{bloch,panyukov,pistolesi,schon,hekking}.

Beginning with the seminal work of Giordano \cite{giordano}, nearly all the experimental efforts to observe evidence for QPS have focused on the region near $T_{\rm C}$ where the stiffness $V_{\rm 1D}$ goes to zero, so we begin our discussion of experiments with this regime. The motivation behind such experiments is the idea that quantum phase slips should become exponentially more frequent as the energy barrier is lowered. Of course, thermally activated phase slips also become exponentially more frequent, so that the objective in such measurements can only be to observe qualitative deviations from simple LAMH thermal activation as the temperature is lowered, in the hope that such deviations can be identified with QPS. A wealth of experimental data now exists in which resistance vs. $(T_{\rm C}-T)$ measurements of superconducting nanowires are compared to LAMH theory, for a range of materials including In \cite{giordano}, Pb \cite{Pbwires}, PbIn \cite{gioPbIn}, Al \cite{mooijQPS,altomare,inhomog,zgirski}, Ti \cite{Tiwires}, MoGe \cite{graybealMoGe,tinkSIT,lau,bezryRev}, Nb \cite{Nbwires}, and NbN \cite{NbNwires}. In many cases deviations are indeed observed, usually in the form of a significantly weaker slope on a plot of $\log{R}$ vs. $T$ (as opposed to the clear crossover in behavior seen in Giordano's original measurements)\footnote[1]{Note that in addition to the superconducting wires discussed in this section, some wires remain resistive all the way to the lowest measurable temperatures, or even appear to become insulating. The latter phenomenon is the subject of section~\ref{s:SIT}.}. This departure from LAMH behavior has been attributed to QPS either using Giordano's model \cite{giordano,lau,zgirski,bezryRev,bezryNP} or a variant of it in which the purely heuristic energy scale $\hbar/\tau_{\rm GL}$ in Giordano's quantum phase-slip-induced resistance is replaced by the GZ result \cite{GZPRL,GZPRB}. Although some reasonable agreement can often be obtained for individual experiments, when all of the available data are considered together, one encounters a problem: the ostensibly quantum-phase-slip induced deviation from LAMH theory does not seem to scale as expected with the predicted QPS action. For example, based on the GZ model, the $T=0$ phase-slip action for Giordano's original 41-nm wide In wire (which exhibited a dramatic departure from LAMH behavior) is $S_{\rm GZ}\approx100$, whereas $S_{\rm GZ}\approx13$ for Bezryadin's 7-nm MoGe wires which showed no anomalous departure from LAMH at all. As we will now show, our model provides a possible explanation for this counterintuitive trend, in terms of thermal fluctuations over the type II phase slip energy barrier.

We cast our problem in a form analogous to the original work of LAMH \cite{LA,MH}, using eq.~\ref{eq:PSRate} to obtain the general expression for a thermal phase-slip-induced effective resistance \cite{MH,lau,giordano,bezryRev} (also used to describe thermal phase slips in JJs \cite{AHattempt,grabert,kautz}):

\begin{equation}
R_{\rm ps}=\frac{\langle V\rangle}{I}=R_{\rm Q}\frac{\hbar\Omega_{\rm ps}}{k_{\rm B}T}\exp{\left(-\frac{\delta E_{\rm ps}}{k_{\rm B}T}\right)}\label{eq:PSR}
\end{equation}

\noindent where $\delta E_{\rm ps}$ is the classical energy barrier, and $\Omega_{\rm ps}$ is the attempt frequency \cite{AHattempt,grabert}. We consider three distinct, simplified regimes: (i) where $\lambda_{\rm E}\gg l$, for which the energy barrier is given by eq.~\ref{eq:deltaEl} and illustrated in fig.~\ref{fig:lowZ}(c); (ii) where $\lambda_{\rm E}\ll l$, so we can neglect entirely the statistical weight of paths that interact with the ends, and:

\begin{equation}
\delta E_{\rm II}(\lambda_{\rm E}\ll l)\approx\hbar\omega_{\rm p}\mathcal{S}_{\rm II}=\frac{1}{2L_\lambda}\left(\frac{\Phi_0}{2}\right)^2K_0\left(\frac{1}{\kappa_{\rm E}}\right)
\end{equation}

\noindent where we have defined the effective total inductance for a type II phase slip: $L_\lambda\equiv\pi\mathcal{L}_{\rm k}\lambda_{\rm E}/4$ (by analogy to eq.~\ref{eq:deltaEl}); and finally (iii), an intermediate regime where $\lambda_{\rm E}\lesssim l$, so that the energy barrier is a saddle point at the wire's center like that shown in fig.~\ref{fig:expt}(c), and we can make the approximation that all phase slips occur at that point:

\begin{equation}
\delta E_{\rm II}(\lambda_{\rm E}\lesssim l)\approx\frac{1}{2L_\lambda}\left(\frac{\Phi_0}{2}\right)^2\left[K_0\left(\frac{1}{\kappa_{\rm E}}\right)+\frac{1}{2}\sum_{k=1}^{N}K_0\left(\frac{k}{\kappa_{\rm E}}\right)\right]
\end{equation}

\noindent truncating the sum at some small $N$ beyond which the additional terms can be neglected.

We model $\Omega_{\rm ps}$ in a simple manner based on well-known results for lumped JJs, where we treat the thermal fluctuations for each length $\lambda_{\rm E}$ of wire as independent if $\lambda_{\rm E}\ll l$ \footnote[1]{This is approximately valid for thermal type II phase slip rates which are low enough that we can neglect the statistical weight of paths in which phase-slips interact with each other substantively.}, or the whole wire as a single fluctuating region if $l\lesssim\lambda_{\rm E}$. We describe each fluctuating region in terms of an effective Josephson inductor $L_{\rm f}$ in parallel with an effective damping resistance $R_{\rm f}$ and shunt capacitance $C_{\rm f}$. For case (i) ($\lambda_{\rm E}\gg l$), these quantities are simply $L_{\rm k}$, $R_{\rm env}$, and $C_{\rm sh}$; for cases (ii) and (iii) ($\lambda_{\rm E}<l$) we take instead: $L_\lambda$, $Z_{\rm L}$ (the effective resistance looking out of the fluctuation region into the plasma modes of the wire), and $C_l(k\lambda_{\rm E}=1)$ [c.f., eq.~\ref{eq:QPScap}]. Strictly speaking this is only correct in case (ii), of course, but we use it here as an estimate also for case (iii). The attempt frequency is given approximately by \cite{grabert}:

\begin{equation}
\Omega_{\rm ps}\approx N_\lambda\omega_{\rm f}\left[\sqrt{1+\frac{1}{4Q_{\rm f}^2}}-\frac{1}{2Q_{\rm f}}\right]\label{eq:omegaps}
\end{equation}

\noindent where $N_\lambda\equiv l/\lambda_{\rm E}$ for $l\gg\lambda_{\rm E}$ and $N_\lambda=1$ otherwise, $\omega_{\rm f}\equiv1/\sqrt{L_{\rm f}C_{\rm f}}$ and $Q_{\rm f}\equiv\omega_{\rm f}R_{\rm f}C_{\rm f}$, and this expression holds in the limit where $k_{\rm B}T\gg\hbar\Omega_{\rm ps}$. In the overdamped regime ($Q_{\rm f}\ll1$) which is relevant in all experimental cases of interest here, $\Omega_{\rm ps}\approx R_{\rm f}/L_{\rm f}$.

\begin{figure}[t]
\begin{center}
\resizebox{0.85\linewidth}{!}{\includegraphics{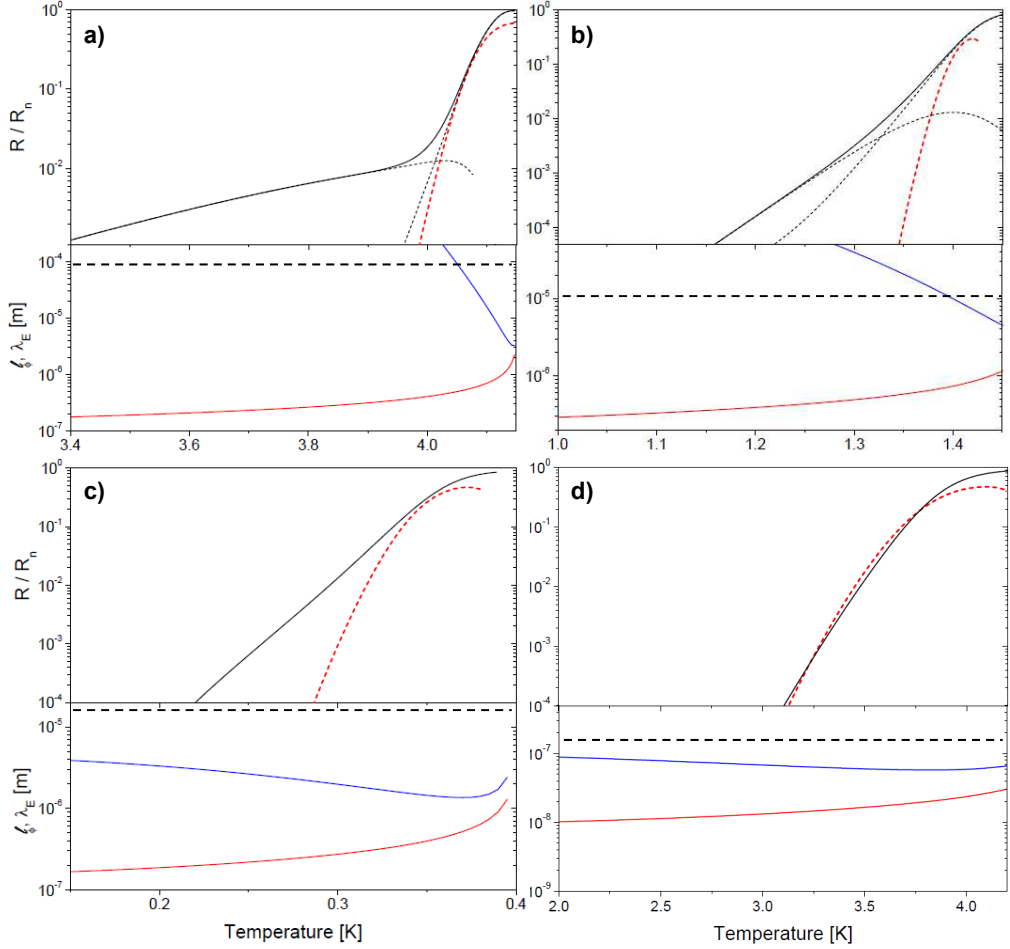}}
\caption{Resistance vs. temperature near $T_{\rm C}$ in our model for four experimental cases. Solid black lines are derived from our model, for parameters relevant to four experimental wires (described in~\ref{a:RvsT}), in order of increasing strength of QPS: (a) 40-nm In wire from ref.~\cite{giordano} ($S_0=100$ [c.f., eq.~\ref{eq:S0eqn}], $S_{\rm GZ}=850$ [c.f., eq.~\ref{eq:GZQPS}]); (b) 15-nm Al wire from ref.~\cite{zgirski} ($S_0=25$, $S_{\rm GZ}=55$); (c) 53-nm Ti wire from ref.~\cite{Tiwires} ($S_0=9.0$, $S_{\rm GZ}=16$); (d) 7.5-nm MoGe wire (S1) from ref.~\cite{bezryNP} ($S_0=5.6$, $S_{\rm GZ}=13$). These curves compare favorably with the experimental results. Dashed black lines are shown in the cases where our model predicts a crossover between two regimes considered in the text, and the solid black line is then a guide to the eye in connecting these smoothly. Predictions of LAMH theory \cite{LA,MH} are shown by red dashed lines. The bottom half of each panel shows the predicted temperature dependence of $\lambda_{\rm E}$ (blue curve) and $l_\phi=1.8\xi$ (red curve). For the In case in (a), with weakest QPS, $\lambda_{\rm E}$ increases sufficiently quickly as $T$ is lowered that a clear crossover is observed when it becomes much larger than the wire length $l$. In the Al (b) and Ti (c) cases which have progressively stronger QPS, $\lambda_{\rm E}$ becomes shorter and the crossover is obscured, such that the qualitative signature is only a reduced slope and change of curvature on the log plot, which in both cases was fit to a Giordano-like model in the experimental references \cite{zgirski,Tiwires}. Finally in the case of MoGe (d), QPS is sufficiently strong that $\lambda_{\rm E}$ does not vary appreciably over the relevant temperature range, and the temperature scaling of the energy barrier becomes very similar to that predicted by LAMH.}
\label{fig:RvsT}
\end{center}
\end{figure}

Figure~\ref{fig:RvsT} shows, for the parameters of four experimental cases (tabulated in~\ref{a:RvsT}), the resulting $R$ vs. $T$ obtained from our model, all of which compare favorably with the corresponding experimental observations\footnote[2]{In fact, for panel (a) the agreement with experiment in the LAMH region of the curve is obtained without the \textit{ad hoc} $4x$ reduction in the energy barrier used by Giordano \cite{giordano} in order to fit LAMH theory to his observations in this region.}. In addition, for each case the corresponding LAMH prediction is shown by a red dashed line. Notice that while QPS gets stronger from (a)-(d), the deviation from LAMH temperature scaling gets \textit{weaker}, just as observed in the experiments. As we will now explain, the reason in our model for this seemingly paradoxical behavior is the crucial role played by the temperature dependence of $\lambda_{\rm E}$ (which has no analog in previous theories for QPS), shown in the bottom graph of each panel in fig.~\ref{fig:RvsT}, relative to $l_\phi$ and the wire length $l$.

First of all, as $T\rightarrow T_{\rm C}$, notice that in all cases we have $l>\lambda_{\rm E}\gtrsim l_\phi$, such that the corresponding energy barrier [c.f., eq.~\ref{eq:typeIItot}] has a similar magnitude and temperature scaling to $\delta E_{\rm LAMH}$ [c.f., eq.~\ref{eq:LAMH}] (in this regime the Bessel function $K_0$ varies only logarithmically). In this limit, then, all of our predictions for the four cases either approximately coincide with or approach that of LAMH\footnote[7]{Note that our treatment of $\delta E_{\rm II}$ is strictly valid only when $\kappa_{\rm E}\gg 1$, since we have neglected the action associated with the phase slip core in comparison to the screening ``currents" ${\bi j}$ in eq.~\ref{eq:typeIIS}. This argument is entirely analogous to that made in the context of magnetic vortices in 2D in the type II limit \cite{orlando}. Very close to $T_{\rm C}$ where typically $\kappa_{\rm E}\ll 1$, the core contribution becomes dominant, our result $\delta E_{\rm II}$ is no longer applicable, and we expect the resulting energy barrier to cross over to $\delta E_{\rm LAMH}$. One might in fact view the LAMH phase slip as the type I analog of our type II phase slips, where the corresponding 2D situation would be a mixed state of a type I superconductor in which a single flux quantum penetrates in a 2D region of linear dimension $\sim\xi$ inside which the gap is suppressed to zero.}. Now, starting with the case of Giordano's In wire where QPS is the weakest, as $T$ is lowered $\lambda_{\rm E}$ increases very quickly, becoming much larger than the wire length already by around $T=4$K. In this limit, eq.~\ref{eq:deltaEl} for the barrier applies, which has the $\sim 1/(T_{\rm C}-T)$ dependence of $L_{\rm k}$, the total inductance of the wire. This scaling is significantly slower than in LAMH theory, resulting in the clear crossover shown in the figure. Thus, in our model the crossover which was previously attributed to a transition from thermal to quantum phase slips is explained instead by a change in the $T$-dependence of the energy barrier for purely thermal phase slips (when $\lambda_{\rm E}$ becomes larger than the total wire length $l$). Extending this interpretation to the different behaviors in panels (b)-(d), we find that our model indeed predicts more and more LAMH-like behavior as the strength of QPS in increased, due to the reduced temperature dependence of $\lambda_{\rm E}$. In the intermediate case of Al ~\cite{zgirski} (b), the crossover is still present but is sufficiently smoothed out that it is also qualitatively consistent with a Giordano-like model, which was used to fit the corresponding data in ref.~\cite{zgirski}. For the Ti wire of panel (c), QPS has become sufficiently strong that there is no longer any crossover, as $\lambda_{\rm E}$ remains well below $l$ over the entire temperature range. For this case the deviation from LAMH scaling that is still present is simply a residual effect of the temperature dependence of $\lambda_{\rm E}$, which although smaller than (a) and (b) is still non-negligible, and causes the barrier height to go up more slowly as temperature is decreased than $\delta E_{\rm LAMH}$. This modified dependence can also be fit with a Giordano-like model, as in ref.~\cite{Tiwires}. Finally, the MoGe wire shown in (d) \cite{bezryNP} has sufficiently strong QPS that $\lambda_{\rm E}$ varies little over the entire relevant temperature range, and there is almost no deviation from LAMH scaling, as shown in the figure. Thus, in a low-Z environment, our model predicts that QPS appears in $R$ vs. $T$ measurements only indirectly, via the phase diffusion \cite{kautz} and associated resistance arising from thermal hopping over the type II phase slip energy barrier.

Similar conclusions arise from our model regarding the more recent experiments of Bezryadin \cite{bezryNP,aref}, in which the \textit{bias current} was increased, with the temperature held fixed, and far below $T_{\rm C}$. These experiments were modeled after the seminal measurements of macroscopic quantum tunneling in JJs \cite{MQTJJ}, in which effective ``escape rates" out of the Josephson potential well were observed as a function of current [c.f., fig.~\ref{fig:MQTQPS}(a)], from which an effective temperature of the phase fluctuations $T_{\rm eff}$ could be inferred. At higher bath temperatures $T$ (still much less than $T_{\rm C}$) it was found that $T_{\rm eff}\approx T$; however, as $T$ was lowered, $T_{\rm eff}$ saturated at a minimum value known as the quantum temperature $T_{\rm Q}$, which could be explained quantitatively in terms of the expected quantum phase fluctuations of the circuit. Similar results were obtained for continuous MoGe nanowires in ref.~\cite{bezryNP,aref}, and this was taken as a signature of quantum phase fluctuations associated with QPS \cite{bezryNP,aref}. However, neither the quantitative values of $T_{\rm Q}$ extracted from these measurements, nor its dependence on wire parameters, was explained. Furthermore, it remained a mystery why the wires which exhibited nonzero apparent $T_{\rm Q}$ also showed no sign of the deviations from LAMH-type temperature scaling of resistance near $T_{\rm C}$ which were previously attributed to QPS.

\begin{figure}[t]
\begin{center}
\resizebox{0.7\linewidth}{!}{\includegraphics{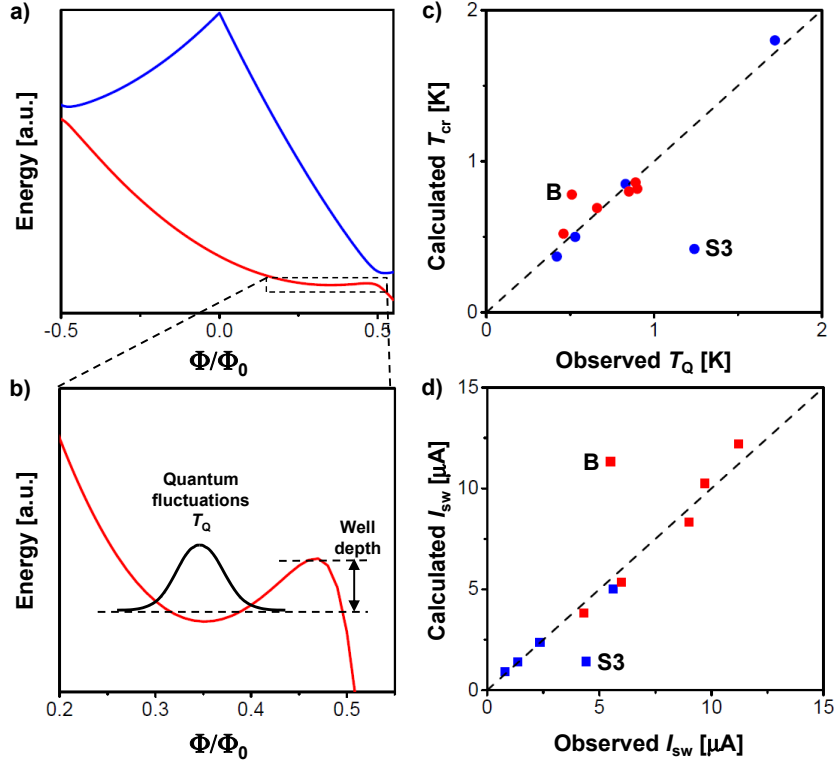}}
\caption{Quantum temperature and switching current in a low-Z environment. (a) lowest two calculated energy bands $U_0(I_{\rm b},\Phi)$ and $U_1(I_{\rm b},\Phi)$ for wire S1 of ref.~\cite{bezryNP} at $I_{\rm b}=$2 $\mu$A. (b) expanded view of the residual potential well in $U_0(I_{\rm b},\Phi)$. Fluctuations of the $L_{\rm k}-R_{\rm env}-C_{\rm sh}$ circuit produced by the wire and its environment can cause the phase particle to escape from this well even when there is still a potential barrier, at which point a voltage appears \cite{MQTJJ,kautz}. (c) calculated quantum temperature, and (d) switching current, for wires S1-5 of ref. \cite{bezryNP} (blue symbols) and A-F of ref.~\cite{aref} (red symbols) vs. the values inferred from measurements. $T_{\rm Q}$ predictions were obtained using ref.~\cite{grabert}, and $I_{\rm sw}$ predictions were derived from eq.~\ref{eq:bands}, assuming that switching occurs at the bias current where the potential well depth is reduced to the experimental $T_{\rm Q}$. With the exception of wire S3 of ref. \cite{bezryNP} and wire B of ref. \cite{aref}, the agreement is good in both cases (c) and (d). The fixed parameters used to obtain this agreement are discussed in~\ref{a:moge}, and the primary adjustable parameter was $R_{\rm env}$. We extract the values: $110\Omega$ for the data of ref.~\cite{bezryNP} and $35\Omega$ for ref.~\cite{aref}. This difference is quite plausible, since the phase-slip plasma frequencies at which $R_{\rm env}$ is to be evaluated are about an order of magnitude higher in the former case (since the wires have significantly smaller $A_{\rm cs}$).}
\label{fig:bezryresults}
\end{center}
\end{figure}

We now show how these phenomena can also be described by our model. We consider the lumped-element case corresponding to the energy band $U_0(\Phi)$ shown in fig.~\ref{fig:lowZ}(b) (since for the parameters of these wires we have $\lambda_{\rm E}>l$ at $T=0$), treating it as a classical potential energy and neglecting transitions to higher bands (in the same manner that the lowest quasicharge band of a lumped JJ in a high-Z environment is often treated \cite{bloch,panyukov,pistolesi,schon,hekking}). The effect of an external bias current $I_{\rm b}$ can be described, just as for a JJ, by the additional potential energy:

\begin{equation}
U_{\rm n}(I_{\rm b},\Phi)=U_{\rm n}(\Phi)-I_{\rm b}\Phi\label{eq:bands}
\end{equation}

\noindent which lowers the energy barrier for phase slips in one direction while raising it in the other \cite{LA,MH,bezryRev,kautz,orlando} [fig. \ref{fig:bezryresults}(a),(b)]. As the barrier is lowered by increasing $I_{\rm b}$, the phase particle has an increasing chance to surmount it per unit time due to a phase fluctuation. If this occurs, it can either be re-trapped in the adjacent potential well by the damping due to $R_{\rm env}$, or it can ``escape" into the voltage state corresponding to a terminal ``velocity" $V=\dot\Phi$ (determined by its effective mass and the damping)\footnote[1]{This appears to be related to the ``deconfinement" predicted in ref.~\cite{buchler}.}. The current at which this occurs then corresponds to the switching current $I_{\rm sw}$ measured in ref.~\cite{bezryNP}. Based on our discussion of case (i) above ($l<\lambda_{\rm E}$), we can adapt the well-known analysis of MQT in JJs to the present purpose, from which we obtain the crossover temperature $T_{\rm cr}$ where the fluctuation energy scale in the exponent of eq.~\ref{eq:PSR} goes over from $k_{\rm B}T$ to $k_{\rm B}T_{\rm Q}$. In the overdamped limit, this is simply: $k_{\rm B}T_{\rm cr}\approx\hbar\Omega_{\rm ps}\approx \hbar R_{\rm env}/L_{\rm k}$. The fact that the capacitance $C_{\rm sh}$ does not appear in $T_{\rm cr}$ in the overdamped limit illustrates that ``quantum temperature" would be a misnomer for this quantity; as discussed in ref.~\cite{grabert}, in the overdamped limit quantum tunneling does not contribute to the escape rate at all. Rather, it is dominated for $T\ll T_{\rm cr}$ by the classical fluctuations that necessarily come with strong damping, via the fluctuation-dissipation theorem\footnote[7]{Note that in the underdamped case, $k_{\rm B}T_{\rm cr}\approx\hbar\omega_{\rm f}$, which \textit{can} be directly identified with quantum zero-point fluctuations.}. Figure~\ref{fig:bezryresults}(c) shows a comparison between the experimental results of refs.~\cite{bezryNP,aref} and our expectations based on the discussion above (the parameters used for this comparison are discussed in~\ref{a:moge}). For nearly all of the reported wires, the agreement is relatively good. We can also compare the average switching current into the voltage state $I_{\rm sw}$ observed in refs.~\cite{bezryNP,aref} with our prediction based on eq.~\ref{eq:bands} (we take the predicted switching current to be that at which the depth of the potential well is equal to the observed quantum temperature). Figure.~\ref{fig:bezryresults}(d) shows that the agreement with experiment is also good for the same wires.

Our discussion also suggests a different explanation for another observation in refs.~\cite{bezryNP,aref} that was was highlighted as direct evidence for QPS: the fact that the width of the stochastic probability distributions $P(I_{\rm sw})$ (obtained from many repeated $I_{\rm sw}$ measurements) increased as $T$ was lowered. Since the system is overdamped, at high $T$ the phase particle moving in the potential $U_0(I_{\rm b},\Phi)$ can be thermally excited over a barrier many times (undergo many phase slips), each time being re-trapped by the damping, before it happens to escape into the voltage state. At low $T$, these excitations are sufficiently rare that in a given time the system is more likely to experience a single fluctuation strong enough to cause escape than it is to experience multiple weaker fluctuations which act together to cause escape. Just as for JJs, this produces a $P(I_{\rm sw})$ that broadens as $T$ is lowered \cite{kautz}, since fewer phase slips are associated with each switching event, and the resulting stochastic fluctuations of $I_{\rm sw}$ are larger. Note that in contrast to ref.~\cite{bezryNP}, where these results were explained by local heating of the wire by individual quantum phase slips, our discussion would suggest that the energy $I_{\rm b}\Phi_0$ released during a type II phase slip is dissipated in the environmental impedance $R_{\rm env}$.

Very recently, in the wake of MS's seminal work \cite{mooijfluxchg}, several experimental groups have pursued entirely new experimental approaches that have allowed more direct observation of QPS phenomena \cite{astafiev,zorin,zorinPRB,arutyunovSR,arutyunovTiQPS}. Astafiev and co-workers \cite{astafiev} have demonstrated the phase-slip qubit of ref.~\cite{PSqubit}, where the nanowire is contained in a closed superconducting loop, using both InO$_x$ and NbN films. This can be viewed as the case of $R_{\rm env}=0$, such that as long as the inductance of the rest of the loop can be neglected, the external flux through the loop corresponds to a fixed-phase boundary condition for the nanowire. When $\Phi_0/2$ threads the loop, the PSJ is then biased right at the avoided crossing of width $E_{\rm S}$ in fig.~\ref{fig:lowZ}(c), such that direct spectroscopic measurement of this splitting becomes possible. For the InO$_x$ wires, $E_{\rm S}/h\sim$ 5-10 GHz \cite{astafiev} was observed, and for the NbN wires $E_{\rm S}/h\sim$ 1-10 GHz \cite{astafievNbN} (note that this particular technique could only measure values in this range due to the microwave bandwidth of the apparatus). It is interesting to note that in our model, the phase-slip qubit biased at $\Phi_0/2$ corresponds to a type II phase slip essentially trapped in the wire, such that a null in the order parameter (of size $\sim l_\phi$) is present somewhere [c.f., fig.~\ref{fig:lowZ}(a)]\footnote[1]{Note that the same is true for any flux qubit when a half-integer number of $\Phi_0$ threads the loop, such that two counter-rotating currents interfere destructively. However, in a conventional flux qubit based on one or more JJs, the corresponding null in the order parameter occurs inside an insulating JJ barrier. This may be an important distinction from the phase-slip qubit of refs.~\cite{PSqubit,astafiev}, because there are no low-lying electronic states in the insulating JJ barrier, while there should be such states inside a region of superconducting wire where the gap is forced to zero by an applied boundary condition (i.e. the flux through a closed loop). The presence of such states might act as a source of dissipation and/or decoherence.}. Another recent pair of experiments, in two different groups, measured NbSi \cite{zorin,zorinPRB} and Ti \cite{arutyunovTiQPS} wires biased through Cr or Bi nanowires with extremely large DC resistances. A clear Coulomb blockade was observed in both cases, with threshold voltages $V_{\rm C}\sim 700\mu$V for the NbSi \cite{zorinPRB}, and $V_{\rm C}\sim 800\mu$V for the Ti \cite{arutyunovTiQPS}.

In table~\ref{tab:ES}, we show that our model can approximately reproduce these observations. Note that although the InO$_x$ and NbN cases fall approximately within the lumped-element regime $\lambda_{\rm E}> l$ where we can use: $V_{\rm C}\approx E_{\rm S}\pi/e$, the opposite is true ($\lambda_{\rm E}\ll l$) for the NbSi and Ti wires. In these two cases, as discussed for 1D JJ arrays in the Coulomb blockade regime \cite{havilandCPS}, the blockade voltage expected when the system is much longer than the soliton length (our $\lambda_{\rm E}$) is given by: $V_{\rm C}\approx E_{\rm C}\lambda_{\rm E}$ where $E_{\rm C}=E_{\rm S}\pi/(el)$ is the critical electric field. This critical voltage for $\lambda_{\rm E}\ll l$ is then defined by the condition that the energy barrier for a single soliton of size $\sim\lambda_{\rm E}$ to enter the array goes to zero, and the subsequent current flow just above $V_{\rm C}$ is carried by a train of these 2$e$-charged objects \cite{havilandCPS}.

The primary unknown physical parameter which enters into these estimates for $E_{\rm S}$ and $V_{\rm C}$ is $\epsilon_{\rm in}$, the chosen values for which are shown in table~\ref{tab:ES}. Also shown are some related values for this quantity derived from various experiments for three of the cases (we were unable to find an experimentally-derived value for Ti). Since the real part of a metal's dielectric constant is nearly always dominated by the strong inductive response of free carriers under typical experimental conditions, it is nontrivial to determine the underlying permittivity due only to bound charges that is relevant for our model of QPS, which we have called $\epsilon_{\rm in}$. For the cases of InO$_x$ and NbSi, we show experimental values obtained \textit{on the insulating side} of the metal-insulator transition in these materials, such that the free carrier response is no longer present. It is plausible that these values provide a useful estimate of the desired quantity on the metallic side of the transition, although this is by no means certain. For the case of NbN, we show a value extracted by fitting to far-infrared absorption spectra; these measurements were made on a film $\sim$10 times thicker than the one used in ref.~\cite{astafievNbN} where QPS was observed, however, so it is likely that this value is an underestimate.

\begin{table}
\caption{\label{tab:ES} Comparison of our model with quantum phase slip observations on several systems. In all cases we take $l_\phi=1.8\xi(0)$ and $\epsilon_{\rm out}=5.5\epsilon_0$. The electric penetration depth was calculated from eq.~\ref{eq:lE}; for InO$_x$ and NbN, where $\lambda_{\rm E}>l$, the critical voltage was calculated using $V_{\rm C}=E_{\rm S}\pi/e$ and eq.~\ref{eq:ES}; for Ti and NbSi where $\lambda_{\rm E}\ll l$, we used $V_{\rm C}\sim E_{\rm C}\lambda_{\rm E}$ as in ref.~\cite{havilandCPS} for blockaded JJ arrays. The last two columns show the GZ result for different values of the coefficient $A$ in eq.~\ref{eq:GZ}, which separately produce agreement with one of the observations.}

\begin{indented}
\lineup
\item[]\begin{tabular}{@{}*{11}{l}}
\bs
\br
&&&&\centre{2}{$\epsilon_{\rm in}[\epsilon_0]$}&&\centre{4}{$E_{\rm S}$[GHz] }\\
\ns\ns\ns
     &  &    &  &  \crule{2} &  &  \crule{4}    \\
\ns
&&&&&&&&&\centre{2}{GZ, $A=$ }\\
\ns\ns\ns\ns
    Wire   & ref. &    \centre{1}{$l$} & \centre{1}{$a$} & \centre{1}{this} & \centre{1}{experi-} & \centre{1}{$\lambda_{\rm E}$} & \centre{1}{this} & \centre{1}{experi-} &\crule{2}\\
\ns
&& [$\mu$m]    &&  \centre{1}{work}&  \centre{1}{ment} & [$\mu$m] & \centre{1}{work} & \centre{1}{ment} & \01.0 & 0.47  \\
\ns
\mr
    InO$_x$ &\cite{astafiev}& 0.4&  1.8 &  \040 & 2-40$^{\rm a}$ & 1.8 & 8.6  & 5-10 & \07.7 & 190 \\
    & &      & 1 & \010 &  & 2.1 & 7.2 &  &  &  \\
    NbN &\cite{astafievNbN}& 0.5   & 4.8  & \090 & \0\030$^{\rm b}$ & 1.9 & 6.9  & 1-12 & 36 kHz & \0\05.0 \\
    & &          & 1 &  \0\01 && 2.7 & 3.7  &&  &   \\
\br
&&&&\centre{2}{$\epsilon_{\rm in}[\epsilon_0]$}&&\centre{4}{$V_{\rm C}$[mV] }\\
\ns\ns\ns
     &  &    &  &  \crule{2} &  &  \crule{4}    \\
\ns
&&&&&&&&&\centre{2}{GZ, $A=$ }\\
\ns\ns\ns\ns
    Wire   & ref. &    \centre{1}{$l$} & \centre{1}{$a$} & \centre{1}{this} & \centre{1}{experi-} & \centre{1}{$\lambda_{\rm E}$} & \centre{1}{this} & \centre{1}{experi-} &\crule{2}\\
\ns
&& [$\mu$m]    &&  \centre{1}{work}&  \centre{1}{ment} & [$\mu$m] & \centre{1}{work} & \centre{1}{ment} & \03.4 & 0.58  \\
\ns
\mr
    Ti  &\cite{arutyunovTiQPS}& 20    & 1 &  \0\05$^{\rm c}$ & \centre{1}{-} & 0.56  & 0.89 & 0.6-0.8 & \0\00.73 & 17 \\
    &  &     &   2 & \0\05$^{\rm c}$ && 0.58  & 0.87 &  &  &  \\
    NbSi &\cite{zorinPRB}& \05 & 1 & \090 & 70-110$^{\rm a}$ & 0.63  & 0.82 & 0.7 & $\sim0$ & \00.76 \\
    & &    &   2  & 220 && 0.62  & 0.84 &  &  &  \\
\br
\bs
\end{tabular}
\item[] $^{\rm a}$Inferred from measurements on the insulating side of a metal-insulator transition: ref.~\cite{InOxepsilon} for InO$_x$ and ref.~\cite{NbSiepsilon} for NbSi.
\item[] $^{\rm b}$Inferred from the plasma frequency extracted from measurements on much thicker NbN films ($\sim$30 nm) \cite{mondal}.
\item[] $^{\rm c}$Chosen by optimizing agreement between fig.~\ref{fig:RvsT}(c) and the experiments of ref.~\cite{Tiwires}. Note that the predictions for this Ti wire are relatively insensitive to the choice of $\epsilon_{\rm in}$ and $a$ because $S_0$ is of order unity due to the small gap.
\end{indented}
\end{table}

For each of the four materials shown in table~\ref{tab:ES}, we list two possible values for the parameter $a$, which is used to obtain the kinetic inductivity $\Lambda=\mu_0\lambda^2$ (which then determines the stiffness $V_{\rm 1D}$) according to the relation:

\begin{equation}
\frac{\Lambda}{\rho_{\rm n}}=a\frac{\hbar}{\pi\Delta}\label{eq:BCSlambda}
\end{equation}

\noindent where $\rho_{\rm n}$ is the normal-state resistivity, $\Delta$ is the superconducting gap, and $a=1$, $\Delta\approx1.78k_{\rm B}T_{\rm C}$ in BCS theory. In the phase-slip qubit experiments on InO$_x$ and NbN, the total kinetic inductance of each wire was extracted from direct measurements, fixing $a=1.8$ for InO$_x$ and $a=4.8$ for NbN. These are significantly different from the BCS value, which may be indicative of proximity to a disorder-driven SIT at which the bulk superfluid stiffness ($\propto\Lambda^{-1}$) goes to zero while the local pairing gap remains finite \cite{gapins}. For these two materials we list also a corresponding $a=1$ case, where we reduce $\epsilon_{\rm in}$ to keep the calculated $E_{\rm S}$ close to the observed value. In the Coulomb blockade measurements (second two rows), the inductance was not measured directly, so we simply show the two cases $a=1$ and $a=2$ in the table for comparison. The question is: near a SIT where the value of $a$ inferred from bulk measurements can be substantially larger than unity (ostensibly due to disorder-driven quantum phase fluctuations), is it appropriate to use the \textit{bulk} kinetic inductivity to calculate the \textit{local} superfluid stiffness $V_{\rm 1D}$ relevant for QPS? This may be an important question, since it has been hypothesized that close proximity to a SIT of this type is a determining factor in the successful observation of nonzero QPS \cite{astafiev}.

Any mechanism for the SIT in these materials which involves only quantum phase fluctuations (in order to explain the observed coexistence of bulk insulating behavior and a local superconducting gap in the insulating state \cite{gapins}) would seem to require the existence of a microscopic phase correlation length, such that the relative phase is well-defined between two points spaced closer together than this, and such that finite superfluid stiffness remains for wavelengths shorter than this \cite{crane}. Furthermore, it would seem unphysical for this length scale to be significantly smaller than the superconductor's coherence length $\xi$, without a corresponding suppression of the gap\footnote[1]{This is apparent in two well-known ``phase-only" models for the SIT: in one, the nominally uniform film is treated as an inhomogeneous system of superconducting islands coupled by tunneling, essentially a JJ array \cite{ghosal,shimshoni}. In this case the phase correlation length cannot be smaller than the island size, and if the island size is much smaller than $\xi$ the Coulomb interaction on the islands will likely suppress the gap \cite{finkelstein}. Alternatively, in the so-called ``dirty boson" model, the quantum phase fluctuations are described in terms of vortex-antivortex pairs \cite{fisher}. In order for such a system to have a phase correlation length shorter than $\xi$, the non-superconducting cores of the vortex fluctuations (with size $\sim\xi$) would need to overlap substantially, and the average gap would be consequently reduced.}. This suggests that the stiffness relevant for QPS, which involves quantum phase fluctuations at the length scale $l_\phi\sim\xi$, is not the \textit{bulk} stiffness inferred from the macroscopic kinetic inductivity, but rather a local stiffness related only to the gap (corresponding to $a=1$). Interestingly, however, as shown in table~\ref{tab:ES} for the NbN case where we set $a=1$, it was necessary to adjust $\epsilon_{\rm in}$ all the way to unity to approach the experimentally observed range of $E_{\rm S}$. Since it is unlikely to be the case that $\epsilon_{\rm in}=1$ in this material, and the value $\epsilon_{\rm in}=90$ obtained using $a=4.8$ is quite plausible, this could be an indication that at least in this case the stiffness is suppressed even on length scales $\sim\xi$ as the SIT is approached from the superconducting side.

The last two columns of table~\ref{tab:ES} show the corresponding predictions of the GZ model in the same four wires, according to ref.~\cite{GZPRB}:

\begin{equation}
E_{\rm S}\approx\Delta S_{\rm GZ}\frac{l}{\xi}e^{-S_{\rm GZ}}\label{eq:GZ}
\end{equation}

\noindent where $\Delta$ is the superconducting gap, and $S_{\rm GZ}$ is given by eq.~\ref{eq:GZQPS}. For these two columns, we have chosen values of the parameter $A$ for which the resulting prediction agrees with one or the other of the observations of a given type ($E_{\rm S}$ or $V_{\rm C}$ measurement). As shown in the table, each case requires a different value for the coefficient $A$ to produce agreement with experiment (given the same material parameters used for our estimates, tabulated in~\ref{a:ES}). The difference is particularly large for the Ti wire, which is extremely long, and therefore requires a large value $A=3.4$ to fit the observed $V_{\rm C}$; by contrast, in our model $V_{\rm C}$ becomes independent of length once the wire is much longer than $\lambda_{\rm E}$, since in this regime it is defined by a vanishing energy barrier for the entry of a single CP soliton of size $\lambda_{\rm E}\ll l$.

\section{Destruction of superconductivity in 1D}\label{s:SIT}

In this final section we consider a possible relationship between our model and the observed destruction of superconductivity all the way down to $T=0$ for short wires with $R_{\rm n}\gtrsim R_{\rm Q}$. Previous theories have predicted insulating or metallic behavior as the wire diameter \cite{GZPRL,GZPRB}, the characteristic impedance $Z_{\rm L}$ \cite{GZPRL,GZPRB,buchler}, or an external shunt resistor \cite{buchler} is tuned through a critical value (our model also makes the latter two predictions, as described in sections~\ref{s:tline} and~\ref{s:expt}). However, none can obviously explain a $T=0$ transition at $R_{\rm n}\sim R_{\rm Q}$ in a low-$Z$ electromagnetic environment. In all of these theories the predicted transition relies on the presence of a form of dissipation which somehow remains even as $T\rightarrow 0$, such as anomalous excited quasiparticles \cite{demler}, a resistive shunt \cite{buchler}, continuum plasmon modes \cite{buchler,GZPRL,GZPRB}, or the quantum phase-slips themselves \cite{meidan}.

Our discussion suggests a possible alternative view, in which a $T=0$ SIT may be driven by \textit{disorder}-induced quantum phase fluctuations, analogous to the SIT observed in some quasi-2D systems \cite{baturinaSIT,samban} when the sheet resistance $R_\square\gtrsim R_{\rm Q}$ \footnote[1]{In these materials, evidence for a nonzero gap is observed even in the insulating state \cite{gapins}, indicating that phase fluctuations drive the transition. A similar disorder-driven SIT at $R_\square\sim R_{\rm Q}$ is also observed in some other materials with higher superfluid density \cite{havilandSIT,graybeal} which is believed to result from a different mechanism not associated with phase fluctuations \cite{finkelstein}.}. This 2D disorder-induced SIT has been interpreted using the ``dirty boson" model of Fisher and co-workers \cite{fisher}, in which disorder nucleates (virtual) unbound vortex-antivortex pairs (VAPs), with sufficient strength that these unpaired vortices themselves form a Bose-condensate, destroying long-range phase coherence and producing a gapped insulator \cite{fisher}. This is closely related to the Berezinskii-Kosterlitz-Thouless (BKT) vortex-unbinding transition in the classical 2D XY model \cite{berezinskii,KT,kosterlitz}.

To connect these ideas to our system, we first recall our discussion above of the BKT-like quantum phase transition expected when $\mathcal{K}$ is decreased from large values down to unity, associated with unbinding of type II phase slip-anti phase slip pairs in 1+1D. This transition is driven in our model by microscopic, homogeneous phase fluctuations associated with the effective permittivity for electric fields along the wire, or equivalently, by zero-point fluctuations of the Cooper pair plasma oscillation at length scales $\sim l_\phi$. As predicted in ref.~\cite{giamarchi}, however, a different kind of transition is also possible, driven by disorder. In the language of the (2+1D) dirty boson model: disorder can nucleate virtual phase slip-anti phase slip pairs in the ground state, which at some critical disorder strength overlap sufficiently to form a ``condensate" (in this case of instantons \cite{susskind,doniach}) with an insulating gap. In the dirty boson model, the $T=0$ critical point at $R_\square\sim R_{\rm Q}=\Phi_0/(2e)$ corresponds to approximately one vortex crossing for every Cooper pair crossing \cite{fisher}. In our 1D case, the corresponding critical point could plausibly be $R_{\rm n}\sim R_{\rm Q}$. In fact, in ref.~\cite{girvin1D} the existence of just such a universal \textit{conductance} $\sim R_{\rm Q}^{-1}$ in 1D at the critical point of a SIT was predicted. Such a disorder-based (as opposed to dissipation-based) mechanism may also be able to explain why the SIT in MoGe nanowires was only clearly evident for short wires with length $\lesssim200$ nm \cite{bezryRev,lau}. Since the logarithmic interaction between type II phase slips is cut off beyond separations $\rho\sim\lambda_{\rm E}$ [c.f. eq.~\ref{eq:typeIIint}] (which effectively functions as the coherence length/time near the transition), we might expect to see a weakening or disappearance of the SIT as the wire becomes significantly longer than $\lambda_{\rm E}$ \cite{VAPsc}; in fact, our theory predicts $\lambda_{\rm E}\sim$100-300 nm for the relevant MoGe wires~\footnote[7]{Note that our analogy to the dirty boson model would not explain the observed reduction in $T_{\rm C}$ near the 1D SIT in refs.~\cite{tinkSIT,bezrySIT}. This reduced $T_{\rm C}$ may be explained by the coexistence in these wires of an unrelated phenomenon: gap suppression due to an enhanced Coulomb interaction \cite{oreg,finkelstein}. This is believed to be the origin of a similar phenomenon observed in thin MoGe films \cite{graybeal} with very similar properties to the wires of refs.~\cite{tinkSIT,bezrySIT}.}.

These ideas may have importance to some recent work on ``honeycomb" bismuth films, consisting essentially of 2D networks of nanowires \cite{valles}. In a remarkable sequence of experiments, a SIT was observed in films with two different network geometries at thicknesses corresponding \textit{not} to a sheet resistance of $R_{\rm Q}$, but instead to thicknesses when $R_{\rm n}$ \textit{of each nanowire} passed through $R_{\rm Q}$, just like the quasi-1D observations of ref.~\cite{bezrySIT}. This may suggest that at the experimentally accessible temperatures, these nanostructured films had not yet reached a 2D universal regime, but were rather in an intermediate regime where quasi-1D behavior of the ``links" in the wire network still dominated the transition. A crossover between these two regimes would be controlled by the coherence between QPS in all of the nanowire links connected to each ``island" node in the network. If the QPS amplitudes for adjacent links is incoherent, the transition would still exhibit quasi-1D behavior. This coherence would be expected to depend, via Aharonov-Casher-like phase shifts, on charge fluctuations on the nodes \cite{glazman,manucharyan2012}. What then would be expected to occur if this coherence existed, such that the film appears uniform from the point of view of QPS?


The original works of LAMH can be used to view the transition in quasi-1D wires from a metallic state to a superconductor as the temperature is lowered in terms of thermally-driven, topological phase fluctuations in 1+1D: phase slips; these can be described formally as passage through the wire of vortices, 1D topological line defects. Mooij and co-workers extended this idea to zero temperature, effectively postulating quantum tunneling of these objects, which we have modelled in our work based on an effectively finite mass and zero-point motion arising from the permittivity for electric fields along the wire. This leads to the following idea: In 2D, one-dimensional line defects (vortices) control the superconducting transition via the BKT mechanism as the temperature is lowered. In 3D, correspondingly, it has long been thought that \textit{vortex rings}, effectively 2D objects, control the analogous transition. This idea has been applied to the lambda transition in $^4$He \cite{williams87,shenoy}, high-$T_{\rm C}$ superconductors \cite{williams99}, ordering in liquid crystals \cite{Lcrystals}, and even to structure formation in the early universe \cite{williams99,antunes}. Starting with such 2D topologically-charged objects, we can imagine a 2D quantum tunneling phenomenon analogous to our 1D QPS, in which a thin film undergoes a quantum fluctuation process that can be viewed formally as \textit{tunneling of vortex rings}. Just as motion of a line defect through a wire creates a ``kink" in some field quantity in 1D, motion of the corresponding 2D ring defect through a film would create a point defect in 2D, inside of which the phase has slipped by one cycle relative to everywhere outside. Coherent tunneling of this kind throughout a very thin film should create a 2D insulating state analogous to what we have discussed here in 1D, and this may have some connection to the so-called ``superinsulating" state suggested in the context of very thin, highly-disordered superconducting films \cite{superins,baturina}.

\section{Conclusion}\label{s:conclusion}

We have described a new alternative to existing theories for quantum phase fluctuations in quasi-1D superconducting wires, built on the hypothesis of flux-charge duality \cite{mooijfluxchg} between these phase fluctuations and the charge fluctuations associated with Josephson tunneling. A crucial aspect of our model is the idea that the electric permittivity due to bound charges both inside and near the wire provides the electrodynamic environment in which quantum phase fluctuations occur. Quantum phase slip can in an abstract sense be viewed as tunneling of ``fluxons" (each carrying flux $\Phi_0$) through the wire, and in our model the permittivity constitutes an effective ``mass" for these objects, whose resulting zero-point ``motion" produces tunneling. In exactly the same way, the kinetic inductance of a superconductor (which arises directly from the finite electron mass) can be viewed as producing the quantum fluctuations responsible for Josephson tunneling. In our model, both QPS and JT arise from zero-point fluctuations of short-wavelength plasma-like oscillations of the Cooper pairs; QPS tends to occur when the impedance of these oscillators and their environment is very high, such that quantum phase fluctuations are only weakly damped and charge tends to be the appropriate well-defined quantum variable; JT on the other hand occurs naturally when the plasma and environment impedances are low, such that charge fluctuations are only weakly damped and phase tends to be the appropriate well-defined quantum variable. This basic model has allowed us to predict the lumped-element phase slip energy $E_{\rm S}$ posited by MN as dual to the Josephson energy \cite{mooijfluxchg}, in terms of measurable physical parameters $\Lambda$, $\epsilon_{\rm in}$, and $\epsilon_{\rm out}$, and one adjustable parameter, the QPS length scale $l_\phi\sim\xi$. Although the latter quantity is an artifact of the discretized form of our model at short length scales, and thus phenomenological in nature, we have been able to use a single, fixed value of $l_\phi=$1.8$\xi$ for all of the comparisons with experiment in this work, with favorable results. In at least some cases our model may suggest qualitatively different conclusions, relative to previous theories, with respect to material parameters favorable for QPS: whereas current experimental efforts are strongly focused on materials relatively close to a metal-insulator transition with extremely high resistances in the normal state (to maximize $R_\xi$), our model would rule out or de-emphasize those which have a very large bound permittivity $\epsilon_{\rm in}$ due to polarizable, localized electronic states which likely appear near such insulating transitions.

Building further on the idea of flux-charge duality, we have constructed a distributed model of quasi-1D wires, dual to the long JJ, which generates $2e$-charged soliton solutions (dual to Josephson vortices) in an infinite wire whose dimensionless admittance $\mathcal{K}\ll1$, and $\Phi_0$-``charged" instanton solutions (dual to Bloch oscillations for short wires) when $\mathcal{K}\gg1$, what we have called ``type II phase slips". A dissipative phase transition at $\mathcal{K}\sim1$ separates these two regimes, which in the short-wire limit is the exact dual of the well-known phase transition for lumped JJs \cite{schon,DPTJJ}. A crucial new element of this distributed model in the context of QPS is the new length scale $\lambda_{\rm E}$, which is dual to the Josephson penetration depth in long JJs. This so-called electric penetration depth determines the size of type II phase slips and their corresponding interaction with each other, and with the circuit environment of a finite wire. Furthermore, the temperature dependence of this length scale provides a mechanism for a richer variety of phenomena in $R$ vs. $T$ measurements than suggested by previous theories, and which can explain a variety of the qualitatively different observations made across multiple materials systems by different research groups. In particular, our model provides an explanation for the observation that qualitative deviations from LAMH temperature scaling of the resistance near $T_{\rm C}$, expected in previous theories to get larger with stronger QPS, in fact appear to get smaller such that the narrowest wires in some cases exhibit the best agreement with simple, thermal LAMH theory with no corrections for quantum fluctuations. Our model also agrees quantitatively with the measurements of so-called ``quantum temperatures" in these narrow wires, previously attributed directly to QPS \cite{bezryNP,aref}. Finally, the involvement of the electric permittivity in our model also provides a very simple and natural mechanism for thermal attempt frequencies of phase-slip processes, in terms of the physics of noise in damped oscillator systems. By contrast, previous theories for such attempt frequencies relied on time-dependent GL theory.

We have compared our model to the results of a new class of experiments in which the quantum phase-slip energy or Coulomb blockade voltage was directly measured at mK temperatures, in InO$_x$ \cite{astafiev}, NbN \cite{astafievNbN}, NbSi \cite{zorinPRB},  and Ti \cite{arutyunovTiQPS} nanowires, and are able to approximately reproduce all four observations with reasonable values for material parameters, and only a single value of the phenomenological parameter ($l_\phi$). By contrast, the GZ theory currently used for most comparisons with experiment evidently requires quite different values of its input parameter $A$ for each material to reproduce the observations. One important reason for this difference is the existence of the additional length scale $\lambda_{\rm E}$ in our model which, as in the $R$ vs. $T$ measurements, results in qualitatively different behavior when $l>\lambda_{\rm E}$. In particular, our model predicts that in this regime the measured blockade voltage should no longer increase with the wire length, as it becomes simply the voltage at which a $2e$-charged soliton (of size $\sim\lambda_{\rm E}$) can enter the wire.

A final topic of some relevance in concluding our work is the relevance of the present model to the prospects for realizing practical QPS devices which are dual to well-known JJ-based circuits, some of which are described in~\ref{a:circuits}, and two of which have already been demonstrated: the phase-slip qubit \cite{astafiev} (dual to the Cooper-pair box), and the phase-slip transistor \cite{zorin} (dual to the DC SQUID). Of particular interest is the prospect of a quantum standard of current dual to the Josephson voltage standard, which would make use of the dual to Shapiro steps \cite{mooijfluxchg,bloch,secondary,hekking}. A device of this kind would have enormous significance to electrical metrology \cite{triangle}, and has been pursued in various forms for many years even before the existence of QPS was contemplated \cite{mooijQPS} and later suggested for this purpose by MN \cite{mooijfluxchg}. Another interesting possibility yet to be discussed is the dual of rapid single-flux quantum digital circuits. This would in principle be a voltage-state logic in which Cooper pairs are shuttled between islands, with no static power dissipation, and possibly a high degree of compatibility with charge-based memory elements.

We can make several qualitative statements about these prospects based on our model. First, we can specify the maximum usable length of a PSJ before non-lumped behavior sets in: the electric penetration length $\lambda_{\rm E}$. Since all of the circuits just mentioned are based on lumped-element behavior, this will constrain how large $E_{\rm S}$ can be. Another interesting observable implication is the dependence of the QPS energy on the permittivity of the dielectric immediately \textit{outside} the wire. This might suggest in some cases a low-permittivity substrate such as glass (or even vacuum if the wire can be suspended) would be preferable to Silicon. Finally, one can show that the quantity $E_{\rm S}/E_{\rm L}$ which determines the extent to which quasicharge can be treated as a classical quantity (dual to $E_{\rm J}/E_{\rm C}$ for a JJ) is simply $Z_{\rm L}/R_{\rm Q}$; that is, all QPS parameters drop out, and only the linear impedance remains. A distributed quasi-1D device with a very large ratio of $Z_{\rm L}/R_{\rm Q}$ has come to be known in the recent literature as a ``superinductor" \cite{masluk,bell}, and is of current interest for a number of quantum superconducting circuit applications.

\ack
We gratefully acknowledge Peter Weichman, Alexey Bezryadin, Jon Fenton, and Franco Nori for helpful discussions and/or comments, and Sergey Tolpygo and Alan Kadin for making the author aware of important references.

This work is sponsored by the Assistant Secretary of Defense for Research \& Engineering under Air Force Contract \#FA8721-05-C-0002.  Opinions, interpretations, conclusions and recommendations are those of the author and are not necessarily endorsed by the United States Government.

\appendix
\newpage
\section{List of selected abbreviations, physical quantities, and variables}\label{a:vars}

\begin{table}
\caption{\label{tab:acro} Selected abbreviations used in the text.}

\footnotesize\begin{tabular}{@{}*{4}{llll}}
\br
   abbreviation  & description & abbreviation & description \\
\mr
    GL  & Ginsburg-Landau &                             MS  & Mooij and Sch\"{o}n  \\
    MN   & Mooij and Nazarov &                    QPS  & Quantum phase slip \\
    JT   & Josephson tunneling &                    JJ  & Josephson junction \\
    MQT   & Macroscopic quantum tunneling &                    PSJ  & Phase slip junction \\
    LAMH   & Langer, Ambegaokar, &                    GZ  & Golubev and Zaikin \\
       & $\;\;\;\;$McCumber, and Halperin &                        SIT   & Superconductor-insulator &\\
    BKT  & Berezinskii, Kosterlitz, &                    & $\;\;\;\;$transition \\
        & and Thouless\\
\br
\end{tabular}

\end{table}

\begin{table}
\caption{\label{tab:vals} Selected quantities used in the text, along with equation numbers.}

\footnotesize\begin{tabular}{@{}*{6}{cclccl}}
\br
    & eq. & description &  & eq. & description \\
\mr
    $\xi$ & & GL coherence length &                             $n$ & & Number of Cooper pairs that have   \\
    $T_{\rm C}$ & & Critical temperature &             &&             $\;\;\;\;$passed through $\Sigma$\\
    $\Delta\phi$  &  & Gauge-invariant phase difference  &      $m$ & & Number of fluxons that have passed\cr
    &  & $\;\;\;\;$across an element  &                         && $\;\;\;\;$ through $\Gamma$\cr
    $\rho_{\rm n}$ &  &  Normal-state resistivity of wire  &          $C_{\rm k}$ & & Kinetic capacitance of a PSJ \cr
    $R_{\rm n}$ &  &  Normal-state resistance of wire &               $\mathcal{C}_{\rm k}$ & & distributed kinetic capacitance (F$\cdot$m) \cr
    $R_{\rm Q}$ &  &  Cooper-pair resistance quantum &                $\mathcal{C}_{\rm k0}$ & & $\mathcal{C}_{\rm k}$ evaluated at $q=0$ \cr
    $L_{\rm J}$ &  & Josephson inductance &                           $C_{\rm J}$ &  & JJ capacitance\\
    $\mathcal{E}_{\rm QPS}$ & \ref{eq:E0} & Ground-state energy per length&       $\mathcal{C}_\perp$ & & distributed shunt capacitance (F/m)\\
    $\mathcal{E}_{\rm S}$ & \ref{eq:ES} & Phase-slip energy per unit length &   $n_{\rm s}$ & & Density of Cooper pairs\\
    $E_{\rm S}$ &  & Phase-slip energy &                              $m_{\rm e}$ & & Electron mass\\
    $V_{\rm C}$ &  & Critical voltage &                               $\Omega_{\rm p}$ & \ref{eq:CPplasma} & Bulk Cooper pair plasma frequency\\
    $E_{\rm C}$ &  & Critical electric field &                        $\tilde\Omega_{\rm p}$ & \ref{eq:S0eqn} & Effective plasma frequency for QPS\\
    $J_{\rm C}$ &  & Depairing current density &                      $l_\phi$ & & Length scale for discrete QPS model\\
    $A_{\rm cs}$ &  & Wire cross-sectional area &                   $\Lambda_{\rm 1D}$ & \ref{eq:lambda1D} & Quasi-1D Coulomb screening length\\
    $l,w,t$ &  & Wire length, width, thickness &                                      $v_{\rm s}$ &  & Mooij-Sch\"{o}n velocity\\
    $Z_{\rm L}$ &  & Linear wire impedance  &                         $\rho_\perp$ &  & Linear charge density on $\mathcal{C}_\perp$\\
    $r_0$ &  & Wire radius in MS model &                        $\Delta\Phi_{\rm J}$ &  & Quasiflux for j$^{\rm th}$ segment\\
    $\Phi_0$ &  & Superconducting flux quantum &                $U(\Delta\Phi)$ & \ref{eq:Uphi} & Potential energy for $\Delta\Phi$\\
    $\lambda$ &  & GL magnetic penetration depth &              $\Lambda_{\rm J}$ &  & Loop charge for j$^{\rm th}$ segment\\
    $Q$ &  & Quasicharge &                                      $C_l$ & \ref{eq:QPScap} & Series capacitance for PSJ segment\\
    $\Phi$ &  & Quasiflux &                                     $\mathcal{L}_{\rm g}$ &  & Geometric inductance of JJ barrier\\
    $\epsilon_{\rm in}$ &  & Electric permittivity due to  &        $\Omega_{\rm ps}$ &  & Phase slip attempt rate  \\
        && $\;\;\;\;$bound charges inside wire&                 $L_\lambda$ & & Type II PS effective inductance\\
    $\epsilon_{\rm out}$ &  & Electric permittivity of insulator &  $V_{\rm 1D}$ &  & Quasi-1D superfluid stiffness\\
        && $\;\;\;\;$outside wire&                              $S_0$ & \ref{eq:S0eqn} & QPS action\\
    $\delta E_{\rm ps}$ &  & Phase-slip energy barrier &            $R_\xi$ &  & Normal-state resistance of length $\xi$\\
    $T_{\rm cr}$ & & Crossover temperature to MQT  &                $u,v$ & \ref{eq:uv} & Euclidean 1+1D coordinates\\
    $I_{\rm b}$ &  & DC bias current &                                $\omega_{\rm p}$ & \ref{eq:wp} & Phase-slip plasma frequency\\
    $I_{\rm sw}$ &  & Switching current&        $\mathcal{K}$ & \ref{eq:K} & Dimensionless plasma admittance\\
    $U_{\rm C}$ &  & Condensation energy density &                    $\lambda_{\rm E}$ & \ref{eq:lE} & Electric penetration length\\
    $\tau_{\rm GL}$ &  & GL relaxation time &          ${\bi j}$ & \ref{eq:j} & Euclidean 1+1D ``current"\\
    $\Delta$ &  & Superconducting energy gap&                   ${\bi q}$ & \ref{eq:d} & 1+1D electric displacement\\
    $T_{\rm Q}$ &  & Quantum temperature &                            ${\bi e}$ & \ref{eq:e} & 1+1D electric field\\
    ${\bi J}_{\rm Q}$ & \ref{eq:JQ} & Quasicharge current density &   $\kappa_{\rm E}$ & \ref{eq:kappa} & 1+1D GL $\kappa$\\
    ${\bi J}_\Phi$ & \ref{eq:Jphi} & Quasiflux current density &$\rho$ &  & 1+1D radial coordinate\\
        && $\;\;\;\;$(total electric field) &                   $\mathcal{S}_{\rm II}$ & \ref{eq:typeIIS} & Type II PS Euclidean action\\
    $\Lambda$ & & London coefficient&                           $\mathcal{S}_{\rm int}$ & \ref{eq:typeIIint} & Type II PS interaction\\
    $L_{\rm k}$ & & Kinetic inductance &                              $\mathcal{S}_{\rm II}^{\rm tot}$ & \ref{eq:typeIItot} & Total type II PS action, \\
    $\rho_{\rm Q}$ & & Free charge density &                          && $\;\;\;\;$with boundary interaction\\
    ${\bi B}_{\rm f}$ & & Free flux density &                         $\delta E_{\rm II}$ & & Classical energy barrier for type II PS\\
    $v_{\rm Q},v_\phi$ & & Free charge and flux velocities&           $R_{\rm env}$ & & High-freq. resistive environment\\
    $C_{\rm sh}$ &  & Lumped shunt capacitance &                    $E_{\rm L}$ & & Inductive energy\\
    &  &  &                                     $\mathcal{L}_{\rm k}$ &  & Kinetic inductance per length\\
\br
\end{tabular}

\end{table}

\section{Thermodynamics of electric flux penetration in 1+1D}\label{a:environment}

Consider the 1+1D electric analog of a magnetic field applied perpendicular to a strongly type II superconducting thin film: a quasi-1D wire (without any external circuit connections) which is subjected to a uniform external \textit{electric} field along its length. In the familiar 2D magnetic case, one has the usual lower critical field $H_{c1}$ below which flux is excluded via the Meissner effect, and above which magnetic vortices enter the sample; the thermodynamics of this transition is governed by the Gibbs free energy:

\begin{equation}
G=F-\int dV{\bi H}_{\rm E}\cdot{\bi B}\label{eq:gibbs}
\end{equation}

\noindent where $F$ is the Helmholtz free energy, ${\bi H}_{\rm E}$ is the external field, and ${\bi B}$ is the actual magnetic flux density. The second term is associated with work done by the field source when flux is excluded from the sample (the overall free energy is lowered when the flux is allowed to penetrate). The condensation energy of the superconductor (contained in $F$) is balanced against this, such that when more free energy is gained by having a uniform superconducting state than the amount of work required from the source were the flux to be expelled, a Meissner state results in which field is excluded from the sample except within a distance from the film edges equal to the so-called ``Pearl length" $\lambda_\perp\approx\lambda^2/2t$ where $t\ll\lambda$ is the film thickness.

It turns out that the additional contribution to the Euclidean action in 1+1D associated with an electric flux source can be written in a completely analogous way:

\begin{equation}
\mathcal{S}_{\rm tot}=\mathcal{S}_{\rm w}-\frac{1}{2\pi\mathcal{K}}\int du\;dv\;{\bi e}\cdot{\bi q}
\end{equation}

\noindent where $\mathcal{S}_{\rm w}$ describes the wire, and the second term describes work done by the source. In a similar manner to eq.~\ref{eq:gibbs}, ${\bi e}$ is the external electric field, and ${\bi q}$ is the resulting electric displacement which contains the system's response to that field. One can get an intuitive feel for the additional work described by the second term in this case by imagining that the external field is produced as shown schematically in fig.~\ref{fig:duality}(d) by a moving source of magnetic flux. In this situation, mechanical work must be done to keep the magnet moving at fixed velocity $v_\phi$ if the wire expels the motional electric field. These considerations imply that external fields below a critical value will be expelled from the wire, except within a spatial distance $\lambda_{\rm E}$ of its ends. Above that critical field, ``lattices" of type II phase slips will occur analogous to magnetic Abrikosov lattices \cite{orlando}, which correspond to a spatially and temporally periodic electric field in the 1+1D case. This analogy also applies to the physics of vortex edge barriers, and in particular to vortex penetration into long, narrow strips \cite{martinisvortex}, which is the 2D case analogous to a finite wire in 1+1D (where the width of the 2D strip is analogous to the length of the wire in our 1+1D case) that we discuss in section~\ref{s:expt}.

\section{Parameters for figure~\ref{fig:RvsT}}\label{a:RvsT}

For all wires we take the single value $l_\phi=1.8\xi$ (which qualitatively produces the best global agreement across all cases considered in this paper), while the rest of the input parameters for each case are shown in table~\ref{tab:RvsT}. The values for $\xi(0)$ are taken from the experimental references, and $\lambda(0)$ are calculated using the BCS relation [eq.~\ref{eq:BCSlambda}] with $a=1$, and $\rho_{\rm n}$ taken from the measured total resistance $R_{\rm n}$ and wire dimensions $A_{\rm cs},l$. The temperature dependence of these quantities was taken from the supplement of ref.~\cite{bezryNP}. The critical temperature $T_{\rm C}$ shown in the table was adjusted to optimize agreement with experiment, and for the In and Al wires, we also adjusted the parameters $R_{\rm env}$ and $C_{\rm sh}$ associated with the electromagnetic environment (for the Ti and MoGe wires these do not enter into our prediction since these cases do not reach the lumped-element limit $\lambda_{\rm E}\gg l$). We took $\epsilon_{\rm in}=5$ for all four cases, which is reasonable for these relatively low-resistivity films. The permittivities $\epsilon_{\rm out}$ describe an effective average experienced by fluctuation electric fields near the wire; for the first three cases we use $\epsilon_{\rm out}\approx (\epsilon_{\rm s}+1)/2$ (where $\epsilon_{\rm s}$ is the substrate permittivity), which is the usual result for a microstrip transmission line with a distant ground plane. We took $\epsilon_{\rm s}=10$ for the Al and Ti wires which were on Si, and $\epsilon_{\rm s}=3$ for the In wire which was on glass. The MoGe wire was deposited on an insulating carbon nanotube suspended in vacuum above its substrate by a distance $\gg l_\phi$. To optimize the agreement with experiment we allowed $\epsilon_{\rm out}=1.5$ (which could plausibly be the case due the effective permittivity of the nanotube). The values for $\mathcal{C}_\perp$ were obtained using Sonnet, a microwave simulation tool, in the first three cases. For the MoGe case, we adjusted $\mathcal{C}_\perp$ upwards from the $15$ fF/m predicted by Sonnet (for a bare, suspended wire) to optimize the agreement; this is again a plausible effect of the nanotube.

\begin{table}
\caption{\label{tab:RvsT} Wire parameters used for figure \ref{fig:RvsT}. In these four cases we took $\epsilon_{\rm in}=5$.}
\begin{indented}
\lineup
\item[]\begin{tabular}{@{}*{11}{l}}
\br
    Wire & ref. & $R_{\rm env}$ & $C_{\rm sh}$ & $\mathcal{C}_\perp$ & $\epsilon_{\rm out}$ & $T_{\rm C}$ & $\sqrt{A_{\rm cs}}$ & $l$ & $\xi(0)$ & $\lambda(0)$\\
    &&[$\Omega$]&[fF]&[pF/m]&  [$\epsilon_0$] &  [K] & [nm] & [$\mu$m] &  [nm] & [$\mu$m]\cr
\mr
    In& \cite{giordano}& 120  & 50 & 25 &  1.5 &  4.2 & 41  & 80   & \040  & 0.15 \cr
    Al& \cite{zgirski}&  \030 & 50 & 48 &  5.5 &  1.5 & 15  & 10   & 100 & 0.21\cr
    Ti& \cite{Tiwires} & -    & -  & 56 &  5.5 &  0.41  & 53  & 20   & \080  & 3.0\cr
    MoGe& \cite{bezryNP} &-    & -  & 25 &  1.5 &  4.0 & \07.5 & \00.11 & \0\05   & 0.71\cr
\br
\end{tabular}
\end{indented}
\end{table}

\section{Parameter values for figure~\ref{fig:bezryresults}}\label{a:moge}

\begin{table}
\caption{\label{tab:moge} MoGe wire parameters used in figs.~\ref{fig:bezryresults}(c)-(d), for wires S1-5 of ref.~\cite{bezryNP} and A-F of ref.~\cite{aref}.}
\begin{indented}
\lineup
\item[]\begin{tabular}{@{}*{8}{l}}
\br
    Wire & $\sqrt{A_{\rm cs}}$& $l$ & $I_{\rm sw}$ & $\Delta$  & $L_{\rm k}$ & $E_{\rm L}$ & $E_{\rm S}$\cr
    &[nm]& [nm] & [$\mu$A]&[meV]& [nH] & [THz]& [GHz]\cr
\mr
    S1 & \08.6 &110& \02.37 & 0.60  & 0.93 & \03.5 & 290 \\
    S2 & \09.3 &195& \01.4 & 0.58  & 1.5 & \02.2 & 260 \\
    S3 & 11.4 &104& \01.42 & 0.49  & 0.62 & \05.2 & \013 \\
    S4 & \09.6 &200& \00.91 & 0.45  & 1.9 & \01.7 & 410 \\
    S5 & 12.2 &120& \04.9 & 0.71  & 0.44 & \07.3 & \0\00.60 \\
    A & 13.4 &115& 10.3 & 0.77  & 0.31 & 10.4 & \0\00.09 \\
    B & 14.6 &221& 11.3 & 0.75  & 0.51 & \06.3 & \0\00.02 \\
    C & 13.5 &100& 12.2 & 0.74  & 0.27 & 12.0 & \0\00.08 \\
    D & 13 &\094& \08.3 & 0.72  & 0.29 & 11.1 & \0\00.22 \\
    E & 11 &\091& \05.3 & 0.69  & 0.41 & \07.88 & \0\06.7 \\
    F & 12.4 &130& \03.8 & 0.49  & 0.63 & \05.13 & \0\06.5 \\
\br
\end{tabular}
\end{indented}
\end{table}

Table~\ref{tab:moge} shows the parameters used to derive the results shown in fig.~\ref{fig:bezryresults} for MoGe wires. In all cases we use the same values $l_\phi=1.8\xi$ with $\xi=5$ nm and $C_{\rm sh}=5$ fF \cite{bezryNP}. The results are insensitive to $C_{\rm sh}$ since the system is overdamped ($R_{\rm env}C_{\rm sh}<\sqrt{L_{\rm k}C_{\rm sh}}$). As before, we infer $L_{\rm k}=\Lambda l/A_{\rm cs}$ using eq.~\ref{eq:BCSlambda} with $a=1$, $\Delta=1.78k_{\rm B}T_{\rm C}$ to obtain $E_{\rm L}\equiv\Phi_0^2/2L_{\rm k}$. Values for $T_{\rm C}$, the wire dimensions, and the switching currents $I_{\rm sw}$ for wires A-F came from the experimental references~\cite{bezryNP,aref}, and the $I_{\rm sw}$ values for wires S1-S5 from ref.~\cite{bezryPC}. The phase-slip energy $E_{\rm S}$ is obtained using eq.~\ref{eq:ES}. For the wires of ref.~\cite{aref}, whose $A_{\rm cs}$ were not published, we infer it from $R_{\rm n}$ and the fixed resistivity $\rho_{\rm n}\approx180\mu\Omega\cdot$cm \cite{bezryPC}. For all wires we use $\epsilon_{\rm in}=5\epsilon_0$, and $\epsilon_{\rm out}=1.5\epsilon_0$, as in table~\ref{tab:RvsT} and fig.~\ref{fig:RvsT}, chosen to optimize agreement with experiment across figs.~\ref{fig:RvsT} and ~\ref{fig:bezryresults}: significantly smaller $\epsilon_{\rm in},\epsilon_{\rm out}$ would degrade the agreement with experiment for wires S1-S5 in fig.~\ref{fig:bezryresults}(d), while larger $\epsilon_{\rm in},\epsilon_{\rm out}$ would degrade the agreement of fig.~\ref{fig:RvsT}(d).

\section{Parameter values for table~\ref{tab:ES}}\label{a:ES}

To produce the values for the four different materials in table~\ref{tab:ES}, in all cases we take $l_\phi=1.8\xi$ and $\epsilon_{\rm out}=5.5\epsilon_0$ (all of these wires were on silicon substrates). All other input parameters are shown in table~\ref{tab:ESparameters}. Wire dimensions, sheet resistance $R_\square$, as well as $\Delta$ and $\xi$ came directly from the experimental references (in some cases using $\Delta=1.78k_{\rm B}T_{\rm C}$). The distributed shunt capacitance $\mathcal{C}_\perp$ was obtained using the Sonnet EM simulation software and the specified experimental geometries. Note that the value for NbN is somewhat larger relative to the other three cases due to the relative proximity of a ground plane in that particular experiment. Values for $\lambda$ were obtained from the BCS relation of eq.~\ref{eq:BCSlambda} with the $a$ values shown in the table.

\begin{table}
\caption{\label{tab:ESparameters} Wire parameters relevant for the comparison of our model with quantum phase slip observations shown in table~\ref{tab:ES}. }
\begin{indented}
\lineup
\item[]\begin{tabular}{@{}*{12}{l}}
\br
    Wire   & ref. &    $w$ & $t$ & $l$ & $R_\square$ & $\mathcal{C}_\perp$ & $a$ & $\lambda$  & $\Delta$&  $\xi$ & $\epsilon_{\rm in}$ \\
    && [nm] & [nm] & [$\mu$m] & [k$\Omega$] & [$\epsilon_0$]   && [$\mu$m] & [meV] &  [nm]&  [$\epsilon_0$]  \\
\mr
    InO$_x$ &\cite{astafiev}& 40 & 35 & 0.4 & 1.7 &  6.3  & 1.8 & 6.6 & 0.41 & 20& 40 \\
    & & & & & &  & 1 & 4.9 & & & 10 \\
    NbN &\cite{astafievNbN}& 30$^a$ & 3 & 0.5 & 2.0 & 8.5   & 4.8 & 1.7 & 1.6 & 4 & 90 \\
    & & & &  &  &   & 1 & 0.79 & & & 1 \\
    Ti  &\cite{arutyunovTiQPS}& 24 & 24 & 20 & 1.1 & 5.8   & 1 & 8.5 & 0.06 & 80& 5  \\
    &  & &  &  & &  &   2 & 12 & & & 5  \\
    NbSi &\cite{zorinPRB}& 20 & 10 & 5 & 0.66 & 5.5 & 1 & 2.5 & 0.18 & 15& 90 \\
    & & & &  & & &  2 & 3.5 & & & 220 \\
\br
\end{tabular}\\
\noindent $^a$ In ref.~\cite{astafievNbN}, the wires for which nonzero $E_{\rm S}$ was observed had an average width ranging from 27-32nm. Also, an appreciable amount of spatial variation of the width was observed along each wire, such that it is possible the measured values are dominated by a ``constriction" much shorter than the total length.
\end{indented}
\end{table}

\section{Flux-charge duality and lumped-element superconducting circuits}\label{a:circuits}

\begin{figure}[t]
\begin{center}
\resizebox{1.0\linewidth}{!}{\includegraphics{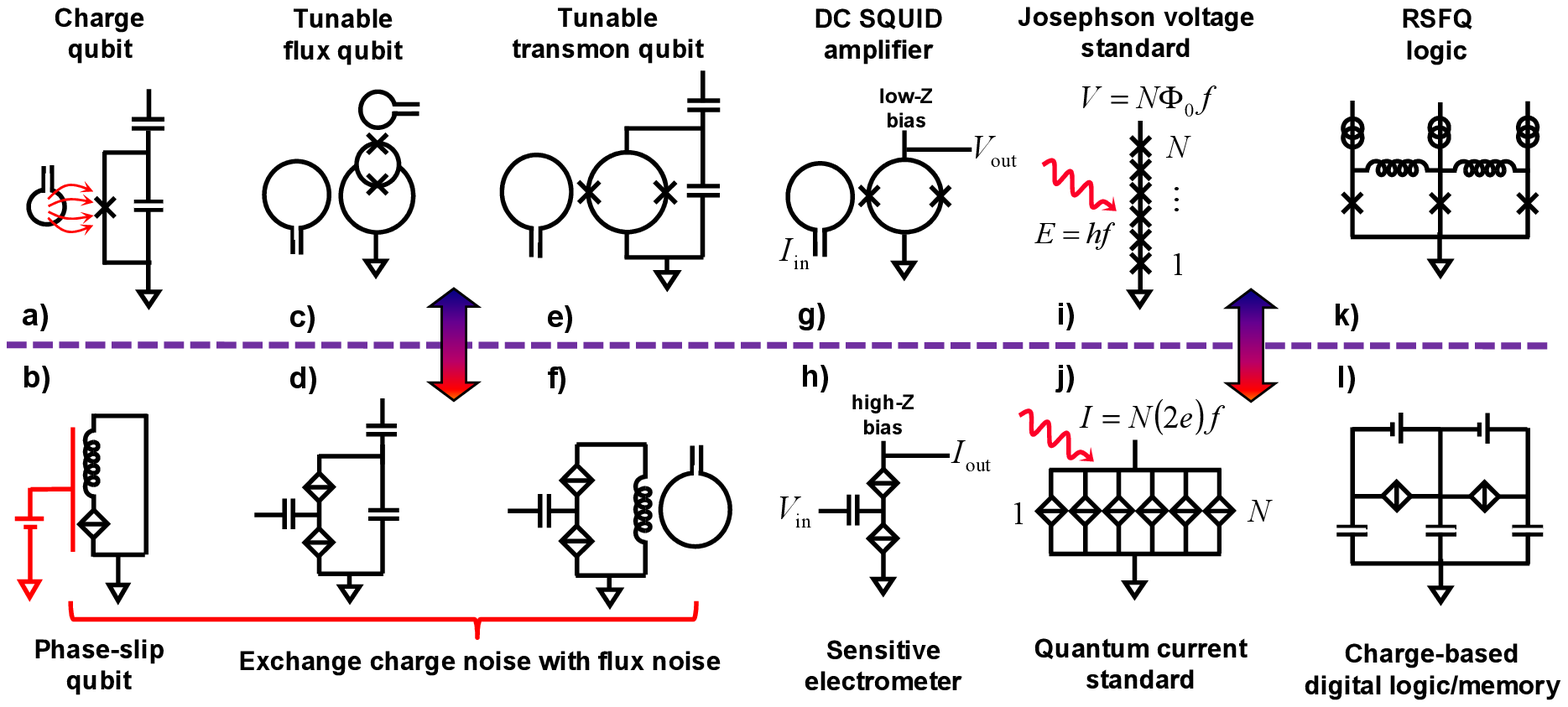}}
\caption{Lumped-element JJ circuits and their duals.}
\label{fig:circuits}
\end{center}
\end{figure}

Figure \ref{fig:circuits} shows specific examples of flux-charge duality applied to more complicated JJ-based circuits. Panels (a) and (b) show the duality between a charge qubit and the phase-slip qubit of ref.~\cite{PSqubit}. PSJ-based superconducting qubits may be of particular interest since flux and charge noise will have their roles interchanged relative to JJ-based qubits. Since the excited-state lifetimes of present-day JJ-based qubits are thought to be limited by high-frequency charge noise, exchanging this for high-frequency flux noise (which is thought to be much weaker \cite{MRFS}) should result in much longer lifetimes. Panels (a) and (b) also illustrate how polarization charge on the nanowire (produced by a nearby gate electrode) is dual to magnetic flux \textit{through} the junction barrier of the JJ. Just as a Fraunhofer interference pattern will be observed in the magnitude of $E_{\rm J}$ vs. flux through the junction (due to the Aharanov-Bohm effect) \cite{orlando}, the same pattern will be observed in the magnitude of $E_{\rm S}$ vs. charge on the nanowire (due to the Aharonov-Casher effect \cite{ABAC}). This may be important for the phase-slip qubit since it implies charge noise on the nanowire would show up as $V_{\rm C}$ noise in the qubit (dual to $I_{\rm C}$ noise commonly observed in JJ-based qubits \cite{ICnoise}). Panels (c)-(f) show two tunable superconducting qubits and their dual circuits. Just as a DC SQUID can be used to implement a flux-tunable composite JJ, the series combination of two PSJs as shown can be used to implement a charge-tunable composite PSJ. Note that (d) is essentially a tunable version of the phase-slip oscillator of Ref. \cite{QPSosc}, and (f) is a tunable version of the phase-slip qubit \cite{PSqubit}.

In addition to qubits, where well-defined, long-lived energy eigenstates are required in which quantum zero-point fluctuations must be kept undisturbed by the environment, the circuits shown in (g)-(l) are intended to function in a regime where either quasiflux (for JJs) or quasicharge (for PSJs) is a \textit{classical} variable (i.e., where quantum fluctuations are small). A well-defined quasiflux requires a low environmental impedance at the Josephson plasma frequency, which is readily obtained using resistively shunted Josephson junctions. A well-defined quasicharge requires a \textit{high} environmental impedance ($\gg R_{\rm Q}$) at the phase-slip plasma frequency, which is much more difficult to realize. In refs.~\cite{zorin,zorinPRB,arutyunovTiQPS}, highly-resistive nanowires were used to bias the device; in ref.~\cite{corlevi}, frustrated DC SQUID arrays in an insulating state were used. Panel (h) shows the ``quantum phase slip transistor" QPST, first suggested in ref.~\cite{QPSdev}, and implemented in ref.~\cite{zorin,zorinPRB}. This device is an electrometer, dual to the DC SQUID amplifier shown in (g). The QPST is similar to a single Cooper-pair transistor (SCPT) \cite{SCPT}; however, it could have a much higher sensitivity than an SCPT, which is limited by the charging energy of the JJs (by how small one can make the junction capacitance). The QPST is instead limited by the kinetic capacitance $C_{\rm k}$, whose ultimate limit is the \textit{series} capacitance of the wires, which can be much smaller. Panel (i) is the Josephson voltage standard, and (j) the quantum current standard proposed in ref. \cite{mooijfluxchg}. Under microwave irradiation, dual features to Shapiro steps would allow locking of the incident frequency $f$ to the applied current $I$ according to $I=Nf2e$, where $N$ is the number of parallel PSJs. Such a device would have enormous impact in electrical metrology, allowing for the first time interconnected fundamental standards of voltage, resistance, and current \cite{triangle}. Finally, panel (k) is a Josephson transmission line, a basic building block of rapid single flux quantum (RSFQ) digital logic; (l) shows the dual to this, in which shunt JJs are replaced by series PSJs, flux stored in loops is replaced by charge stored on islands, and current bias is replaced by voltage bias. Such circuits could be of practical interest, both because unlike RSFQ they have no static power dissipation, and also because voltage-state logic could be significantly easier to integrate with memory elements than flux-state logic.

\section*{References}

\bibliography{QPS_NJP}

\end{document}